\documentclass[aps,prd,reprint,preprintnumbers,superscriptaddress,showpacs,floatfix,nofootinbib]{revtex4-2}
\usepackage[utf8]{inputenc}
\usepackage{parskip}
\usepackage{amssymb}
\usepackage{hhline}
\usepackage{amsmath}
\usepackage{mathtools}
\usepackage{bm}
\usepackage[dvipsnames]{xcolor}
\usepackage{xspace}
\usepackage{multirow,tabularx}
\usepackage{siunitx}
\usepackage{textgreek}
\usepackage{upgreek}
\usepackage{multirow}
\usepackage{graphicx}
\usepackage{pifont}
\usepackage{xstring}
\usepackage{etoolbox}
\usepackage{stmaryrd}
\usepackage{notoccite}
\usepackage{natbib}
\usepackage{mathrsfs}
\usepackage{lineno}
\usepackage{tensor}
\usepackage[thinlines,thicklines]{easybmat}
\usepackage{bbm}
\usepackage{accents}
\DeclareSIUnit\parsec{pc} %define a parsec unit
\DeclareSIUnit\littleh{\mathsf{h}} %define normalised Hubble number
\DeclareSIUnit\ccs{{m_{\text{p}}}^2} %use of reduced Planck mass as unit
\DeclareSIUnit\nothing{\relax} %enable use of dimensionless quantities via siunitx package

\def\prd{\rm{Phys.~Rev.~D}}

\newcommand{\ifeq}{\mathrel{\smash[t]{\stackrel{{\text{(e.g.)}}}{=}}}} %analogue equality

\newrobustcmd{\stall}[1]{%
\IfEqCase{#1}{%
  {1}{State I\xspace}%
  {2}{State II\xspace}%
}[\packageError{cosmicclass}{Unidentified Critical Case: #1}{}]%
}
\newrobustcmd{\numbercosmicclass}{14}%
\newrobustcmd{\cosmicclass}[1]{%
\IfEqCase{#1}{%
{1}{Class~\textsuperscript{3}E\xspace}%
{2}{Class~\textsuperscript{2}A\xspace}%
{8}{Class~\textsuperscript{4}H\xspace}%
{9}{Class~\textsuperscript{5}M\xspace}%
{10}{Class~\textsuperscript{5}M\xspace}%
{11}{Class~\textsuperscript{4}I\xspace}%
{12}{Class~\textsuperscript{4}J\xspace}%
{14}{Class~\textsuperscript{4}H\xspace}%
{13}{Class~\textsuperscript{4}K\xspace}%
{16}{Class~\textsuperscript{3}C\xspace}%
{15}{Class~\textsuperscript{3}D\xspace}%
{21}{Class~\textsuperscript{3}F\xspace}%
{20}{Class~\textsuperscript{3}F\xspace}%
{23}{Class~\textsuperscript{4}L\xspace}%
{22}{Class~\textsuperscript{3}F\xspace}%
{26}{Class~\textsuperscript{4}N\xspace}%
{27}{Class~\textsuperscript{3}E\xspace}%
{24}{Class~\textsuperscript{3}F\xspace}%
{25}{Class~\textsuperscript{3}F\xspace}%
{30}{Class~\textsuperscript{3}E\xspace}%
{31}{Class~\textsuperscript{3}G\xspace}%
{28}{Class~\textsuperscript{4}N\xspace}%
{29}{Class~\textsuperscript{2}A\xspace}%
{35}{Class~\textsuperscript{3}E\xspace}%
{34}{Class~\textsuperscript{3}G\xspace}%
{33}{Class~\textsuperscript{4}N\xspace}%
{32}{Class~\textsuperscript{4}N\xspace}%
{39}{Class~\textsuperscript{3}G\xspace}%
{38}{Class~\textsuperscript{3}G\xspace}%
{37}{Class~\textsuperscript{2}A\xspace}%
{36}{Class~\textsuperscript{2}B\xspace}%
{40}{Class~\textsuperscript{2}B\xspace}%
{41}{Class~\textsuperscript{2}A\xspace}%
{null}{Class~\textsuperscript{3}C*\xspace}%
{cnull}{Class~\textsuperscript{2}A*\xspace}%
}[\packageError{cosmicclass}{Unidentified Critical Case: #1}{}]%
}
\newrobustcmd{\criticalcase}[1]{%
\IfEqCase{#1}{%
{9}{Case~\textsuperscript{*1}9\xspace}%
{10}{Case~\textsuperscript{*3}10\xspace}%
{11}{Case~\textsuperscript{*4}11\xspace}%
{13}{Case~\textsuperscript{*2}13\xspace}%
{2}{Case~2\xspace}%
{16}{Case~16\xspace}%
{3}{Case~3\xspace}%
{17}{Case~17\xspace}%
{20}{Case~20\xspace}%
{26}{Case~\textsuperscript{*6}26\xspace}%
{24}{Case~24\xspace}%
{25}{Case~\textsuperscript{*5}25\xspace}%
{28}{Case~28\xspace}%
{32}{Case~32\xspace}%
}[\packageError{criticalcase}{Unidentified Critical Case: #1}{}]%
}
\newrobustcmd{\planck}{%
  {m_{\text{p}}}%
}

\newrobustcmd{\caligR}{%
  {\mathcal{R}}%
}
\newrobustcmd{\caligT}{%
  {\mathcal{T}}%
}

\newrobustcmd{\pgt}{%
  PGT\textsuperscript{q,+}\ %
}

\newrobustcmd{\ovl}[1]{%
\overline{#1}%
}

\newrobustcmd{\indiq}[2][placeholder]{%
\IfEqCase{#1}{%
  {placeholder}{%
    \IfEqCase{#2}{%
      {1}{\ovl{k}}%
      {2}{\ovl{kl}}%
      {3}{\ovl{klm}}%
    }%
  }%
}[#1]%
}%

\newrobustcmd{\indaq}[2][placeholder]{%
\IfEqCase{#1}{%
  {placeholder}{%
    \IfEqCase{#2}{%
      {1}{\overline{k}}%
      {2}{\overline{kl}}%
      {3}{\overline{klm}}%
    }%
  }%
}[#1]%
}%

\newrobustcmd{\indeq}[2][placeholder]{%
\IfEqCase{#1}{%
  {placeholder}{%
    \IfEqCase{#2}{%
      {1}{k}%
      {2}{kl}%
      {3}{klm}%
    }%
  }%
}[#1]%
}%

\newrobustcmd{\indoq}[2][placeholder]{%
\IfEqCase{#1}{%
  {placeholder}{%
    \IfEqCase{#2}{%
      {1}{\alpha}%
      {2}{\alpha\beta}%
      {3}{\alpha\beta\gamma}%
    }%
  }%
}[#1]%
}%

\newrobustcmd{\covderl}[1]{%
\tensor{\mathcal{D}}{^{\flat}_{\indiq[#1]{1}}}%
}

\newrobustcmd{\deltal}[2]{%
  \tensor*{\delta}{_{\phantom{\flat}}^{\flat}_{#1}^{#2}}%
}

\newrobustcmd{\etau}[1]{%
\tensor{\eta}{^{\indiq[#1]{2}}}%
}
\newrobustcmd{\etaul}[1]{%
\tensor{\eta}{^{\flat}^{\indiq[#1]{2}}}%
}
\newrobustcmd{\etad}[1]{%
\tensor{\eta}{_{\indiq[#1]{2}}}%
}
\newrobustcmd{\etadl}[1]{%
\tensor{\eta}{^{\flat}_{\indiq[#1]{2}}}%
}

\newrobustcmd{\epsul}[1]{%
\tensor{\epsilon}{^{\flat}^{\indiq[#1]{3}}^{\perp}}
}
\newrobustcmd{\epsdl}[1]{%
\tensor{\epsilon}{^{\flat}_{\indiq[#1]{3}}_{\perp}}
}
\newrobustcmd{\epsd}[1]{%
\tensor{\epsilon}{_{\indiq[#1]{3}}_{\perp}}
}
\newrobustcmd{\epsu}[1]{%
\tensor{\epsilon}{^{\indiq[#1]{3}}^{\perp}}
}

\newrobustcmd{\hfl}[2]{%
  \tensor{h}{^{\flat}_{#1}^{#2}}
}

\newrobustcmd{\bet}[1]{%
  \tensor{\hat{\beta}}{_{#1}}
}
\newrobustcmd{\alp}[1]{%
  \tensor{\hat{\alpha}}{_{#1}}
}

\newrobustcmd{\haml}[2][placeholder]{%
\IfEqCase{#2}{%
{mom0p}{\tensor{\mathcal{H}}{^{\flat}_{\perp}}}%
{mom1m}{\tensor{\mathcal{H}}{^{\flat}_{\indoq[#1]{1}}}}%
{rot1p}{\tensor{\mathcal{H}}{^{\flat}_{\indaq[#1]{2}}}}%
{rot1m}{\tensor{\mathcal{H}}{^{\flat}_{\perp}_{\indaq[#1]{1}}}}%
}[\packageError{cosmicclass}{Unidentified Critical Case: #1}{}]%
}

\newrobustcmd{\pic}[2][placeholder]{%
\IfEqCase{#2}{%
{B0p}{\varphi}%
{B1p}{\tensor{\overset{\wedge}{\varphi}}{_{\indiq[#1]{2}}}}%
{B1m}{\tensor{\varphi}{_{\perp\indiq[#1]{1}}}}%
{B2p}{\tensor{\overset{\sim}{\varphi}}{_{\indiq[#1]{2}}}}%
{A0p}{\tensor{\varphi}{_\perp}}%
{A0m}{\tensor[^{\text{P}}]{\varphi}{}}%
{A1p}{\tensor{\overset{\wedge}{\varphi}}{_{\perp\indiq[#1]{2}}}}%
{A1m}{\tensor{\overset{\rightharpoonup}{\varphi}}{_{\indiq[#1]{1}}}}%
{A2p}{\tensor{\overset{\sim}{\varphi}}{_{\perp\indiq[#1]{2}}}}%
{A2m}{\tensor[^{\text{T}}]{\varphi}{_{\indiq[#1]{3}}}}%
}[\packageError{cosmicclass}{Unidentified Critical Case: #1}{}]%
}

\newrobustcmd{\picu}[2][placeholder]{%
\IfEqCase{#2}{%
{B0p}{\varphi}%
{B1p}{\tensor{\smash{\overset{\wedge}{\varphi}}}{^{\indiq[#1]{2}}}}%
{B1m}{\tensor{\varphi}{^{\perp\indiq[#1]{1}}}}%
{B2p}{\tensor{\smash{\overset{\sim}{\varphi}}}{^{\indiq[#1]{2}}}}%
{A0p}{\tensor{\varphi}{_\perp}}%
{A0m}{\tensor[^{\text{P}}]{\varphi}{}}%
{A1p}{\tensor{\smash{\overset{\wedge}{\varphi}}}{^{\perp\indiq[#1]{2}}}}%
{A1m}{\tensor{\smash{\overset{\rightharpoonup}{\varphi}}}{^{\indiq[#1]{1}}}}%
{A2p}{\tensor{\smash{\overset{\sim}{\varphi}}}{^{\perp\indiq[#1]{2}}}}%
{A2m}{\tensor[^{\text{T}}]{\varphi}{^{\indiq[#1]{3}}}}%
}[\packageError{cosmicclass}{Unidentified Critical Case: #1}{}]%
}

\newrobustcmd{\picl}[2][placeholder]{%
\IfEqCase{#2}{%
{B0p}{\tensor{\varphi}{^{\flat}}}%
{B1p}{\tensor{\smash{\overset{\wedge}{\varphi}}}{^{\flat}_{\indiq[#1]{2}}}}%
{B1m}{\tensor{\varphi}{^{\flat}_{\perp}_{\indiq[#1]{1}}}}%
{B2p}{\tensor{\smash{\overset{\sim}{\varphi}}}{^{\flat}_{\indiq[#1]{2}}}}%
{A0p}{\tensor{\varphi}{_\perp}^{\flat}}%
{A0m}{\tensor[^{\text{P}}]{\varphi}{^{\flat}}}%
{A1p}{\tensor{\smash{\overset{\wedge}{\varphi}}}{^{\flat}_{\perp\indiq[#1]{2}}}}%
{A1m}{\tensor{\smash{\overset{\rightharpoonup}{\varphi}}}{^{\flat}_{\indiq[#1]{1}}}}%
{A2p}{\tensor{\smash{\overset{\sim}{\varphi}}}{^{\flat}_{\perp\indiq[#1]{2}}}}%
{A2m}{\tensor[^{\text{T}}]{\varphi}{^{\flat}_{\indiq[#1]{3}}}}%
}[\packageError{cosmicclass}{Unidentified Critical Case: #1}{}]%
}

\newrobustcmd{\mull}[2][placeholder]{%
\IfEqCase{#2}{%
{B0p}{\tensor{u}{^{\flat}}}%
{B1p}{\tensor{\smash{\overset{\wedge}{u}}}{^{\flat}_{\indiq[#1]{2}}}}%
{B1m}{\tensor{u}{^{\flat}_{\perp}_{\indiq[#1]{1}}}}%
{B2p}{\tensor{\smash{\overset{\sim}{u}}}{^{\flat}_{\indiq[#1]{2}}}}%
{A0p}{\tensor{u}{_\perp}^{\flat}}%
{A0m}{\tensor[^{\text{P}}]{u}{^{\flat}}}%
{A1p}{\tensor{\smash{\overset{\wedge}{u}}}{^{\flat}_{\perp\indiq[#1]{2}}}}%
{A1m}{\tensor{\smash{\overset{\rightharpoonup}{u}}}{^{\flat}_{\indiq[#1]{1}}}}%
{A2p}{\tensor{\smash{\overset{\sim}{u}}}{^{\flat}_{\perp\indiq[#1]{2}}}}%
{A2m}{\tensor[^{\text{T}}]{u}{^{\flat}_{\indiq[#1]{3}}}}%
}[\packageError{cosmicclass}{Unidentified Critical Case: #1}{}]%
}

\newrobustcmd{\PiP}[2][placeholder]{%
\IfEqCase{#2}{%
{B0p}{\hat{\pi}}%
{B1p}{\tensor{\overset{\wedge}{\hat{\pi}}}{_{\indiq[#1]{2}}}}%
{B1m}{\tensor{\hat{\pi}}{_{\perp\indiq[#1]{1}}}}%
{B2p}{\tensor{\overset{\sim}{\hat{\pi}}}{_{\indiq[#1]{2}}}}%
{A0p}{\tensor{\hat{\pi}}{_\perp}}%
{A0m}{\tensor[^{\text{P}}]{\hat{\pi}}{}}%
{A1p}{\tensor{\overset{\wedge}{\hat{\pi}}}{_{\perp\indiq[#1]{2}}}}%
{A1m}{\tensor{\overset{\rightharpoonup}{\hat{\pi}}}{_{\indiq[#1]{1}}}}%
{A2p}{\tensor{\overset{\sim}{\hat{\pi}}}{_{\perp\indiq[#1]{2}}}}%
{A2m}{\tensor[^{\text{T}}]{\hat{\pi}}{_{\indiq[#1]{3}}}}%
}[\packageError{cosmicclass}{Unidentified Critical Case: #1}{}]%
}

\newrobustcmd{\PiPu}[2][placeholder]{%
\IfEqCase{#2}{%
{B0p}{\hat{\pi}}%
{B1p}{\tensor{\smash{\overset{\wedge}{\hat{\pi}}}}{^{\indiq[#1]{2}}}}%
{B1m}{\tensor{\smash{\hat{\pi}}}{^{\perp\indiq[#1]{1}}}}%
{B2p}{\tensor{\smash{\overset{\sim}{\hat{\pi}}}}{^{\indiq[#1]{2}}}}%
{A0p}{\tensor{\smash{\hat{\pi}}}{^\perp}}%
{A0m}{\tensor[^{\text{P}}]{\smash{\hat{\pi}}}{}}%
{A1p}{\tensor{\smash{\overset{\wedge}{\hat{\pi}}}}{^{\perp\indiq[#1]{2}}}}%
{A1m}{\tensor{\smash{\overset{\rightharpoonup}{\hat{\pi}}}}{^{\indiq[#1]{1}}}}%
{A2p}{\tensor{\smash{\overset{\sim}{\hat{\pi}}}}{^{\perp\indiq[#1]{2}}}}%
{A2m}{\tensor[^{\text{T}}]{\smash{\hat{\pi}}}{^{\indiq[#1]{3}}}}%
}[\packageError{cosmicclass}{Unidentified Critical Case: #1}{}]%
}

\newrobustcmd{\sicl}[2][placeholder]{%
\IfEqCase{#2}{%
{B0p}{\tensor{\chi}{^{\flat}}}%
{B1p}{\tensor{\smash{\overset{\wedge}{\chi}}}{^{\flat}_{\indiq[#1]{2}}}}%
{B1m}{\tensor{\chi}{^{\flat}_{\perp}_{\indiq[#1]{1}}}}%
{B2p}{\tensor{\smash{\overset{\sim}{\chi}}}{^{\flat}_{\indiq[#1]{2}}}}%
{A0p}{\tensor{\chi}{^{\flat}_\perp}}%
{A0m}{\tensor[^{\text{P}}]{\chi}{^{\flat}}}%
{A1p}{\tensor{\smash{\overset{\wedge}{\chi}}}{^{\flat}_{\perp\indiq[#1]{2}}}}%
{A1m}{\tensor{\smash{\overset{\rightharpoonup}{\chi}}}{^{\flat}_{\indiq[#1]{1}}}}%
{A2p}{\tensor{\smash{\overset{\sim}{\chi}}}{^{\flat}_{\perp\indiq[#1]{2}}}}%
{A2m}{\tensor[^{\text{T}}]{\chi}{^{\flat}_{\indiq[#1]{3}}}}%
}[\packageError{cosmicclass}{Unidentified Critical Case: #1}{}]%
}

\newrobustcmd{\ticl}[2][placeholder]{%
\IfEqCase{#2}{%
{B0p}{\tensor{\zeta}{^{\flat}}}%
{B1p}{\tensor{\smash{\overset{\wedge}{\zeta}}}{^{\flat}_{\indiq[#1]{2}}}}%
{B1m}{\tensor{\zeta}{^{\flat}_{\perp}_{\indiq[#1]{1}}}}%
{B2p}{\tensor{\smash{\overset{\sim}{\zeta}}}{^{\flat}_{\indiq[#1]{2}}}}%
{A0p}{\tensor{\zeta}{^{\flat}_\perp}}%
{A0m}{\tensor[^{\text{P}}]{\zeta}{^{\flat}}}%
{A1p}{\tensor{\smash{\overset{\wedge}{\zeta}}}{^{\flat}_{\perp\indiq[#1]{2}}}}%
{A1m}{\tensor{\smash{\overset{\rightharpoonup}{\zeta}}}{^{\flat}_{\indiq[#1]{1}}}}%
{A2p}{\tensor{\smash{\overset{\sim}{\zeta}}}{^{\flat}_{\perp\indiq[#1]{2}}}}%
{A2m}{\tensor[^{\text{T}}]{\zeta}{^{\flat}_{\indiq[#1]{3}}}}%
}[\packageError{cosmicclass}{Unidentified Critical Case: #1}{}]%
}

\newrobustcmd{\PiPl}[2][placeholder]{%
\IfEqCase{#2}{%
{B0p}{\tensor{\hat{\pi}}{^{\flat}}}%
{B1p}{\tensor{\smash{\overset{\wedge}{\hat{\pi}}}}{^{\flat}_{\indiq[#1]{2}}}}%
{B1m}{\tensor{\hat{\pi}}{^{\flat}_{\perp}_{\indiq[#1]{1}}}}%
{B2p}{\tensor{\smash{\overset{\sim}{\hat{\pi}}}}{^{\flat}_{\indiq[#1]{2}}}}%
{A0p}{\tensor{\hat{\pi}}{_\perp}^{\flat}}%
{A0m}{\tensor[^{\text{P}}]{\hat{\pi}}{^{\flat}}}%
{A1p}{\tensor{\smash{\overset{\wedge}{\hat{\pi}}}}{^{\flat}_{\perp\indiq[#1]{2}}}}%
{A1m}{\tensor{\smash{\overset{\rightharpoonup}{\hat{\pi}}}}{^{\flat}_{\indiq[#1]{1}}}}%
{A2p}{\tensor{\smash{\overset{\sim}{\hat{\pi}}}}{^{\flat}_{\perp\indiq[#1]{2}}}}%
{A2m}{\tensor[^{\text{T}}]{\hat{\pi}}{^{\flat}_{\indiq[#1]{3}}}}%
}[\packageError{cosmicclass}{Unidentified Critical Case: #1}{}]%
}

\newrobustcmd{\sic}[2][placeholder]{%
\IfEqCase{#2}{%
{B0p}{\chi}%
{B1p}{\tensor{\overset{\wedge}{\chi}}{_{\indiq[#1]{2}}}}%
{B1m}{\tensor{\chi}{_{\perp\indiq[#1]{1}}}}%
{B2p}{\tensor{\overset{\sim}{\chi}}{_{\indiq[#1]{2}}}}%
{A0p}{\tensor{\chi}{_\perp}}%
{A0m}{\tensor[^{\text{P}}]{\chi}{}}%
{A1p}{\tensor{\overset{\wedge}{\chi}}{_{\perp\indiq[#1]{2}}}}%
{A1m}{\tensor{\overset{\rightharpoonup}{\chi}}{_{\indiq[#1]{1}}}}%
{A2p}{\tensor{\overset{\sim}{\chi}}{_{\perp\indiq[#1]{2}}}}%
{A2m}{\tensor[^{\text{T}}]{\chi}{_{\indiq[#1]{3}}}}%
}[\packageError{cosmicclass}{Unidentified Critical Case: #1}{}]%
}

\newrobustcmd{\Tl}[2][placeholder]{%
\IfEqCase{#2}{%
{B0p}{\tensor{\chi}{^{\flat}}}%
{B1p}{\tensor{\smash{\overset{\wedge}{\chi}}}{^{\flat}_{\indiq[#1]{2}}}}%
{B1m}{\tensor{\chi}{^{\flat}_{\perp}_{\indiq[#1]{1}}}}%
{B2p}{\tensor{\smash{\overset{\sim}{\chi}}}{^{\flat}_{\indiq[#1]{2}}}}%
{A0p}{\tensor{\chi}{^{\flat}_\perp}}%
{A0m}{\tensor[^{\text{P}}]{\mathcal{T}}{^{\flat}}}%
{A1p}{\tensor{\smash{\overset{\wedge}{\chi}}}{^{\flat}_{\perp\indiq[#1]{2}}}}%
{A1m}{\tensor{\smash{\overset{\rightharpoonup}{\mathcal{T}}}}{^{\flat}_{\indiq[#1]{1}}}}%
{A2p}{\tensor{\smash{\overset{\sim}{\chi}}}{^{\flat}_{\perp\indiq[#1]{2}}}}%
{A2m}{\tensor[^{\text{T}}]{\mathcal{T}}{^{\flat}_{\indiq[#1]{3}}}}%
}[\tensor{\mathcal{T}}{^{\flat}_{\indiq[#1]{3}}}]%
}

\newrobustcmd{\cT}[2][placeholder]{%
\IfEqCase{#2}{%
{B1p}{\tensor{\mathcal{T}}{_{\perp\indiq[#1]{2}}}}%
{B1m}{\tensor{\overset{\rightharpoonup}{\mathcal{T}}}{_{\indiq[#1]{1}}}}%
{A0m}{\tensor[^{\text{P}}]{\mathcal{T}}{}}%
{A2m}{\tensor[^{\text{T}}]{\mathcal{T}}{_{\indiq[#1]{3}}}}%
}[\packageError{cosmicclass}{Unidentified Critical Case: #1}{}]%
}

\newrobustcmd{\cTl}[2][placeholder]{%
\IfEqCase{#2}{%
{B1p}{\tensor{\mathcal{T}}{^{\flat}_{\perp\indiq[#1]{2}}}}%
{B1m}{\tensor{\overset{\rightharpoonup}{\mathcal{T}}}{^{\flat}_{\indiq[#1]{1}}}}%
{A0m}{\tensor[^{\text{P}}]{\mathcal{T}}{^{\flat}}}%
{A2m}{\tensor[^{\text{T}}]{\mathcal{T}}{^{\flat}_{\indiq[#1]{3}}}}%
}[\packageError{cosmicclass}{Unidentified Critical Case: #1}{}]%
}

\newrobustcmd{\cTu}[2][placeholder]{%
\IfEqCase{#2}{%
{B1p}{\tensor{\mathcal{T}}{^{\perp\indiq[#1]{2}}}}%
{B1m}{\tensor{\smash{\overset{\rightharpoonup}{\mathcal{T}}}}{^{\indiq[#1]{1}}}}%
{A0m}{\tensor[^{\text{P}}]{\mathcal{T}}{}}%
{A2m}{\tensor[^{\text{T}}]{\mathcal{T}}{^{\indiq[#1]{3}}}}%
}[\packageError{cosmicclass}{Unidentified Critical Case: #1}{}]%
}

\newrobustcmd{\ncT}[2][placeholder]{%
\IfEqCase{#2}{%
{B0p}{\tensor{\mathcal{T}}{_{\indiq[#1]{2}\perp}}}%
{B1p}{\tensor{\mathcal{T}}{_{[\indiq[#1]{2}]\perp}}}%
{B1m}{\tensor{\mathcal{T}}{_{\perp\indiq[#1]{1}\perp}}}%
{B2p}{\tensor{\mathcal{T}}{_{\langle\indiq[#1]{2}\rangle\perp}}}%
}[\packageError{cosmicclass}{Unidentified Critical Case: #1}{}]%
}

\newrobustcmd{\cR}[2][placeholder]{%
\IfEqCase{#2}{%
{A0p}{\tensor{\underline{\mathcal{R}}}{}}%
{A0m}{\tensor[^{\text{P}}]{\mathcal{R}}{_{\perp\circ}}}%
{A1p}{\tensor{\underline{\mathcal{R}}}{_{[\indiq[#1]{2}]}}}%
{A1m}{\tensor{\mathcal{R}}{_{\perp\indiq[#1]{1}}}}%
{A2p}{\tensor{\underline{\mathcal{R}}}{_{\langle\indiq[#1]{2}\rangle}}}%
{A2m}{\tensor[^{\text{T}}]{\mathcal{R}}{_{\perp\indiq[#1]{3}}}}%
}[\packageError{cosmicclass}{Unidentified Critical Case: #1}{}]%
}

\newrobustcmd{\cRl}[2][placeholder]{%
\IfEqCase{#2}{%
  {A0p}{\tensor{\underline{\mathcal{R}}}{^{\flat}}}%
{A0m}{\tensor[^{\text{P}}]{\mathcal{R}}{^{\flat}_{\perp\circ}}}%
{A1p}{\tensor{\underline{\mathcal{R}}}{^{\flat}_{[\indiq[#1]{2}]}}}%
{A1m}{\tensor{\mathcal{R}}{^{\flat}_{\perp\indiq[#1]{1}}}}%
{A2p}{\tensor{\underline{\mathcal{R}}}{^{\flat}_{\langle\indiq[#1]{2}\rangle}}}%
{A2m}{\tensor[^{\text{T}}]{\mathcal{R}}{^{\flat}_{\perp\indiq[#1]{3}}}}%
}[\packageError{cosmicclass}{Unidentified Critical Case: #1}{}]%
}

\newrobustcmd{\cRu}[2][placeholder]{%
\IfEqCase{#2}{%
{A0p}{\tensor{\underline{\mathcal{R}}}{}}%
{A0m}{\tensor[^{\text{P}}]{\mathcal{R}}{_{\perp\circ}}}%
{A1p}{\tensor{\underline{\mathcal{R}}}{^{[\indiq[#1]{2}]}}}%
{A1m}{\tensor{\mathcal{R}}{^{\perp\indiq[#1]{1}}}}%
{A2p}{\tensor{\underline{\mathcal{R}}}{^{\langle\indiq[#1]{2}\rangle}}}%
{A2m}{\tensor[^{\text{T}}]{\mathcal{R}}{^{\perp\indiq[#1]{3}}}}%
}[\packageError{cosmicclass}{Unidentified Critical Case: #1}{}]%
}

\newrobustcmd{\ncR}[2][placeholder]{%
\IfEqCase{#2}{%
  {A0p}{\tensor{\mathcal{R}}{_{\perp\perp}}}%
{A0m}{\tensor[^{\text{P}}]{\mathcal{R}}{_{\circ\perp}}}%
{A1p}{\tensor{\mathcal{R}}{_{\perp[\indiq[#1]{2}]\perp}}}%
{A1m}{\tensor{\mathcal{R}}{_{\indiq[#1]{1}\perp}}}%
{A2p}{\tensor{\mathcal{R}}{_{\perp\langle\indiq[#1]{2}\rangle\perp}}}%
{A2m}{\tensor[^{\text{T}}]{\mathcal{R}}{_{\indiq[#1]{3}\perp}}}%
}[\packageError{cosmicclass}{Unidentified Critical Case: #1}{}]%
}

\newcommand{\BPiP}{\hat{\pi}}
\usepackage{hyperref}
\hypersetup{%
     colorlinks = true,%
     linkcolor = Blue,%
     citecolor = Blue,%
     filecolor = Blue,%
     urlcolor = Blue% 
     }%
\usepackage[capitalize]{cleveref} %always load this last in preamble

\begin{document}
%\setpagewiselikenumbers
%\modulolinenumbers[5]
\preprint{Prepared for submission to Phys. Rev. D.}

\title{Nonlinear Hamiltonian analysis of new quadratic torsion theories\\
I. Cases with curvature-free constraints}

\author{W.E.V. Barker}
\email{wb263@cam.ac.uk}
\affiliation{Astrophysics Group, Cavendish Laboratory, JJ Thomson Avenue, Cambridge CB3 0HE, UK}
\affiliation{Kavli Institute for Cosmology, Madingley Road, Cambridge CB3 0HA, UK}
\author{A.N. Lasenby}
\email{a.n.lasenby@mrao.cam.ac.uk}
\affiliation{Astrophysics Group, Cavendish Laboratory, JJ Thomson Avenue, Cambridge CB3 0HE, UK}
\affiliation{Kavli Institute for Cosmology, Madingley Road, Cambridge CB3 0HA, UK}
\author{M.P. Hobson}
\email{mph@mrao.cam.ac.uk}
\affiliation{Astrophysics Group, Cavendish Laboratory, JJ Thomson Avenue, Cambridge CB3 0HE, UK}
\author{W.J. Handley}
\email{wh260@cam.ac.uk}
\affiliation{Astrophysics Group, Cavendish Laboratory, JJ Thomson Avenue, Cambridge CB3 0HE, UK}
\affiliation{Kavli Institute for Cosmology, Madingley Road, Cambridge CB3 0HA, UK}

%\date{}

\begin{abstract}
  It was recently found that, when linearised in the absence of matter, 58 cases of the general gravitational theory with quadratic curvature and torsion are (i) free from ghosts and tachyons and (ii) power-counting renormalisable. We inspect the nonlinear Hamiltonian structure of the eight cases whose primary constraints do not depend on the curvature tensor. 
  We confirm the particle spectra and unitarity of all these theories in the linear regime. We uncover qualitative dynamical changes in the nonlinear regimes of all eight cases, suggesting at least a broken gauge symmetry, and possibly the activation of negative kinetic energy spin-parity sectors and acausal behaviour. Two of the cases propagate a pair of massless modes at the linear level, and were interesting as candidate theories of gravity. However, we identify these modes with vector excitations, rather than the tensor polarisations of the graviton. Moreover, we show that these theories do not support a viable cosmological background.
\end{abstract}


\maketitle

\section{Introduction}
In light of both theoretical minimalism, and experimental and observational verification, the preferred effective theory of gravity is that of Einstein and Hilbert
\begin{equation}
  L_{\text{T}}=-\frac{1}{2}{m_{\text{p}}}^2 R+L_{\text{M}}.
  \label{GR}
\end{equation}
The gravitational portion ${L_\text{G}\equiv L_\text{T}-L_\text{M}}$ of the total Lagrangian $L_\text{T}$ is powered by the scalar part of the Riemann curvature tensor $R\equiv\tensor{R}{^{\mu\nu}_{\mu\nu}}$, which is the de facto gravitational field strength and contains \emph{second} derivatives of the metric gravitational potential ${R\sim\partial^2 g+(\partial g)^2}$. The matter Lagrangian $L_\text{M}$ is taken to be minimally coupled.

Two approaches to generalising \eqref{GR} have proven especially popular
\begin{enumerate}
  \item The artificially imposed symmetry of the Levi--Civita connection could be relaxed. 
  \item Higher-order geometric invariants could be added to the Lagrangian. 
\end{enumerate}
The first approach leads to a non-vanishing torsion $\tensor{\mathcal{  T}}{^i_{jk}}$, and corresponding non-Riemann curvature $\tensor{\mathcal{  R}}{^i_{jkl}}$. The Roman indices refer to a local Lorentz basis, mediated by \emph{tetrads} (vierbein), or equivalent translational gauge fields $\tensor{b}{^i_\mu}$. The now independent \emph{spin connection} may likewise be cast as a rotational gauge field $\tensor{A}{^{ij}_\mu}$. In terms of these new potentials, the gravitational field strengths ${\mathcal{  T}\sim\partial b+bA}$ and ${\mathcal{  R}\sim\partial A+A^2}$ are closer to the Yang--Mills form familiar from the strong and electroweak sectors of the standard model: they are linear in \emph{first} derivatives and the structure constants of the Poincar{\'e} group.
By demanding positive parity and freedom from Ostrogradsky ghosts\footnote{Note that Ostrogradsky's theorem forbids all terms quadratic in the second-order Riemann tensor except for the Gauss--Bonnet term; this does not apply to the first-order Riemann--Cartan or torsion tensors.} in combination with the second approach, one arrives at the general quadratic $L_\text{G}\sim\planck^2\mathcal{  R}+\mathcal{  R}^2+\planck^2\mathcal{  T}^2$ theory 
\begin{equation}
  \begin{aligned}
    L_{\text{T}}& = -\frac{1}{2}\alpha_0 {m_\text{p}}^2\mathcal{  R}+{m_\text{p}}^2\tensor{\mathcal{  T}}{_{ijk}}\big(\beta_1\tensor{\mathcal{  T}}{^{ijk}}
    +\beta_2\tensor{\mathcal{  T}}{^{jik}}\big)\\
      &+\beta_3{m_\text{p}}^2\tensor{\mathcal{  T}}{_{i}}\tensor{\mathcal{  T}}{^{i}}+
      \alpha_1\mathcal{  R}^2+\tensor{\mathcal{  R}}{_{ij}}\big(\alpha_2\tensor{\mathcal{  R}}{^{ij}}+\alpha_3\tensor{\mathcal{  R}}{^{ji}}\big)\\
      &+\tensor{\mathcal{  R}}{_{ijkl}}\big(\alpha_4\tensor{\mathcal{  R}}{^{ijkl}}+\alpha_5\tensor{\mathcal{  R}}{^{ikjl}}+\alpha_6\tensor{\mathcal{  R}}{^{klij}}\big)+L_{\text{M}},
  \label{lagrangian_soft}
  \end{aligned}
\end{equation}
where $\tensor{\mathcal{  R}}{_{ij}}\equiv\tensor{\mathcal{  R}}{^{l}_{ilk}}$, $\mathcal{  R}\equiv\tensor{\mathcal{  R}}{^l_l}$ and $\tensor{\mathcal{  T}}{_i}\equiv\tensor{\mathcal{  T}}{^l_{il}}$.
These quadratic terms are added to the scalar curvature invariant for a total of ten dimensionless couplings in the theory. Note that no scalar invariant can be formed from the torsion.

The theory \eqref{lagrangian_soft} has been deeply studied over four decades.
When linearised on a Minkowski background, the theory is capable of propagating six massive torsion modes (\emph{rotons} or \emph{tordions}) of spin-parity ${J^P=0^{\pm}, 1^{\pm}, 2^{\pm}}$, in addition to the $2^+$ mode of the massless graviton~\cite{10.1143/PTP.64.2222}. 
Following early studies by Neville~\cite{1978PhRvD..18.3535N,1980PhRvD..21..867N}, Sezgin and van Nieuwenhuizen found $12$ cases of the theory whose propagator poles have positive residues and real masses, i.e. unitary theories~\cite{1980PhRvD..21.3269S,1981PhRvD..24.1677S}. An exhaustive survey by~\citet*{2019PhRvD..99f4001L,2020PhRvD.101f4038L} recently found that there are $450$ unitary cases in total. Of these, $58$ are also power-counting renormalisable (PCR), such that the graviton and roton propagators tend to $p^{-4}$ and $p^{-2}$ respectively in the ultraviolet limit\footnote{With the exception of \criticalcase{9}, \criticalcase{10}, \criticalcase{11} and \criticalcase{13} as labelled in~\cite{2020PhRvD.101f4038L}, $J^P$ sectors propagate which violate these rules. However, these `bad' modes are understood to decouple at high energies without producing divergent loops~\cite{2020PhRvD.101f4038L}.}.

Perhaps surprisingly, the restriction from unitary to PCR-unitary cases of \eqref{lagrangian_soft} universally switches off the Einstein--Hilbert term
\begin{equation}
  \alpha_0=0.
  \label{bizarre}
\end{equation}
In the context of \eqref{lagrangian_soft}, this term in isolation constitutes the Einstein--Cartan theory of gravity. Einstein--Cartan theory is dynamically equivalent to GR when the spin tensor of the matter sector vanishes; the linearised theory contains only the $2^+$ graviton. Usually, the quadratic $\mathcal{  R}^2+\planck^2\mathcal{  T}^2$ terms are viewed as corrections to the scalar $\planck^2\mathcal{  R}$, which may be motivated by analogy to Einstein's theory at one loop, ${L_\text{G}\sim \planck^2R+R^2}$. A pure ${L_\text{G}\sim\mathcal{  R}^2+\planck^2\mathcal{  T}^2}$ theory is hard to reconcile with this picture, and necessitates a great deal of work since one cannot appeal to the viable Einstein--Cartan limit at low energies. In some sense, a purely quadratic gravity is actually quite natural.
Einstein gravity amounts to a $\mathbb{R}^{1,3}$ gauge theory of diffeomorphisms, whose gauge potential is $\tensor{g}{_{\mu\nu}}$. However, the fields $\tensor{b}{^i_\mu}$ and $\tensor{A}{^{ij}_\mu}$ additionally gauge rotations and, by extension, the whole Poincar{\'e} group ${\mathbb{R}^{1,3}\rtimes\mathrm{SO}^+(1,3)}$. The theory \eqref{lagrangian_soft} is more properly known as the \emph{quadraic, parity-preserving Poincar{\'e} gauge theory} (PGT\textsuperscript{q+}) of gravity, as pioneered by Kibble~\cite{1961JMP.....2..212K}, Utiyama~\cite{PhysRev.101.1597}, Sciama~\cite{RevModPhys.36.463} and others\footnote{For an excellent series on the dynamical structure of PGT\textsuperscript{q+} from the Lagrangian perspective, see \cite{10.1143/PTP.64.866,10.1143/PTP.64.1435,10.1143/PTP.64.2222,10.1143/PTP.65.525}.}. In this context, a Lagrangian quadratic in Yang--Mills field strengths would make an appealing addition to the standard model -- should it prove viable in the nonlinear regime.

The purpose of this series is to test the nonlinear viability of the 58 novel theories by probing their Hamiltonian structure. As a higher-spin gauge theory, the PGT\textsuperscript{q+} \eqref{lagrangian_soft} is always \emph{singular}: this degeneracy of the kinetic Hessian greatly complicates the Lagrangian analysis, incentivising the Hamiltonian approach. By implementing the algorithm of Dirac and Bergmann, we are guaranteed to obtain all propagating degrees of freedom (D.o.F), along with all constraints~\cite{Henneaux:1992ig}. In the linearised theory, this is especially easy, and allows us to verify the particle spectra and unitarity of the cases obtained in~\cite{2019PhRvD..99f4001L,2020PhRvD.101f4038L}. In the nonlinear case, the algorithm allows us to flag potentially fatal pathologies which develop under significant departures from Minkowski spacetime -- if this spacetime is taken to be a vacuum, then the nonlinear regime is equivalent to that of strong fields. 
In particular, we rely on the simple `health indicator' of modified gravity set out by Chen, Nester and Yo: \emph{the number and type of constraints should not change in passing from the linear to nonlinear regimes}~\cite{1998AcPPB..29..961C,2002IJMPD..11..747Y}. The motivation for this criterion is twofold. Generically, a decrease in the number of constraints involves the activation of potentially \emph{ghostly} fields~\cite{2002IJMPD..11..747Y}. Moreover, it may be that the nonlinear constraint structure is itself field dependent: this is thought to be associated with the propagation of \emph{acausal} degrees of freedom~\cite{1998AcPPB..29..961C}. Neither of these qualities is necessarily fatal unless shown to incur a physical ghostly or acausal D.o.F, but for the purposes of this particular study we will take the avoidance of them as being desirable.

In this paper we will test \criticalcase{3}, \criticalcase{17}, \criticalcase{20}, \criticalcase{24}, \criticalcase{25}, \criticalcase{26}, \criticalcase{28} and \criticalcase{32}, using the numbering of~\cite{2020PhRvD.101f4038L}, with the numbering of cases previously discovered in~\cite{2019PhRvD..99f4001L} indicated by (*).
These eight cases are most conducive to the Hamiltonian analysis. Specifically, these are the only cases whose primary constraints are not functions of the curvature. To our knowledge, this practical restriction does no more than to ease the evaluation of commutators. 
We therefore tentatively view the eight cases to be an \emph{representative sample} of the 58 novel theories\footnote{We mention that \emph{none} of the eight cases are PCR in the conventional sense of~\cite{1980PhRvD..21.3269S}, i.e. all of them feature a $J^P$ propagator whose momentum power is non-PCR in the IR, and which decouples in the UV.}.

All eight cases fail the prescribed strong-field tests. In some sense, they do so more dramatically than those `minimal' cases of PGT\textsuperscript{q+} which were previously tested, due to the vanishing of mass parameters~\cite{2002IJMPD..11..747Y}. Based on these results, we find no evidence that the simultaneous imposition of the weak-field PCR and unitarity criteria remedy the questionable health of PGT\textsuperscript{q+} in the strong-field regime, as observed in~\cite{1998AcPPB..29..961C,1999IJMPD...8..459Y,2002IJMPD..11..747Y}. If these findings turn out to be general, it would seem more efficient to perform future surveys of PGT\textsuperscript{q+} in the strong-field regime \emph{from the outset}.

We are also able to rule the cases out on cosmological grounds, using the scalar-tensor analogue theory which replicates the background cosmology of the general ten-parameter PGT\textsuperscript{q+}~\cite{2020arXiv200603581B}. Out of the eight cases, only \criticalcase{3} and \criticalcase{17} propagate massless modes consistent with long-range gravitational forces, yet their nonlinear cosmological equations are non-dynamical. However, we do show that these cases are the degenerate limit of an otherwise viable and interesting class of torsion theories obtained by imposing two very simple constraints on the couplings of \eqref{lagrangian_soft}, whose background cosmology perfectly replicates that of Einstein's torsion-free gravity \eqref{GR}, conformally coupled to a scalar inflaton $\xi$
\begin{equation}
  \tensor{L}{_{\text{T}}}= -\frac{1}{2}{m_{\text{p}}}^2 R+\frac{1}{12}\xi^2 R+\tensor{X}{^{\xi\xi}}-\frac{1}{2}{m_\xi}^2\xi^2+L_{\text{M}}.
  \label{mafin}
\end{equation}
Here, the inflaton has kinetic term ${\tensor{X}{^{\xi\xi}}\equiv\tfrac{1}{2}\tensor{\nabla}{_\mu}\xi\tensor{\nabla}{^\mu}\xi}$ and mass $m_\xi$. The cosmology resulting from \eqref{mafin} is not scale-invariant due to the mass term, which is fortunate for minimal coupling to cosmological matter. However, it is an interesting surprise that the non-minimal coupling should be exactly scale-invariant.
The failure of \criticalcase{3} and \criticalcase{17} certainly is \emph{not} a necessary consequence of the linearised unitarity and power-counting. Indeed, one of the 58 cases has an \emph{excellent} cosmological background~\cite{2020arXiv200302690B,2020arXiv200603581B}, though an analysis of its Hamiltonian structure is deferred to the companion paper, since its primary constraints depend on curvature.

Despite our concerns about the strong-field regime, we are able to confirm the weak-field unitarity of all eight cases. We also obtain linearised dynamics which are consistent with the particle spectra found in~\cite{2019PhRvD..99f4001L,2020PhRvD.101f4038L}. We also offer tighter bounds on the massless particle spectra, identifying the massless modes of \criticalcase{3} and \criticalcase{17} as \emph{vector} excitations, rather than the expected \emph{tensor} polarisations of the graviton.

The remainder of this paper is set out as follows. In \cref{constrained_Hamiltonian} we develop the Hamiltonian formulation of the ten-parameter theory \eqref{lagrangian_soft}. In \cref{massive_only,massless_theories} we apply the Dirac--Bergmann algorithm to each of the linearised cases, and compare with the constraint structure of the nonlinear theories. In \cref{phenomenology} we use efficient methods to show that even the cases with massless modes cannot support any Friedmann-like cosmological equation.
Conclusions follow in \cref{conclusions}.
Following the conventions of~\cite{blagojevic2002gravitation} we will use Roman and Greek indices from the middle of the alphabet $i$, $j$\ldots\ $\mu$, $\nu$\ldots\ to refer to general Lorentz and coordinate indices running from $0$ to three, while $a$, $b$\ldots\ and $\alpha$, $\beta$\ldots\ strictly run from one to three. We use the `West Coast' signature $(+,-,-,-)$. Our potentially nonstandard acronyms are detailed in \cref{acro}.

\section{Constrained Hamiltonian}\label{constrained_Hamiltonian}

\begin{table}
  \caption{\label{acro} Nonstandard abbreviations.}
\begin{center}
\begin{tabularx}{\linewidth}{c|X}
\hline\hline
PGT\textsuperscript{q+} & quadratic, parity-preserving Poincar{\'e} gauge theory\\
PCR & power-counting renormalisable\\
D.o.F & degrees of freedom \\
PPM & primary Poisson matrix \\
(P/S/T)iC & (primary/secondary/tertiary) if-constraint \\
(F/S)C & (first/second)-class \\
i(P/S/T)(F/S)C & (P/S/T)iC which is (F/S)C on the final shell \\
s(P/S)FC & sure (primary/secondary) constraint, always FC \\
\hline\hline
\end{tabularx}
\end{center}
\end{table}

\DeclareRobustCommand{\particle}[1]{%
  \begingroup\normalfont
  \includegraphics[height=\fontcharht\font`\B]{#1}%
  \endgroup
}
\begin {table*}[htp]
  \caption{\label{table-1} From the 58 unitary, power-counting renormalisable cases of \eqref{lagrangian_soft}, we consider the eight cases whose primary constraints do not depend on the Riemann--Cartan curvature. The numbering follows~\cite{2020PhRvD.101f4038L}, with original numbering of cases first identified in~\cite{2019PhRvD..99f4001L} indicated by (*). The definitive constraints on the couplings are given, along with extra conditions which fix the unitarity, separated by a caret ($\wedge$). The methods of~\cite{2019PhRvD..99f4001L,2020PhRvD.101f4038L} provide partial information about the particle spectrum. Propagators may be of $\tensor{A}{^{ij}_\mu}$ (blue), or the antisymmetric (green) or symmetric (red) parts of $\tensor{b}{^i_\mu}$. Propagator poles are massless (empty circles) or massive (filled circles). Within each $J^P$ sector, $\tensor{A}{^{ij}_\mu}$ and $\tensor{b}{^i_\mu}$ modes may be coupled (multiple colours) or transmuted by gauge transformations (multiple circles). Ultimately, the propagator spectrum is only indicative of the particle spectrum: the actual number of propagating degrees of freedom are shown in the final column, but their field character or $J^P$ is \emph{indeterminate} in the massless case, because poles from multiple $J^P$ sectors coincide at the origin of momentum-space.}
\begin{center}
  \begin{tabularx}{\textwidth}{l|l|l|c!{\vrule width 0.1pt}c!{\vrule width 0.1pt}c!{\vrule width 0.1pt}c!{\vrule width 0.1pt}c!{\vrule width 0.1pt}c|X} 
\hline\hline
\# & criticality equalities & unitary inequalities & $0^-$ & $0^+$ & $1^-$ & $1^+$ & $2^-$ & $2^+$ & D.o.F\\
\hline
\criticalcase{3}&${{\alp{1}= \alp{2}= \alp{4}= \alp{6}= \bet{2}= \bet{1}+2\bet{3}}=0}$&${{\alp{3} < 0\wedge \alp{5} < 0\wedge \bet{1} < 0}}$& \includegraphics[width=0.5cm]{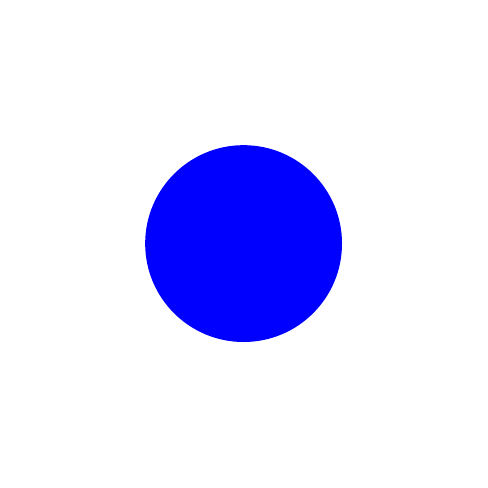}& \includegraphics[width=0.5cm]{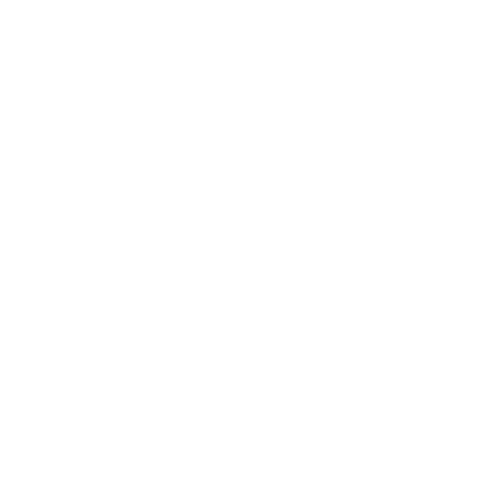}& \includegraphics[width=0.5cm]{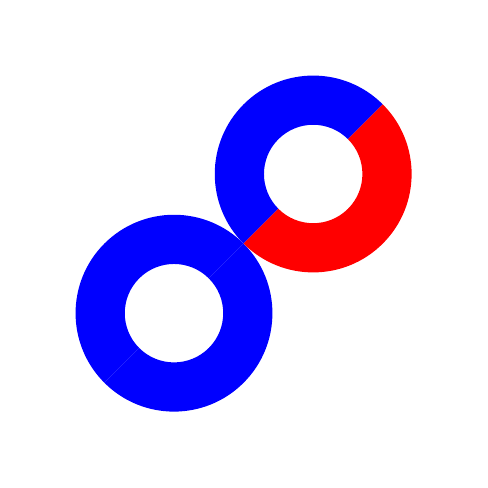}& \includegraphics[width=0.5cm]{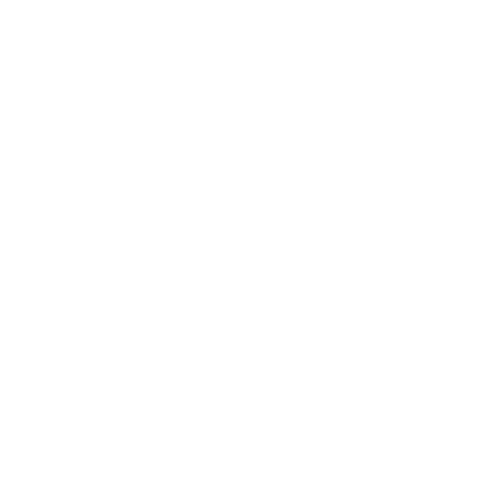}& \includegraphics[width=0.5cm]{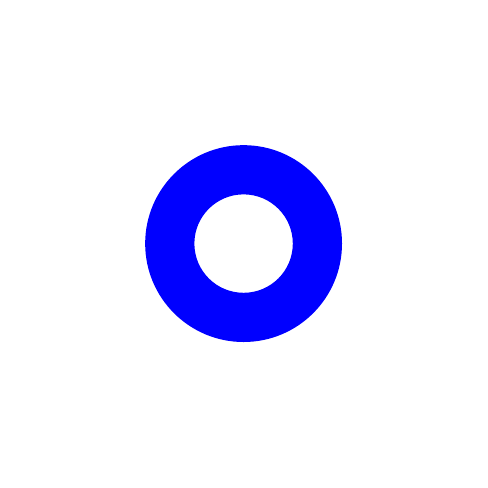}& \includegraphics[width=0.5cm]{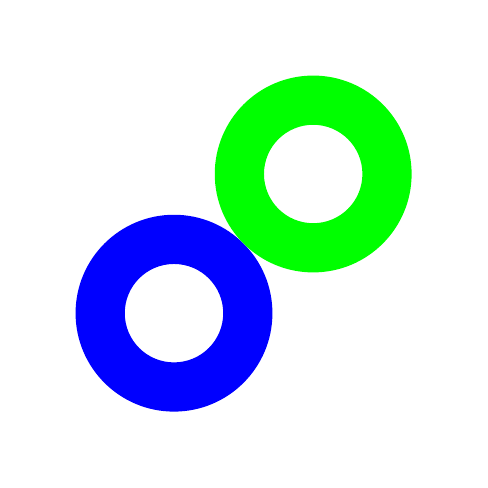}& \includegraphics[width=0.5cm]{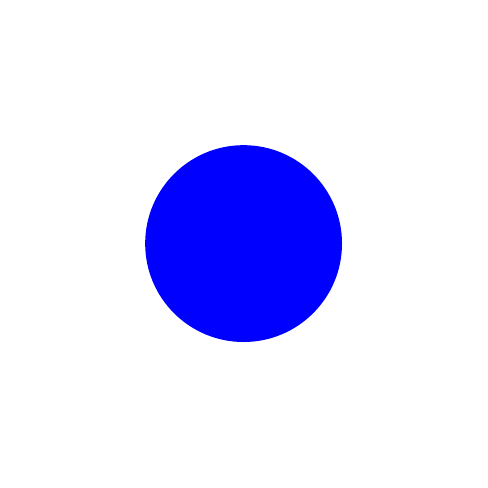}\includegraphics[width=0.5cm]{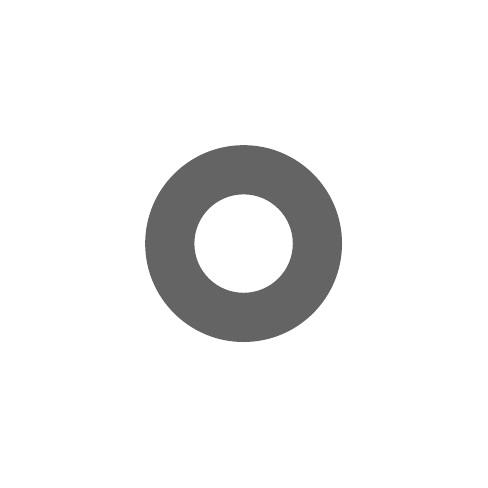}\includegraphics[width=0.5cm]{massless.pdf}\\ 
\hline 
\criticalcase{17}&${{\alp{1}= \alp{2}= \alp{3}= \alp{4}= \alp{6}= \bet{2}= \bet{1}+2\bet{3}}=0}$&${{\alp{5} < 0}}$& \includegraphics[width=0.5cm]{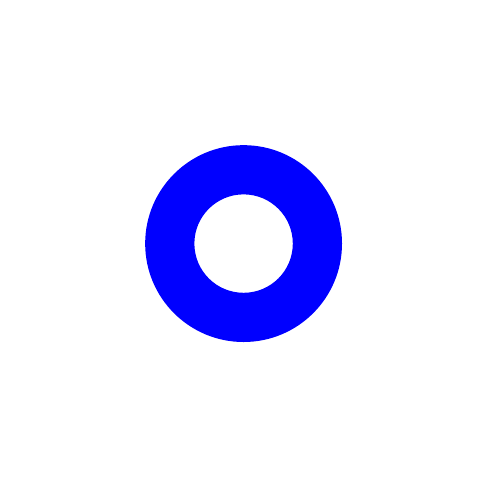}& \includegraphics[width=0.5cm]{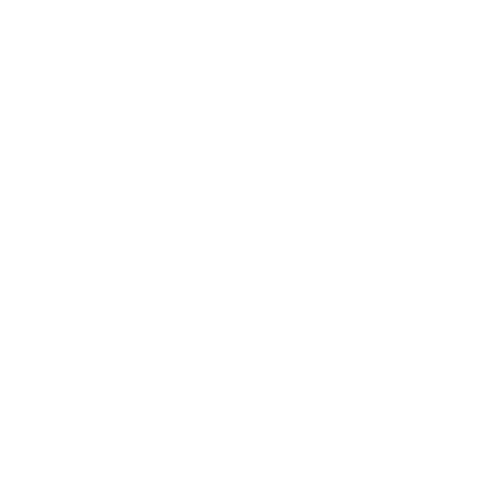}& \includegraphics[width=0.5cm]{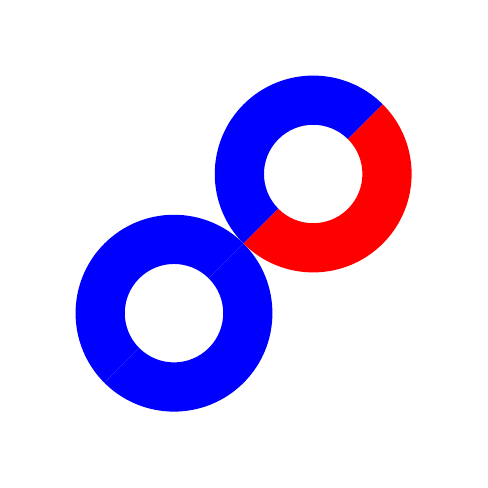}& \includegraphics[width=0.5cm]{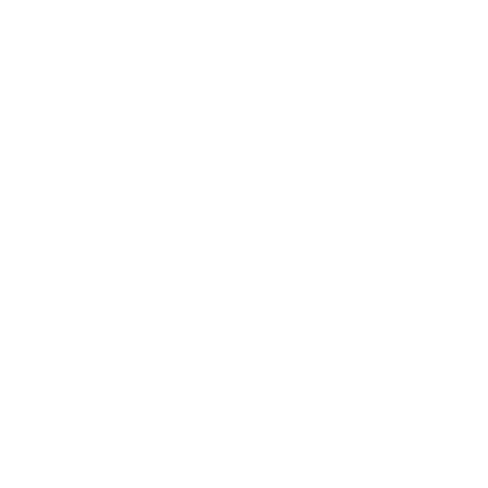}& \includegraphics[width=0.5cm]{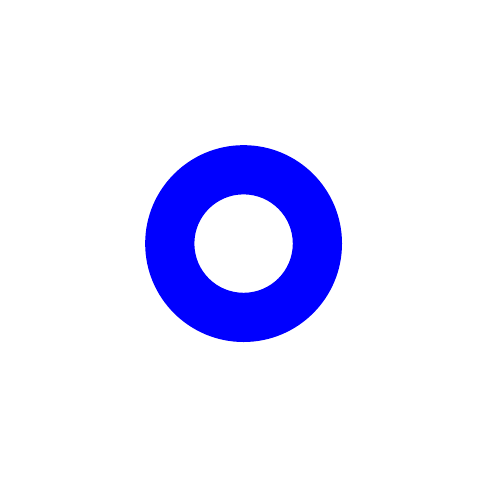}& \includegraphics[width=0.5cm]{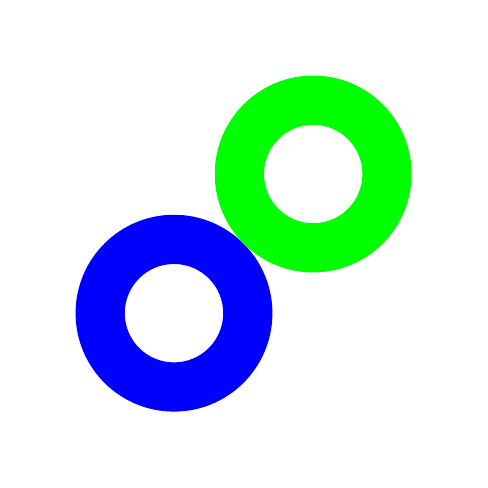}& \includegraphics[width=0.5cm]{massless.pdf}\includegraphics[width=0.5cm]{massless.pdf}\\ 
\hline 
\criticalcase{20}&${{\alp{1}= \alp{2}= \alp{4}= \alp{5}= \alp{6}}=0}$&${{0 < \bet{3}\wedge \alp{3} < 0}}$& \includegraphics[width=0.5cm]{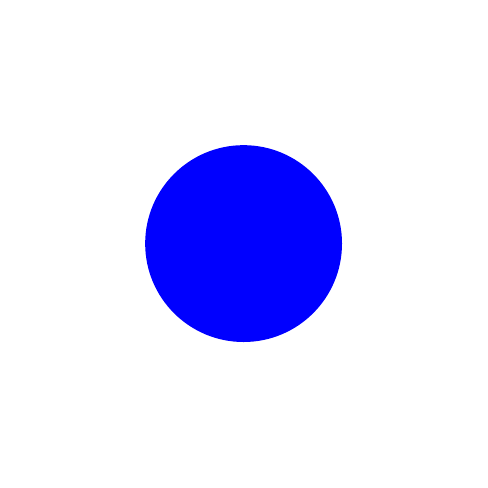}& \includegraphics[width=0.5cm]{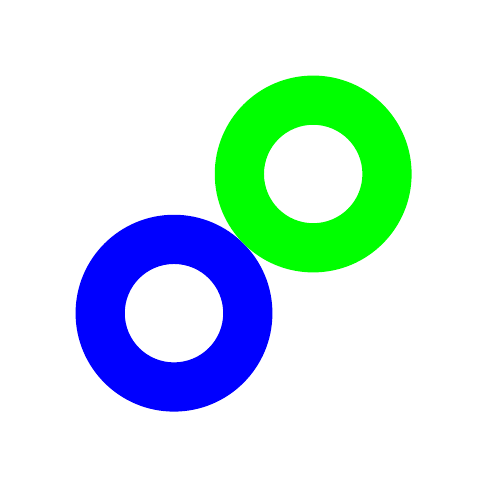}& \includegraphics[width=0.5cm]{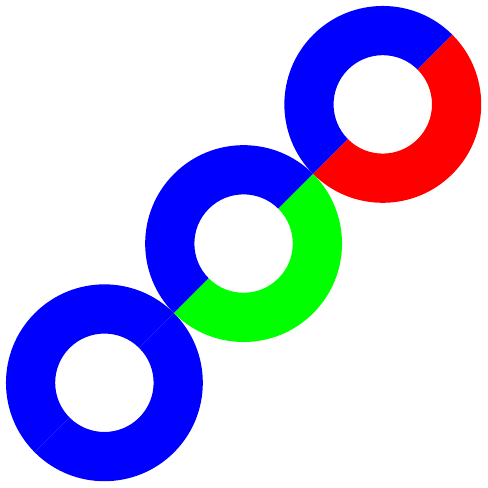}& \includegraphics[width=0.5cm]{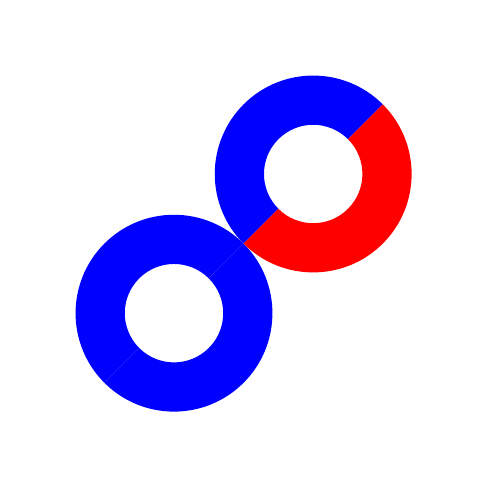}& \includegraphics[width=0.5cm]{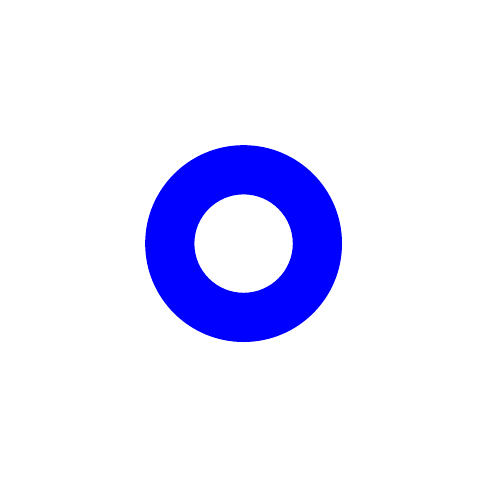}& \includegraphics[width=0.5cm]{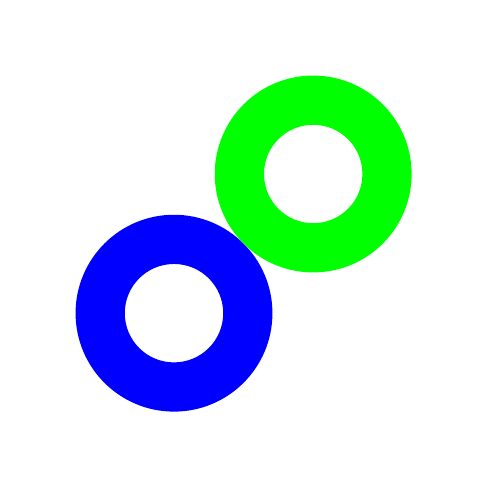}& \multirow{6}{*}{\includegraphics[width=0.5cm]{massive.pdf}}\\ 
\criticalcase{24}&${{\alp{1}= \alp{2}= \alp{4}= \alp{6}= \bet{1}}=0}$&${{0 < \bet{3}\wedge \alp{3} < 0}}$& \includegraphics[width=0.5cm]{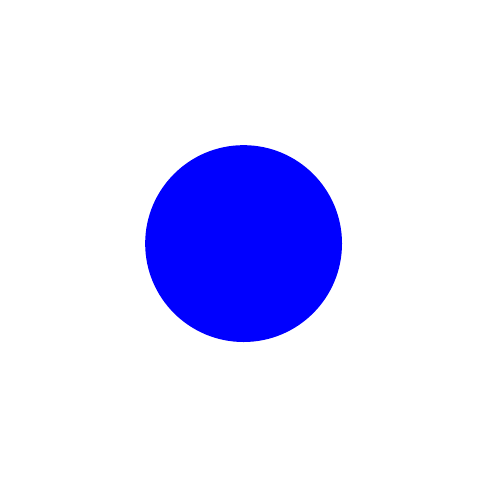}& \includegraphics[width=0.5cm]{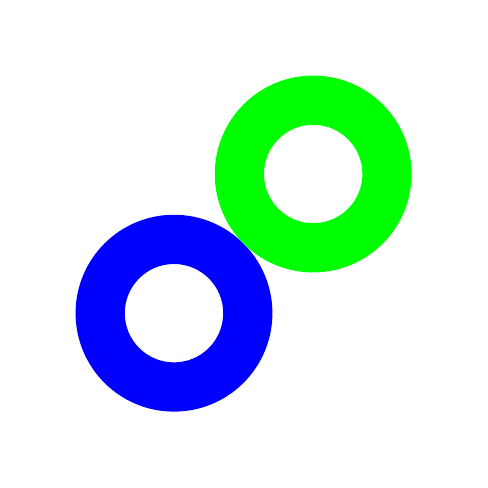}& \includegraphics[width=0.5cm]{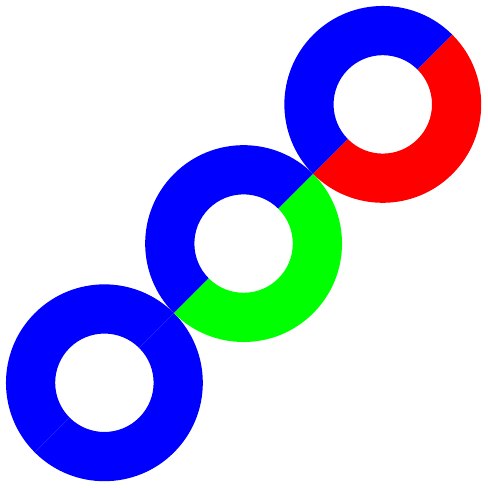}& \includegraphics[width=0.5cm]{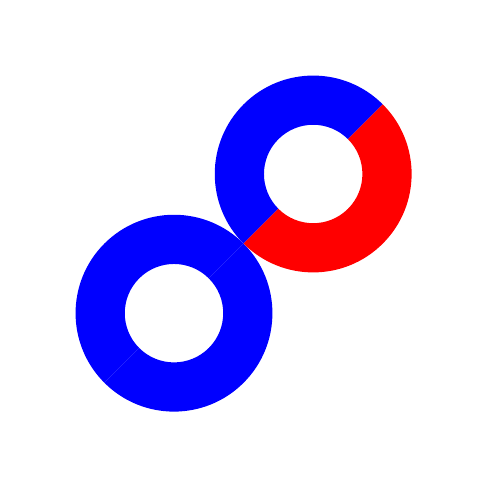}& \includegraphics[width=0.5cm]{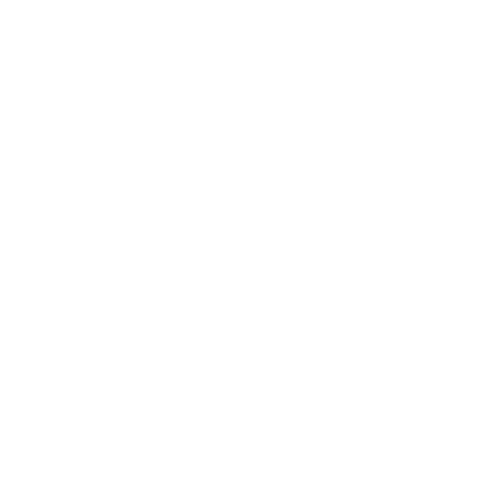}& \includegraphics[width=0.5cm]{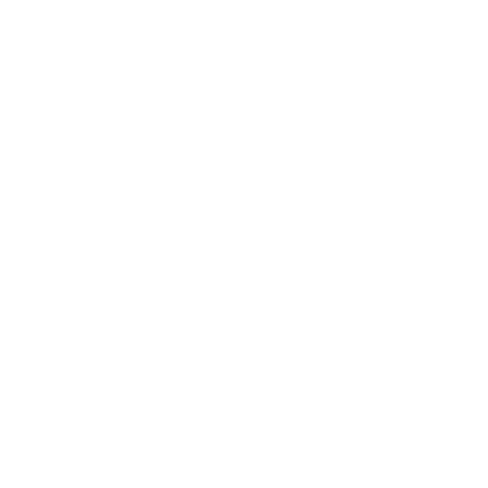}&\\ 
\criticalcase{25}&${{\alp{1}= \alp{2}= \alp{4}= \alp{5}= \alp{6}= \bet{1}}=0}$&${{0 < \bet{3}\wedge \alp{3} < 0}}$& \includegraphics[width=0.5cm]{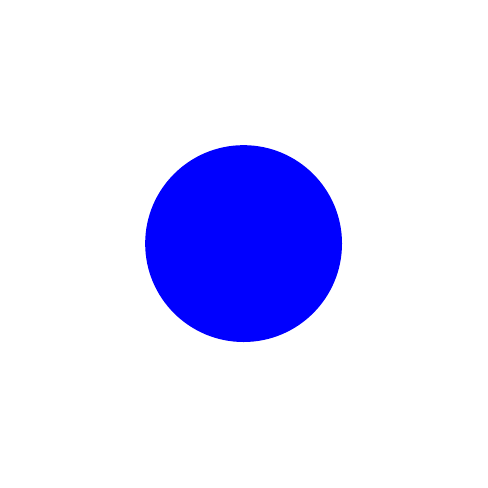}& \includegraphics[width=0.5cm]{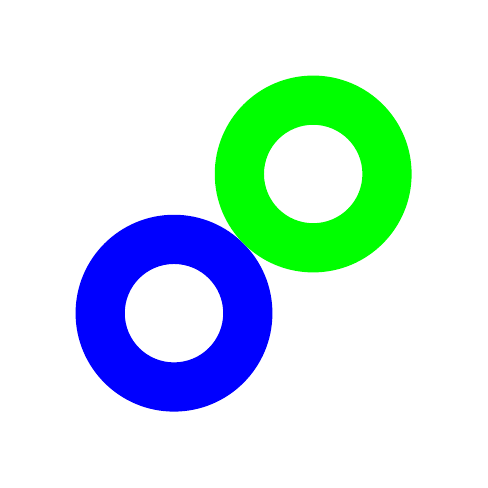}& \includegraphics[width=0.5cm]{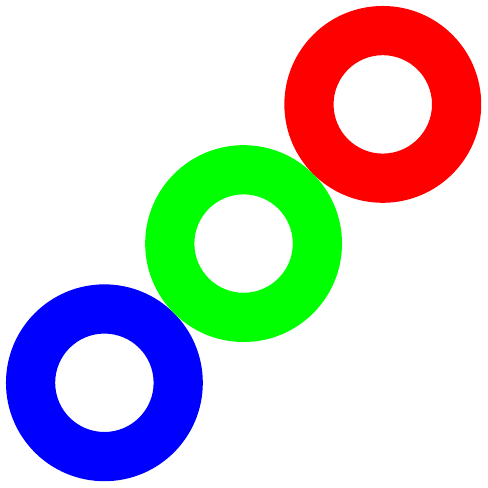}& \includegraphics[width=0.5cm]{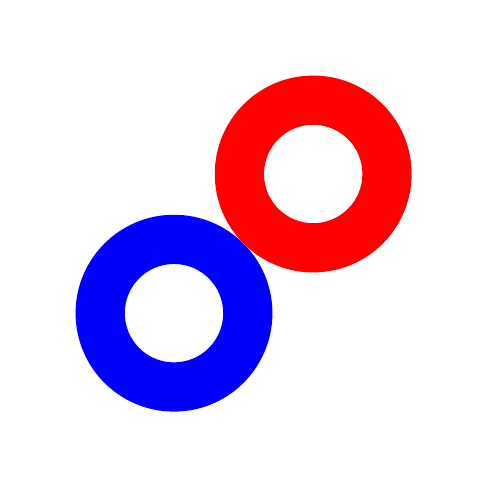}& \includegraphics[width=0.5cm]{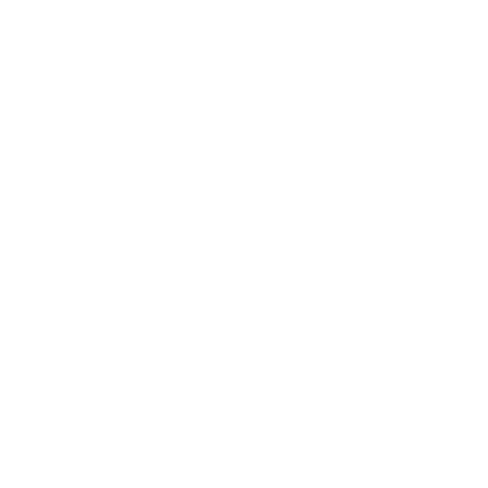}& \includegraphics[width=0.5cm]{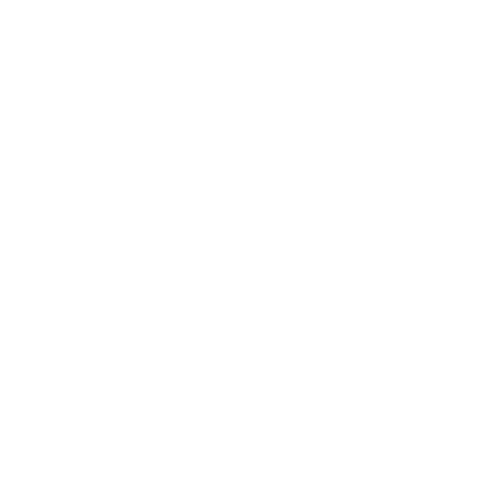}&\\ 
\criticalcase{26}&${{\alp{1}= \alp{2}= \alp{4}= \alp{5}= \alp{6}= \bet{1}= \bet{2}}=0}$&${{0 < \bet{3}\wedge \alp{3} < 0}}$& \includegraphics[width=0.5cm]{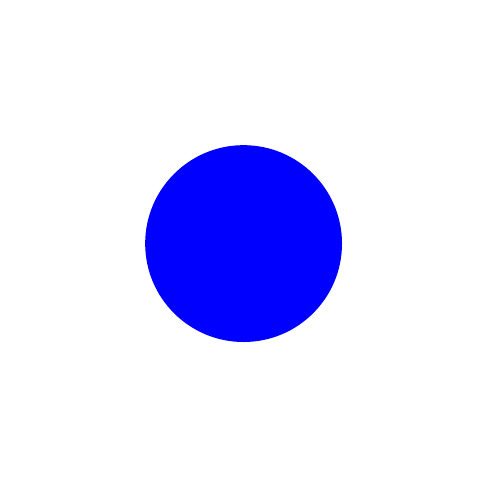}& \includegraphics[width=0.5cm]{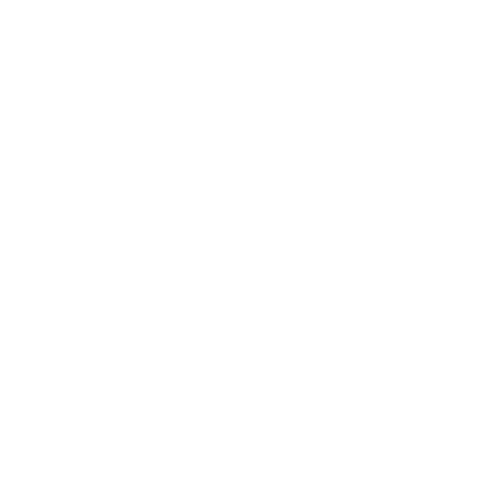}& \includegraphics[width=0.5cm]{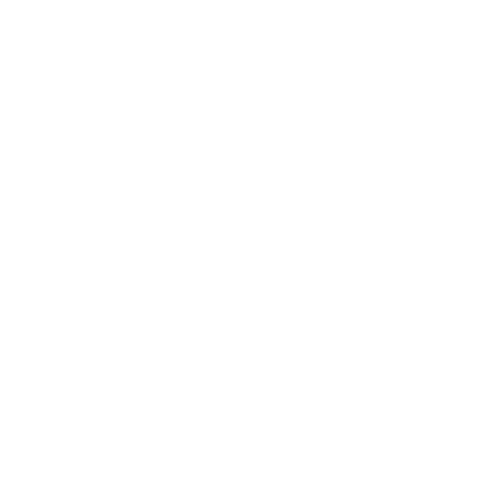}& \includegraphics[width=0.5cm]{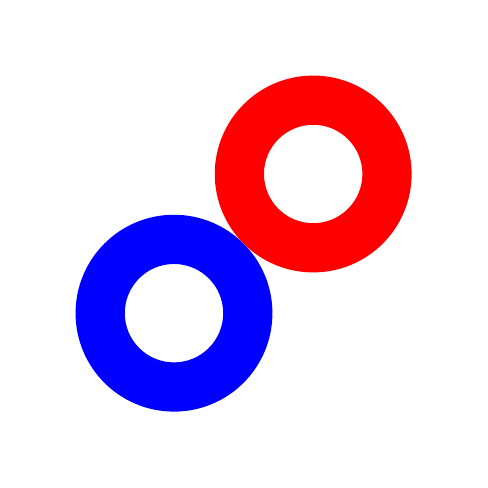}& \includegraphics[width=0.5cm]{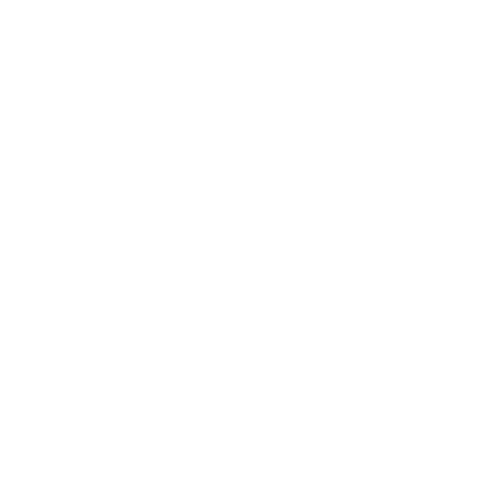}& \includegraphics[width=0.5cm]{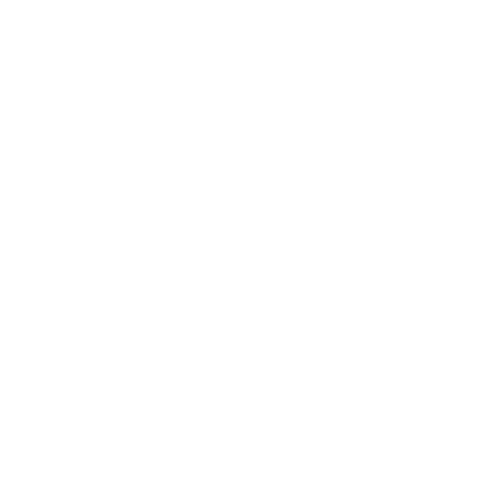}&\\ 
\criticalcase{28}&${{\alp{1}= \alp{2}= \alp{4}= \alp{6}= \bet{1}= \bet{2}}=0}$&${{0 < \bet{3}\wedge \alp{3} < 0}}$& \includegraphics[width=0.5cm]{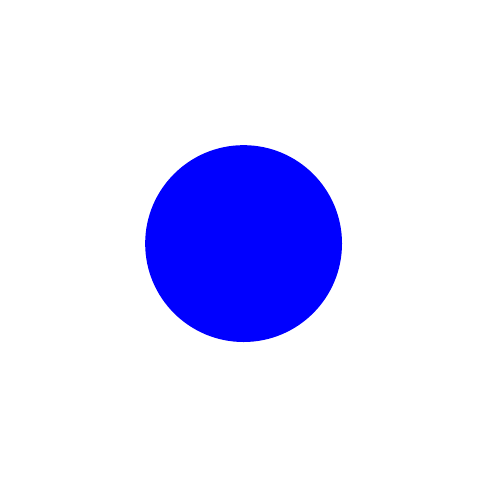}& \includegraphics[width=0.5cm]{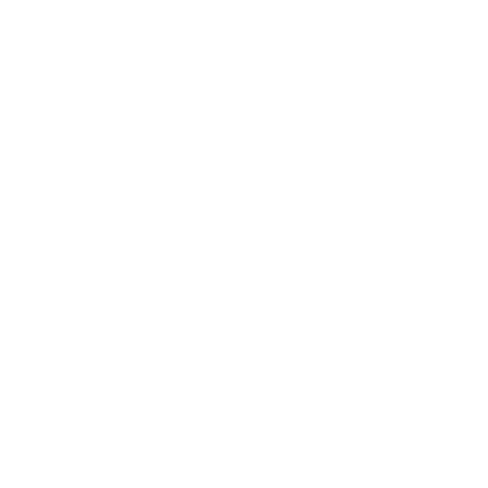}& \includegraphics[width=0.5cm]{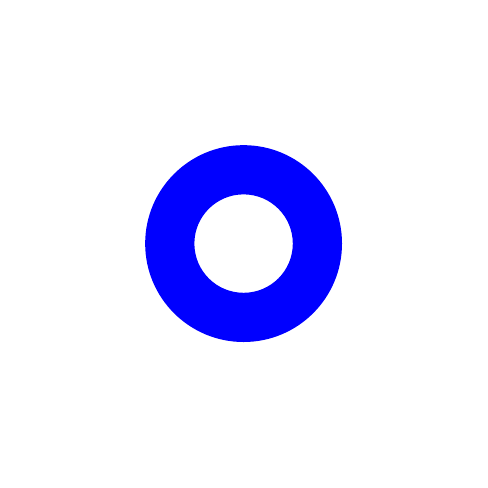}& \includegraphics[width=0.5cm]{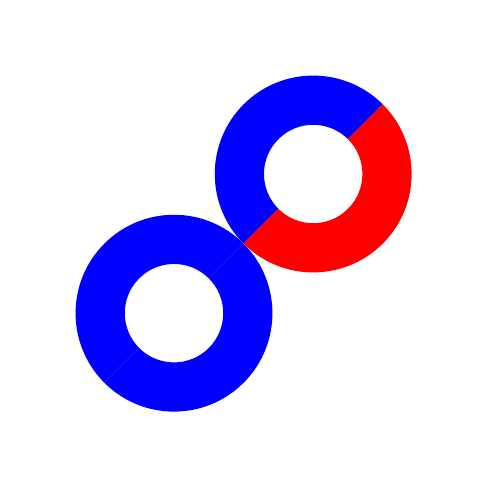}& \includegraphics[width=0.5cm]{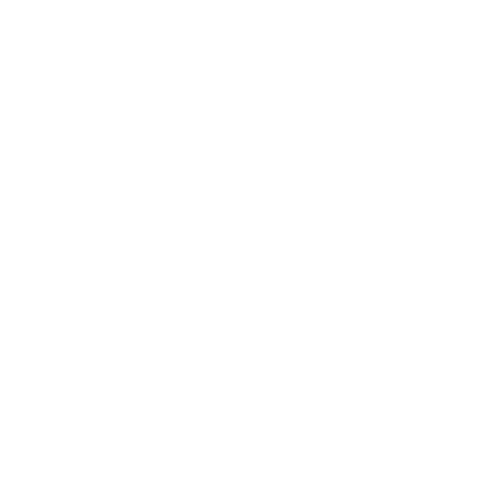}& \includegraphics[width=0.5cm]{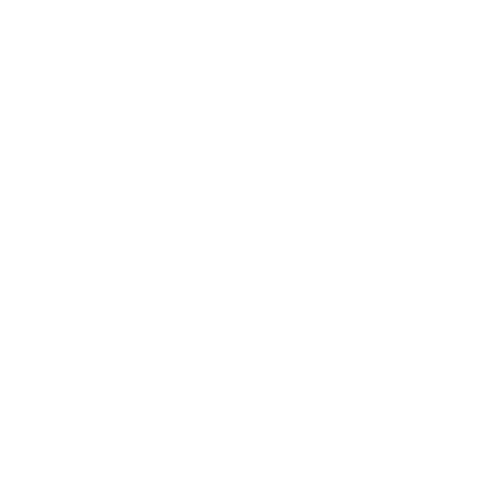}&\\ 
\criticalcase{32}&${{\alp{1}= \alp{2}= \alp{4}= \alp{5}= \alp{6}= \bet{2}}=0}$&${{0 < \bet{3}\wedge \alp{3} < 0}}$& \includegraphics[width=0.5cm]{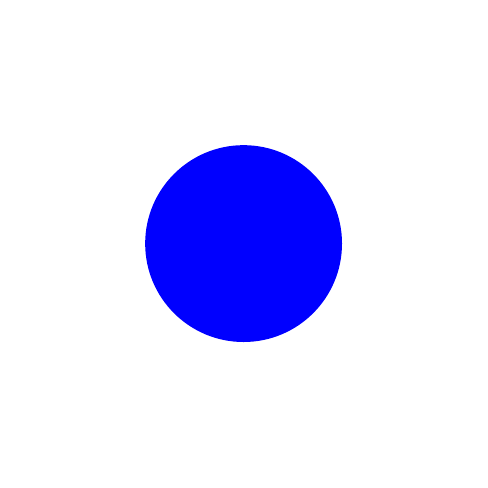}& \includegraphics[width=0.5cm]{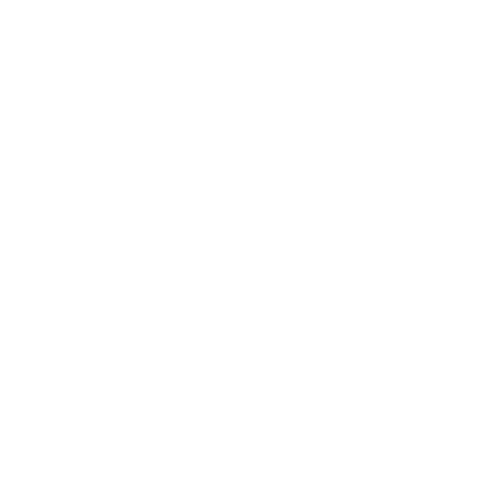}& \includegraphics[width=0.5cm]{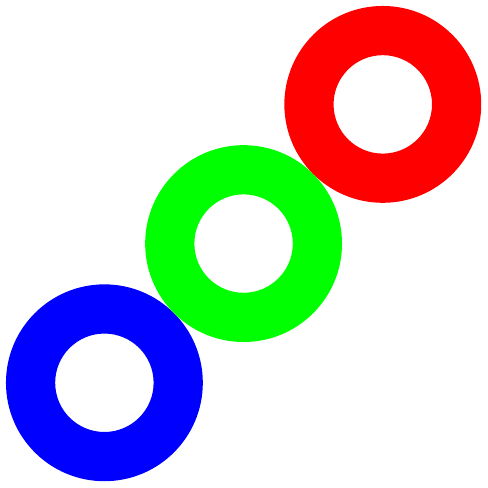}& \includegraphics[width=0.5cm]{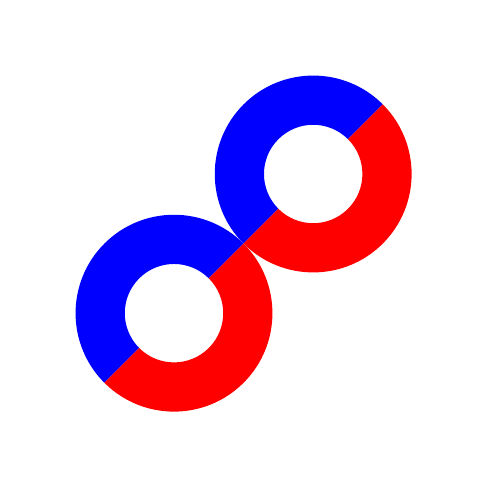}& \includegraphics[width=0.5cm]{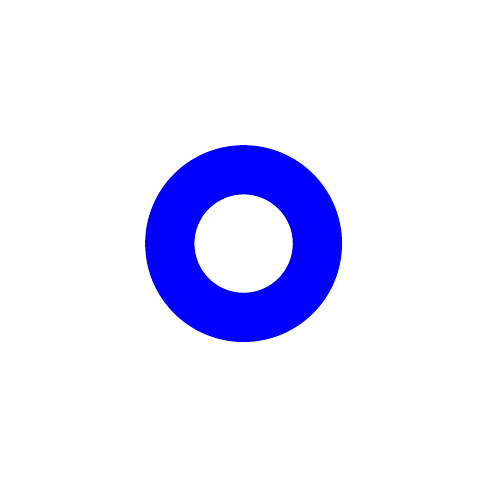}& \includegraphics[width=0.5cm]{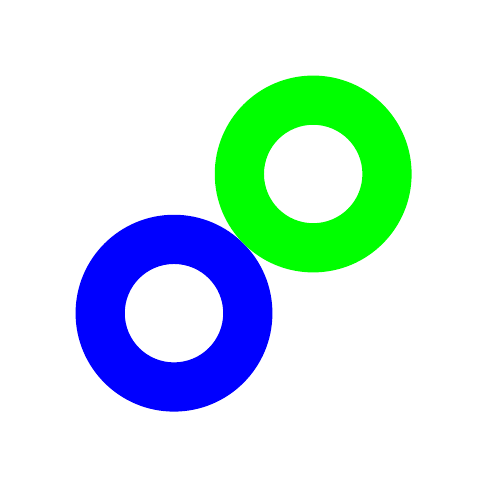}&

\\
\hline\hline
\end{tabularx}
\end{center}
\end{table*}
\subsection{Gauge theory formulation}
Recall that in Einstein's theory, the covariant derivative acting on a vector $\tensor{V}{^\mu}$ is simply
\begin{equation}
  \tensor{\nabla}{_\nu}\tensor{V}{^\mu}\equiv\tensor{\partial}{_\nu}\tensor{V}{^\mu}+\tensor{\Gamma}{^\mu_{\lambda\nu}}\tensor{V}{^\lambda},
  \label{<+label+>}
\end{equation}
where ${\tensor{\Gamma}{^\mu_{\nu\lambda}}\equiv\tfrac{1}{2}\tensor{g}{^{\mu\sigma}}(\tensor{\partial}{_\nu}\tensor{g}{_{\lambda\sigma}}+\tensor{\partial}{_\lambda}\tensor{g}{_{\nu\sigma}}-\tensor{\partial}{_\sigma}\tensor{g}{_{\nu\lambda}})}$ is the Levi--Civita connection with respect to some metric, itself defined by tangent vectors of the coordinate functions ${\tensor{g}{_{\mu\nu}}\equiv\tensor{\bm{e}}{_\mu}\cdot{\bm{e}}{_\nu}}$, on a \emph{curved} manifold $\mathcal{  M}$. The Riemann curvature tensor is then given by
\begin{equation}
  \tensor{R}{_{\alpha\beta\mu}^{\nu}}\equiv 2(\tensor{\partial}{_{[\beta}}\tensor{\Gamma}{^\nu_{\alpha]\mu}}+\tensor{\Gamma}{^\lambda_{[\alpha|\mu}}\tensor{\Gamma}{^\nu_{|\beta]\lambda}}).
  \label{riemann}
\end{equation}
This geometric interpretation of gravity is not strictly necessary. In generalising to theories with both curvature and torsion, we adopt the setup more familiar from the standard model, where the underlying manifold is always flat Minkowski space $\check{\mathcal{  M}}$. The metric for generally curvillinear coordinates is then ${\tensor{\gamma}{_{\mu\nu}}\equiv\tensor{\bm{e}}{_\mu}\cdot{\bm{e}}{_\nu}}$, and we note that this metric is strictly flat in the sense of \eqref{riemann}. Besides the coordinate basis, we define a Lorentz basis whose inner product always returns the Minkowski metric components ${\tensor{\eta}{_{ij}}\equiv\tensor{\hat{\bm{e}}}{_i}\cdot{\hat{\bm{e}}}{_j}}$. Under this condition, the Lorentz basis is completely free to rotate under the proper, orthochronous Lorentz rotations at each point in $\check{\mathcal{  M}}$, and is not presumed to follow from any particular coordinates (i.e. it is non-holonomic). A vector $\tensor{\mathcal{  V}}{^i}$ referred to this basis has a covariant derivative
\begin{equation}
  \tensor{\mathcal{  D}}{_j}\tensor{\mathcal{  V}}{^i}\equiv\tensor{h}{_j^\mu}(\tensor{\partial}{_\mu}\tensor{\mathcal{  V}}{^i}+\tensor{A}{^{i}_{k\mu}}\tensor{\mathcal{  V}}{^k}),
  \label{poin_covd}
\end{equation}
in which the (inverse) translational gauge field $\tensor{h}{_i^\mu}$ and rotational gauge field ${\tensor{A}{^{ij}_\mu}\equiv\tensor{A}{^{[ij]}_\mu}}$ are introduced to maintain invariance under general coordinate transformations on $\check{\mathcal{  M}}$ (i.e. passively interpreted diffeomorphisms) and rotations of the Lorentz basis. In this way, the Poincar{\'e} group is gauged. 
The inverse translational gauge field satisfies ${\tensor{b}{^i_\mu}\tensor{h}{_i^\nu}\equiv\tensor*{\delta}{_\mu^\nu}}$ and ${\tensor{b}{^i_\mu}\tensor{h}{_j^\mu}\equiv\tensor*{\delta}{_j^i}}$.
The metric components in the curved space of Einstein's theory (such as the flat cosmological metric we will consider in \cref{phenomenology}) can be recovered using ${\tensor{g}{_{\mu\nu}}\equiv\tensor{\eta}{_{ij}}\tensor{b}{^i_\mu}\tensor{b}{^j_\nu}}$ and ${\tensor{g}{^{\mu\nu}}\equiv\tensor{\eta}{^{ij}}\tensor{h}{_i^\mu}\tensor{h}{_j^\nu}}$. The gauge invariant measures on $\mathcal{  M}$ and $\check{\mathcal{  M}}$ are respectively $\sqrt{-g}$, where ${g\equiv\det \tensor{g}{_{\mu\nu}}}$, and ${b\equiv h^{-1}\equiv\det\tensor{b}{^i_\mu}}$.
Two field strength tensors are motivated by commuting the derivative \eqref{poin_covd}
\begin{subequations}
  \begin{align}
    \tensor{\mathcal{R}}{^{ij}_{kl}}&\equiv 2\tensor{h}{_k^\mu}\tensor{h}{_l^\nu}\big(\tensor{\partial}{_{[\mu}}\tensor{A}{^{ij}_{\nu]}}+\tensor{A}{^i_{m[\mu}}\tensor{A}{^{mj}_{\beta]}}\big),\label{riemanndef}\\
    \tensor{\mathcal{T}}{^i_{kl}}&\equiv 2\tensor{h}{_k^\mu}\tensor{h}{_l^\nu}\big(\tensor{\partial}{_{[\mu}}\tensor{b}{^i_{\nu]}}+\tensor{A}{^i_{m[\mu}}\tensor{b}{^m_{\nu]}}\big)\label{torsiondef}.
  \end{align}
\end{subequations}
These are the Riemann--Cartan curvature and torsion tensors, but are not understood to imbue $\check{\mathcal{  M}}$ with any geometry.
While the `particle physics' picture is consistent with our earlier treatments on this topic~\cite{2016JMP....57i2505L,2020arXiv200302690B,2020arXiv200603581B}, we concede that it is far more usual to treat torsion and Riemann--Cartan curvature as complimentary geometric qualities. In the geometric picture, the Riemann spacetime $\mathcal{  M}$ is generalised to a Riemann--Cartan spacetime, rather than specialised to a Minkowski spacetime. The interpretations are dynamically indistinguishable, and translations between the two are provided in~\cite{2016JMP....57i2505L,blagojevic2002gravitation}.
\subsection{Primary constraints and $3+1$}\label{primary_constraints}
In order to transition to the constrained Hamiltonian picture~\cite{blagojevic2002gravitation,Henneaux:1992ig,1813910}, we first define the \emph{canonical} momenta as follows
\begin{equation}
  \tensor{\pi}{_i^\mu}\equiv\frac{\partial bL_{\text{G}}}{\partial(\partial_0\tensor{b}{^i_{\mu}})}, \quad \tensor{\pi}{_{ij}^{\mu}}\equiv\frac{\partial bL_{\text{G}}}{\partial(\partial_0\tensor{A}{^{ij}_{\mu}})}.
  \label{canonicalmomenta}
\end{equation}
Following~\cite{2019PhRvD..99f4001L,2020PhRvD.101f4038L}, we will consider only the gravitational part of the Lagrangian, i.e. without any matter $L_\text{M}$.
Since the field strengths \eqref{riemanndef} and \eqref{torsiondef} from which \eqref{lagrangian_soft} is constructed make no reference to the velocities of $\tensor{b}{^k_0}$ and $\tensor{A}{^{ij}_0}$, the definitions \eqref{canonicalmomenta} incur $10$ primary constraints
\begin{equation}
  \tensor{\varphi}{_k^0}\equiv\tensor{\pi}{_k^0}\approx 0, \quad \tensor{\varphi}{_{ij}^0}\equiv\tensor{\pi}{_{ij}^0}\approx 0,
  \label{sureprimaries}
\end{equation}
so that the conjugate fields $\tensor{b}{^k_0}$ and $\tensor{A}{^{ij}_0}$ are non-physical. Notice that the \emph{weak} equality is denoted by ($\approx$). The constraints \eqref{sureprimaries} are a consequence of Poincar{\'e} gauge symmetry; their presence is independent of the couplings, and they are first class (FC). We refer to them as `sure' primary, first class (sPFC) constraints.
In order to systematically isolate the `sure' non-physical fields, we introduce the $3+1$ (ADM) splitting of the spacetime, in which a spacelike foliation is characterised by timelike unit normal $\tensor{n}{_k}$. Any vector which refers to the local Lorentz basis may be split into components ${\tensor{\mathcal{  V}}{^i}=\tensor{\mathcal{  V}}{^\perp}\tensor{n}{^i}+\tensor{\mathcal{  V}}{^{\ovl i}}}$ which are respectively perpendicular and parallel to the foliation: parallel indices are always denoted with an overbar. In what follows, it is very useful to note the identities $\smash{\tensor{b}{^{\ovl k}_\alpha}\tensor{h}{_{\ovl{l}}^\alpha}=\tensor*{\delta}{_{\ovl l}^{\ovl k}}}$ and $\smash{\tensor{b}{^{\ovl k}_\alpha}\tensor{h}{_{\ovl{k}}^\beta}=\tensor*{\delta}{_\alpha^\beta}}$.
The \emph{lapse} function and \emph{shift} vector are defined with reference to the non-physical part of the translational gauge field using this normal
\begin{equation}
  N\equiv\tensor{n}{_k}\tensor{b}{^k_0}, \quad \tensor{N}{^\alpha}\equiv\tensor{h}{_{\ovl{k}}^\alpha}\tensor{b}{^{\ovl{k}}_0}.
  \label{<+label+>}
\end{equation}
The remaining momenta can be expressed in the `parallel' form $\smash{\tensor{\hat{\pi}}{_{i}^{\overline{k}}}\equiv\tensor{\pi}{_{i}^\alpha}\tensor{b}{^{k}_\alpha}}$ and $\smash{\tensor{\hat{\pi}}{_{ij}^{\overline{k}}}\equiv\tensor{\pi}{_{ij}^\alpha}\tensor{b}{^{k}_\alpha}}$. 
In order to reveal the Hamiltonian structure of the theory as naturally as possible, the Lagrangian in \eqref{lagrangian_soft} is best written in the irreducible form 
\begin{equation}
\begin{aligned}
  L_{\text{T}}=-\frac{1}{2}\alpha_{0}{m_\text{p}}^2&\tensor{\mathcal{  R}}{}+\sum_{I=1}^{6}\tensor{\hat{\alpha}}{_I}\tensor{\mathcal{  R}}{^{ij}_{kl}}\tensor[^I]{\mathcal{  P}}{_{ij}^{kl}_{nm}^{pq}}\tensor{\mathcal{  R}}{^{nm}_{pq}}\\
  +{m_\text{p}}^2&\sum_{I=1}^{3}\tensor{\hat{\beta}}{_I}\tensor{\mathcal{  T}}{^{i}_{jk}}\tensor[^I]{\mathcal{  P}}{_{i}^{jk}_{l}^{nm}}\tensor{\mathcal{  T}}{^{l}_{nm}}+L_{\text{M}}.
  \label{lagrangian}
\end{aligned}
\end{equation}
where the nine operators $\tensor[^I]{\mathcal{  P}}{^{\dots}_{\dots}}$ project out all the irreducible representations of $\mathrm{SO}(1,3)$, in no particaular order, from the field strengths. For the details of these projections, including the linear translation between the quadratic couplings of \eqref{lagrangian_soft} and \eqref{lagrangian}, see~\cref{irreducible_decomposition}.
Within the field strengths, the $3+1$ splitting is used again to separate out the fields $\tensor{b}{^k_0}$ and $\tensor{A}{^{ij}_0}$ (which are non-physical) and the velocities of all fields (which are non-canonical)
\begin{subequations}
  \begin{align}
    \tensor{\mathcal{T}}{^i_{kl\vphantom{\ovl{l}}}}&=\tensor{\mathcal{T}}{^i_{\ovl{kl}}}+2\tensor{n}{_{[k\vphantom{\ovl{l}}}}\tensor{\mathcal{  T}}{^i_{\perp\ovl{l}]}},\label{torsionsplit}\\
    \tensor{\mathcal{R}}{^{ij}_{kl\vphantom{\ovl{l}}}}&=\tensor{\mathcal{R}}{^{ij}_{\ovl{kl}}}+2\tensor{n}{_{[k\vphantom{\ovl{l}}}}\tensor{\mathcal{  R}}{^{ij}_{\perp\ovl{l}]}},\label{riemannsplit}
  \end{align}
\end{subequations}
where such variables are confined to the second term in each case.
We are concerned with theories of the quadratic $L_\text{G}\sim\mathcal{  R}^2+\planck^2\mathcal{  T}^2$ form (i.e. $\alpha_0=0$), under the source-free condition $L_{\text{M}}=0$. Substituting \eqref{lagrangian} into \eqref{canonicalmomenta} and using \cref{torsionsplit,riemannsplit}, we find that the parallel momenta can be neatly expressed as functions of the field strengths
\begin{subequations}
  \begin{align}
    \frac{\tensor{\hat{\pi}}{_{i}^{\overline{k}}}}{J}=&\ \frac{\partial L_{\text{T}}}{\partial \tensor{\mathcal{  T}}{^i_{\perp\overline{k}}}}=4{m_\text{p}}^2\sum_{I=1}^{3}\tensor{\hat{\beta}}{_{I}}\tensor[^I]{\mathcal{  P}}{_i^{\perp\overline{k}}_n^{ml}}\tensor{\mathcal{  T}}{^n_{ml}},\label{translationalmomenta}\\
    \frac{\tensor{\hat{\pi}}{_{ij}^{\overline{k}}}}{J}=&\ \frac{\partial L_{\text{T}}}{\partial \tensor{\mathcal{  R}}{^{ij}_{\perp\overline{k}}}}=8\sum_{I=1}^{6}\tensor{\hat{\alpha}}{_{I}}\tensor[^I]{\mathcal{  P}}{_{ij}^{\perp\overline{k}}_{mn}^{pq}}\tensor{\mathcal{  R}}{^{mn}_{pq}},\label{rotationalmomenta}
  \end{align}
\end{subequations}
where the measure $J=b/N$ on the foliation is strictly physical, since $\tensor{b}{^k_0}$ is divided out by $N$.

Writing the parallel momenta in this form facilitates the identification of further primary constraints.  
Beginning with \eqref{translationalmomenta}, we find that the $12$ translational parallel momenta decompose into four irreducible representations of $\mathrm{O}(3)$. Using the spin-parity notation of~\cite{1999IJMPD...8..459Y,2002IJMPD..11..747Y} we write these as
\begin{subequations}
\begin{align}
  \tensor{\hat{\pi}}{_{k\ovl{l}}}&=\tensor{\hat\pi}{_{\ovl{kl}}}+\tensor{n}{_k}\PiP[\ovl{l}]{B1m},\\
  \tensor{\hat{\pi}}{_{\ovl{kl}}}&=\frac{1}{3}\etad{\ovl{kl}}\PiP{B0p}+\PiP[\ovl{kl}]{B1p}+\PiP[\ovl{kl}]{B2p}.
  \label{<+label+>}
\end{align}
\end{subequations}
In this expansion we identify the $0^+$ scalar $\PiP{B0p}$, the antisymmetric $1^+$ vector $\PiP{B1p}$, the $1^-$ vector $\PiP{B1m}$ and symmetric-traceless $2^+$ tensor $\PiP{B2p}$. Applying this decomposition to \eqref{translationalmomenta} as a whole, we obtain four functions which, with the aid of \eqref{torsionsplit}, are simultaneously defined both in terms of canonical and non-canonical variables 
\begin{subequations}
\begin{align}
  \pic{B0p}&\equiv J^{-1}\PiP{B0p}=\bet{2}{m_\text{p}}^2\etau{\ovl{kl}}\ncT{B0p},\label{PiCB0p}\\
  \pic{B1p}&\equiv J^{-1}\PiP{B1p}-\frac{4}{3}(\bet{1}-\bet{3} ){m_\text{p}}^2\cT{B1p}\nonumber\\
  &\ \ \ =\frac{1}{3}(\bet{1}+2\bet{3} ){m_\text{p}}^2\ncT{B1p},\label{PiCB1p}\\
  \pic{B1m}&\equiv J^{-1}\PiP{B1m}-\frac{4}{3}(\bet{1}-\bet{2} ){m_\text{p}}^2\cT{B1m}\nonumber\\
  &\ \ \ =\frac{1}{3}(2\bet{1}+\bet{2} ){m_\text{p}}^2\ncT{B1m},\label{PiCB1m}\\
  \pic{B2p}&\equiv J^{-1}\PiP{B2p}=\bet{1}{m_\text{p}}^2\ncT{B2p},\label{PiCB2p}
\end{align}
\end{subequations}
where the vector and symmetric-traceless torsion are 
\begin{equation}
  \cT[\ovl{k}]{B1m}\equiv \tensor{\mathcal{T}}{^{\ovl i}_{\ovl{ki}}}, \quad \ncT[\ovl{kl}]{B2p}\equiv\ncT[{(\ovl{kl})}]{B0p}-\frac{1}{3}\etad{\ovl{kl}}\etau{\ovl{ij}}\ncT[\ovl{ij}]{B0p}.
\end{equation}
In each case, \emph{if} the combination of coupling constants appearing in the non-canonical RHS definition vanishes, the canonically defined function on the LHS becomes a primary `if' constraint (PiC). An analogous construction is available for \eqref{rotationalmomenta}, with the $18$ remaining momenta decomposing as follows
\begin{subequations}
\begin{align}
  \tensor{\hat{\pi}}{_{kl\ovl{m}}}&=\tensor{\hat{\pi}}{_{\ovl{klm}}}+2\tensor{n}{_{[k}}\tensor{\hat{\pi}}{_{\perp\ovl{l}]\ovl{m}}},\\
  \tensor{\hat{\pi}}{_{\perp\ovl{kl}}}&=\frac{1}{3}\etad{\ovl{kl}}\PiP{A0p}+\PiP[\ovl{kl}]{A1p}+\PiP[\ovl{kl}]{A2p},\\
  \tensor{\hat{\pi}}{_{\ovl{klm}}}&=\frac{1}{6}\epsd{\ovl{klm}}\PiP{A0m}+\PiP[[\ovl{k}]{A1m}\etad{{{\ovl{l}]\ovl{m}}}}+\frac{4}{3}\PiP[\ovl{klm}]{A2m}.
  \label{<+label+>}
\end{align}
\end{subequations}
These are the $0^+$ scalar $\PiP{A0p}$, antisymmetric $1^+$ vector $\PiP{A1p}$, symmetric $2^+$ tensor $\PiP{A2p}$, and then the $0^-$ pseudoscalar $\PiP{A0m}$, the $1^-$ vector $\PiP{A1m}$ and $2^-$ tensor $\PiP{A2m}$. We will use ($\tensor[^{\text{P}}]{\ \cdot\ }{}$) to refer to the pseudoscalar part of general tensors, and ($\tensor[^{\text{T}}]{\ \cdot\ }{_{\ovl{klm}}}$) to refer to the tensor part (with antisymmetry implicit in the \emph{first} pair of indices, even if $\tensor{\cdot\ }{_{\ovl{klm}}}\equiv\tensor{\cdot\ }{_{\ovl{k}[\ovl{lm}]}}$).
The six PiC functions from \eqref{rotationalmomenta} are then
\begin{subequations}
\begin{align}
  \pic{A0p}&\equiv J^{-1}\PiP{A0p}+2\left(\alp{4}-\alp{6} \right)\cR{A0p}\nonumber\\
  &\ \ \ = \left(\alp{4}+\alp{6} \right)\ncR{A0p},\label{PiCA0p}\\
  \pic{A0m}&\equiv J^{-1}\PiP{A0m}+4\left(\alp{2}-\alp{3} \right)\cR{A0m}\nonumber\\
  &\ \ \ = \left(\alp{2}+\alp{3} \right)\ncR{A0m},\label{PiCA0m}\\
  \pic{A1p}&\equiv J^{-1}\PiP{A1p}+4\left(\alp{2}-\alp{5} \right)\cR{A1p}\nonumber\\
  &\ \ \ = \left(\alp{2}+\alp{5} \right)\ncR{A1p},\label{PiCA1p}\\
  \pic{A1m}&\equiv J^{-1}\PiP{A1m}+4\left(\alp{4}-\alp{5} \right)\cR{A1m}\nonumber\\
  &\ \ \ = \left(\alp{4}+\alp{5} \right)\ncR{A1m},\label{PiCA1m}\\
  \pic{A2p}&\equiv J^{-1}\PiP{A2p}+4\left(\alp{1}-\alp{4} \right)\cR{A2p}\nonumber\\
  &\ \ \ = \left(\alp{1}+\alp{4} \right)\ncR{A2p},\label{PiCA2p}\\
  \pic{A2m}&\equiv J^{-1}\PiP{A2m}-4\left(\alp{1}-\alp{2} \right)\cR{A2m}\nonumber\\
  &\ \ \ = \left(\alp{1}+\alp{2} \right)\ncR{A2m},\label{PiCA2m}
\end{align}
\end{subequations}
where we make further notational definitions
\begin{equation}
  \begin{gathered}
  \cR{A0m}\equiv\epsu{\ovl{ijk}}\tensor{\mathcal{R}}{_{\ovl{ijk}\perp}}, \quad \ncR{A0m}\equiv\epsu{\ovl{ijk}}\tensor{\mathcal{R}}{_{\perp\ovl{ijk}}},\\ 
  \tensor{\underline{\mathcal{R}}}{_{\ovl{kl}}}\equiv\tensor{\mathcal{R}}{^{\ovl i}_{\ovl{kli}}}, \quad \tensor{\underline{\mathcal{R}}}{}\equiv\tensor{\underline{\mathcal{R}}}{^{\ovl i}_{\ovl{i}}}.
\end{gathered}
\end{equation}
By this analysis, the possible occurence of primary constraints is systematically exhausted.

\subsection{Secondary constraints and the Hamiltonian}\label{poisson_section}

In order to be consistent, a primary constraint should not have any velocity within the final mass shell, so its commutator with the \emph{total} Hamiltonian should weakly vanish
\begin{equation}
  \dot{\varphi}(x_1)\equiv \int \mathrm{d}^3 x_2\Big\{\varphi(x_1),\tensor{\mathcal{H}}{_{\text{T}}}(x_2)\Big\}\approx 0.
  \label{consistency}
\end{equation}
The total Hamiltonian is related to the \emph{canonical} Hamiltonian, the Legendre-transformed Lagrangian, by the constraints and their multiplier fields
\begin{equation}
  \tensor{\mathcal{H}}{_{\text{T}}}\equiv \tensor{\mathcal{H}}{_{\text{C}}}+\tensor{u}{^k_0}\tensor{\varphi}{_k^0}+\frac{1}{2}\tensor{u}{^{ij}_0}\tensor{\varphi}{_{ij}^0}+(u\cdot\varphi),
  \label{total_hamiltonian_def}
\end{equation}
where the last term schematically represents any PiCs which may arise.
The canonical Hamiltonian may generally be collected into the insightful \emph{Dirac} form~\cite{1987PhRvD..35.3748B,PhysRevD.30.2508}
\begin{equation}
  \tensor{\mathcal{H}}{_{\text{C}}}\equiv N\tensor{\mathcal{H}}{_\perp}+\tensor{N}{^\alpha}\tensor{\mathcal{H}}{_\alpha}-\frac{1}{2}\tensor{A}{^{ij}_0}\tensor{\mathcal{H}}{_{ij}}+\tensor{\partial}{_\alpha}\tensor{\mathscr{D}}{^\alpha},
  \label{canonicalhamiltonian}
\end{equation}
i.e. as a linear function of the non-physical fields up to a surface term. The remaining functions which appear in \eqref{canonicalhamiltonian} are defined as follows
\begin{subequations}
\begin{align}
  \tensor{\mathcal{H}}{_\perp} &\equiv\tensor{\hat{\pi}}{_i^{\overline{k}}}\tensor{\mathcal{  T}}{^i_{\perp\overline{k}}}+\frac{1}{2}\tensor{\hat{\pi}}{_{ij}^{\overline{k}}}\tensor{\mathcal{  R}}{^{ij}_{\perp\overline{k}}}-JL-\tensor{n}{^k}\tensor{D}{_\alpha}\tensor{\pi}{_k^\alpha},\label{total_hamiltonian}\\
  \tensor{\mathcal{H}}{_\alpha} & \equiv\tensor{\pi}{_i^\beta}\tensor{T}{^i_{\alpha\beta}}+\frac{1}{2}\tensor{\pi}{_{ij}^\beta}\tensor{R}{^{ij}_{\alpha\beta}}-\tensor{b}{^k_\alpha}\tensor{D}{_\beta}\tensor{\pi}{_k^\beta}, \\
  \tensor{\mathcal{H}}{_{ij}}& \equiv 2\tensor{\pi}{_{[i}^\alpha}\tensor{b}{_{j]\alpha}}-\tensor{D}{_\alpha}\tensor{\pi}{_{ij}^\alpha},\\
  \tensor{\mathscr{D}}{^\alpha}& \equiv \tensor{b}{^i_0}\tensor{\pi}{_i^\alpha}+\frac{1}{2}\tensor{A}{^{ij}_0}\tensor{\pi}{_{ij}^\alpha}.
\end{align}
\end{subequations}
It is clear from \eqref{consistency} and the Dirac form \eqref{canonicalhamiltonian} that the consistency of \eqref{sureprimaries} invokes $10$ `sure' \emph{secondary} first class (sSFC) constriants
\begin{equation}
  \tensor{\mathcal{H}}{_\perp}\approx 0, \quad \tensor{\mathcal{H}}{_\alpha}\approx 0, \quad \tensor{\mathcal{H}}{_{ij}}\approx 0.
  \label{suresecondaries}
\end{equation}
It is important to note that while the sSFCs in \eqref{suresecondaries} are always enforced, it does not always follow that $2\times 10$ D.o.F are removed from the theory, as is the case with the sPFCs in \eqref{sureprimaries}. The functions involved are quite complicated, and may degenerately express a reduced number of FCs, or FCs which only appear at deeper levels in the consistency chain. Indeed, while this is very rare in the literature, we will find that it occurs for all eight novel theories, as a consequence of vanishing mass parameters.

The linear and rotational super-momenta $\tensor{\mathcal{H}}{_\alpha}$ and $\tensor{\mathcal{H}}{_{ij}}$ are kinematic generators which do not impinge on the dynamics. Thus, in the evaluation of \eqref{consistency}, it is sufficient to work purely with the super-Hamiltonian $\tensor{\mathcal{H}}{_\perp}$, which is, at length, expanded out using \cref{lagrangian,riemannsplit,torsionsplit} to give
\begin{widetext}
\begin{equation}
\begin{aligned}
  \mathcal{  H}_{\perp}=&\ {m_\text{p}}^2J\sum_{I=1}^{3}\tensor{\hat{\beta}}{_{I}}\Big[4\tensor{\mathcal{  T}}{^i_{\perp\overline{k}}}\tensor[^I]{\mathcal{  P}}{_i^{\perp\overline{k}}_j^{\perp\overline{l}}}\tensor{\mathcal{  T}}{^j_{\perp\overline{l}}}-\tensor{\mathcal{  T}}{^i_{\overline{mk}}}\tensor[^I]{\mathcal{  P}}{_i^{\overline{mk}}_j^{\overline{nl}}}\tensor{\mathcal{  T}}{^j_{\overline{nl}}}\Big]\\
  &+J\sum_{I=1}^{6}\tensor{\hat{\alpha}}{_{I}}\Big[4\tensor{\mathcal{  R}}{^{ip}_{\perp\overline{k}}}\tensor[^I]{\mathcal{  P}}{_{ip}^{\perp\overline{k}}_{jq}^{\perp\overline{l}}}\tensor{\mathcal{  R}}{^{jq}_{\perp\overline{l}}}-\tensor{\mathcal{  R}}{^{ip}_{\overline{mk}}}\tensor[^I]{\mathcal{  P}}{_{ip}^{\overline{mk}}_{jq}^{\overline{nl}}}\tensor{\mathcal{  R}}{^{jq}_{\overline{nl}}}\Big]-\tensor{n}{^k}\tensor{D}{_\alpha}\tensor{\pi}{_k^\alpha}\\
  =&\ \frac{J}{16}\Bigg[\frac{2\pic{B0p}^2}{3\bet{2}{m_\text{p}}^2}+\frac{6\pic{B1p}\picu{B1p}}{(\bet{1}+2\bet{3}){m_\text{p}}^2}+\frac{6\pic{B1m}\picu{B1m}}{(2\bet{1}+\bet{2}){m_\text{p}}^2}+\frac{2\pic{B2p}\picu{B2p}}{\bet{1}{m_\text{p}}^2}+\frac{2\pic{A0p}^2}{3(\alp{4}+\alp{6})}+\frac{\pic{A0m}^2}{6(\alp{2}+\alp{3})}+\frac{2\pic{A1p}\picu{A1p}}{\alp{2}+\alp{5}}\\
  &+\frac{\pic{A1m}\picu{A1m}}{\alp{4}+\alp{5}}+\frac{2\pic{A2p}\picu{A2p}}{\alp{1}+\alp{4}}+\frac{16\pic{A2m}\picu{A2m}}{9(\alp{1}+\alp{2})}\Bigg]+J\bigg[\frac{1}{3}(2\bet{1}+\bet{3}){m_\text{p}}^2\cT{B1p}\cTu{B1p}+\frac{1}{3}(\bet{1}+2\bet{2}){m_\text{p}}^2\cT{B1m}\cTu{B1m}\\
  &-\frac{1}{6}\bet{3}{m_\text{p}}^2\cT{A0m}^2+\frac{16}{9}\bet{1}{m_\text{p}}^2\cT{A2m}\cTu{A2m}+\frac{1}{6}(\alp{4}+\alp{6})\cR{A0p}^2-\frac{1}{6}(\alp{2}+\alp{3})\cR{A0m}^2+2(\alp{2}+\alp{5})\cR{A1p}\cRu{A1p}\\
&+(\alp{4}+\alp{5})\cR{A1m}\cRu{A1m}+2(\alp{1}+\alp{4})\cR{A2p}\cRu{A2p}+\frac{16}{9}(\alp{1}+\alp{2})\cR{A2m}\cRu{A2m}\bigg]-\tensor{n}{^k}\tensor{D}{_\alpha}\tensor{\pi}{_k^\alpha}.
  \label{superexpansion}
\end{aligned}
\end{equation}
\end{widetext}
To arrive at the second equality in \eqref{superexpansion}, the non-canonical `perpendicular' field strengths appearing in the first equality are canonicalised at length by the dual PiC definitions in \crefrange{PiCB0p}{PiCB2p} and \crefrange{PiCA0p}{PiCA2m}, resulting in terms quadratic in the PiC functions, and in the canonical `parallel' field strengths. The resulting expression is quite lengthy, but can be simplified for any given theory by safely eliminting those PiC functions which become constraints. The signs of the remaining quadratic PiC terms are then instrumental in the identification of unconstrained ghosts, since the PiC functions are schematically of the form ${\varphi\sim \pi+\mathcal{  R}}$ or ${\varphi\sim \pi+\mathcal{T}}$.

The consistency of the PiCs is less straightforward. Generally, the PiCs may be FC or SC within their own mass shell. In the case that a PiC is FC, \eqref{consistency} provides a secondary if-constraint (SiC). Possibly, the PiC and SiC do not commute; in that case both become SC within the new mass shell and the consistency of the SiC allows a multiplier to be determined
\begin{align}
  \begin{split}
  \dot{\chi}(x_1)\equiv \int \mathrm{d}^3 x_2 \Big(& N\Big\{\chi(x_1),\tensor{\mathcal{H}}{_{\perp}}(x_2)\Big\}\\
  &+u\cdot\Big\{\chi(x_1),\varphi(x_2)\Big\}\Big)\approx 0.
  \label{consistency2}
\end{split}
\end{align}
Otherwise, a tertiary if-constraint (TiC) may be found, and the process continues until the contraint chain from the PiC is absorbed by another chain, or by the sSFCs. In the case that a PiC is already SC within the PiC mass shell, its chain terminates immediately and two multipliers are determined. We note that occasionally, a constraint may be encountered at some deep level which retroactively terminates the chain at a shallower point. Only once the algoritm has terminated is it safe to categorise the if-constraints as FC or SC.

\begin{table}
  \caption{\label{table-2} Spin-parity sectors and associated PiCs, along with their kinetic and mass parameters. For completeness, we include the $\planck^2\mathcal{  R}$ term, mediated by $\alpha_{0}$.}
\begin{center}
  \def\arraystretch{1.3}
\begin{tabularx}{\linewidth}{c|c|l|X}
\hline\hline
 $J^P$ &PiC &  Kinetic parameter & Mass parameter \\
 \hline
 \multirow{2}{*}{$0^+$} &$\pic{B0p}$ &  $\bet{2}$ &\multirow{2}{*}{$\alpha_0(2\alpha_0+\bet{2})$} \\
 \cline{2-3}
  &$\pic{A0p}$ & $\alp{4}+\alp{6}$ & \\
 \hline
  $0^-$ & $\pic{A0m}$ &$\alp{2}+\alp{3}$ & $\alpha_0+2\bet{3}$ \\
 \hline
 \multirow{2}{*}{$1^+$} &$\pic{B1p}$ &  $\bet{1}+2\bet{3}$ & \multirow{2}{*}{$(\alpha_0+2\bet{3})(\alpha_0-\bet{1})$} \\
 \cline{2-3}
   &$\pic{A1p}$ & $\alp{2}+\alp{5}$ &  \\
 \hline
 \multirow{2}{*}{$1^-$} &$\pic{B1m}$ &  $2\bet{1}+\bet{2}$ & \multirow{2}{*}{$(2\alpha_0+\bet{2})(\alpha_0-\bet{1})$} \\
 \cline{2-3}
   &$\pic{A1m}$ & $\alp{4}+\alp{5}$ &  \\
 \hline
 \multirow{2}{*}{$2^+$} &$\pic{B2p}$ &  $\bet{1}$ & \multirow{2}{*}{$\alpha_0(\alpha_0-\bet{1})$} \\
 \cline{2-3}
  &$\pic{A2p}$ &  $\alp{1}+\alp{4}$ &  \\
 \hline
 $2^-$ &$\pic{A2m}$ &  $\alp{1}+\alp{2}$ & $\alpha_0-\bet{1}$ \\
\hline\hline
\end{tabularx}
\end{center}
\end{table}

In the linearised theory~\cite{10.1143/PTP.64.1435}, the analysis is greatly simplified by an understanding of the mass spectrum~\cite{10.1143/PTP.64.2222}. Only the $\mathcal{O}(1)$ parts of the PiC commutators contribute to the evaluation of the multipliers. Such commutators are possible only between pairs of PiCs which belong to the same $\mathrm{O}(3)$ irrep, and which are known as \emph{conjugate pairs}~\cite{1983PhRvD..28.2455B}. Conjugate PiCs will fail to commute in the linear theory only when their common mass parameter is non-vanishing. In this case, if only one PiC in a pair is present, it will still fail to commute with the SiC that maintains its consistency. 
Particularly, the rotational $\pic{A0m}$ and $\pic{A2m}$ have no $\mathrm{O}(3)$ counterparts in the translational sector, and are conjugate with their secondaries $\sic{A0m}$ and $\sic{A2m}$ a priori.
In the case of \emph{vanishing} mass parameters, the PiCs are FC, and a new gauge symmetry is invoked. The PiCs belonging to various $\mathrm{O}(3)$ irreps, along with their kinetic parameters and linearised mass parameters are listed in \cref{table-2}. 

Note that up to this point, our discussion has been fully general, and lays the theoretical foundations and conventions for the forthcoming series.
The evaluation of Poisson brackets is made tedious by the dependence of various quantities on the translational gauge field, as illustrated by the following useful identities 
\begin{equation}
  \begin{gathered}
    \frac{\partial\tensor{n}{_l}}{\partial\tensor{b}{^k_\mu}}\equiv -\tensor{n}{_k}\tensor{h}{_{\overline{l}}^\mu}, \quad \frac{\partial\tensor{h}{_l^\nu}}{\partial\tensor{b}{^k_\mu}}\equiv -\tensor{h}{_k^\nu}\tensor{h}{_l^\mu},\\
    \frac{\partial b}{\partial\tensor{b}{^k_\nu}}\equiv b\tensor{h}{_k^\nu}, \quad \frac{\partial J}{\partial\tensor{b}{^k_\nu}}\equiv J\tensor{h}{_{\overline{k}}^\nu}, \quad \frac{\partial N}{\partial\tensor{b}{^k_\nu}}\equiv N\tensor{n}{_k}\tensor{h}{_{\perp}^\nu}.
  \end{gathered}
  \label{<+label+>}
\end{equation}
As a crude measure to simplify the calculations, we artificially restrict our analysis in this paper to theories whose PiCs among \crefrange{PiCA0p}{PiCA2m} do not depend on $\tensor{\mathcal{  R}}{^{ij}_{\ovl{kl}}}$. It must be emphasised that this does not (to our knowledge) translate into any useful restriction on the physics. Of the 58 novel theories in~\cite{2019PhRvD..99f4001L,2020PhRvD.101f4038L}, eight satisfy our criterion: \criticalcase{3}, \criticalcase{17}, \criticalcase{20}, \criticalcase{24}, \criticalcase{25}, \criticalcase{26}, \criticalcase{28} and \criticalcase{32}. 
For most of these cases, we are fortunate that the remaining PiCs among \crefrange{PiCB0p}{PiCB2p} also do not depend on $\tensor{\mathcal{  T}}{^i_{\ovl{kl}}}$. \criticalcase{3} and \criticalcase{17} are exceptions to this rule.
We detail in \cref{table-1} our prior understanding of these theories, as encoded by the saturated graviton and roton propagators, linearised on Minkowski spacetime in the absence of matter. Aside from having torsion-dependent PiCs, \criticalcase{3} and \criticalcase{17} are particularly interesting as candidate theories of gravity, as they contain two massless D.o.F with power in the $2^+$ part of the propagator -- we will return to this point in \cref{massless_theories}. 

From our discussion in \cref{poisson_section}, we see that the constraint structure of the theory depends partially on the commutators between the PiCs, which form the \emph{primary Poisson matrix} (PPM). In \cref{massive_only,massless_theories} we will use the structure of the nonlinear PPM as a proxy for the health of each theory.

\section{Massive-only results}\label{massive_only}

\subsection{\criticalcase{26}}\label{section_26}
Conveniently, the PiCs of the massive theories depend on neither $\tensor{\mathcal{  T}}{^i_{\ovl{jk}}}$ nor $\tensor{\mathcal{  R}}{^{ij}_{\ovl{kl}}}$, so we will have schematically $\varphi\sim\pi$ for both translational and rotational sectors. By substituting the definition of \criticalcase{26} from \cref{table-1} into \eqref{superexpansion} and \eqref{total_hamiltonian_def}, the total Hamiltonian is seen to take the form
\begin{equation}
  \mathcal{  H}_\text{T}=\frac{b}{96}\left( \frac{18\pic{B1p}\picu{B1p}}{\bet{3}}-\frac{\pic{A0m}^2}{\alp{3}} \right)+\text{fields},
  \label{ham26}
\end{equation}
where we include only the part quadratic in the momenta. The remaining eight PiC functions that do not appear in \eqref{ham26} are primarily constrained, and give rise to the following nonvanishing commutators within the PiC shell
\begin{subequations}
  \begin{align}
    \Big\{\pic[\ovl{i}]{B1m},\pic[\ovl{l}]{B1m}\Big\}&\approx \frac{2}{J^2}\PiP[\ovl{il}]{B1p}\delta^3,\label{comm1}\\
    \Big\{\pic[\ovl{i}]{B1m},\pic[\ovl{lm}]{A1p}\Big\}&\approx-\frac{1}{6J^2}\epsd{\ovl{ilm}}\PiP{A0m}\delta^3,\label{comm1a}\\
    \Big\{\pic[\ovl{ij}]{B2p},\pic[\ovl{lm}]{B2p}\Big\}&\approx\frac{1}{J^2}\Big[\etad{\ovl{i}{(}\ovl{l}|}\PiP[\ovl{j}|\ovl{m}{)}]{B1p}+\etad{\ovl{j}{(}\ovl{l}|}\PiP[\ovl{i}|\ovl{m}{)}]{B1p}\Big]\delta^3,\label{comm1b}\\
    \Big\{\pic[\ovl{ij}]{B2p},\pic[\ovl{lmn}]{A2m}\Big\}&\approx\frac{1}{24J^2}\Big[2\etad{ {(}\ovl{i}|\ovl{n}}\epsd{|\ovl{j}{)}\ovl{lm}}-\etad{ {(}\ovl{i}|\ovl{l}}\epsd{|\ovl{j}{)}\ovl{mn}}\nonumber\\
    &\ \ \  +\etad{ {(}\ovl{i}|\ovl{m}}\epsd{|\ovl{j}{)}\ovl{ln}} \Big]\PiP{A0m}\delta^3,\label{comm2}
  \end{align}
\end{subequations}
where $\delta^3$ represents the equal-time Dirac function.
The nonlinear PPM of \criticalcase{26} is then written:
\begin{equation}
  \begin{BMAT}{rrccccccccl}{ccccccccccc}
 & & & & & & & & & & \\
 & & \scriptstyle{\pic{B0p}} & \scriptstyle{\pic{B1m}} & \scriptstyle{\pic{B2p}} & \scriptstyle{\pic{A0p}} & \scriptstyle{\pic{A1p}} & \scriptstyle{\pic{A1m}} & \scriptstyle{\pic{A2p}} & \scriptstyle{\pic{A2m}} & \\
  &\scriptstyle{\pic{B0p}}  & \cdot & \cdot & \cdot & \cdot & \cdot & \cdot & \cdot & \cdot& \scriptstyle{1} \\
  &\scriptstyle{\pic{B1m}}  & \cdot & \BPiP & \cdot & \cdot & \BPiP! & \cdot & \cdot & \cdot& \scriptstyle{3} \\
  &\scriptstyle{\pic{B2p}}  & \cdot & \cdot & \BPiP & \cdot & \cdot & \cdot & \cdot & \BPiP!& \scriptstyle{5} \\
  &\scriptstyle{\pic{A0p}}  & \cdot & \cdot & \cdot & \cdot & \cdot & \cdot & \cdot & \cdot& \scriptstyle{1} \\
  &\scriptstyle{\pic{A1p}}  & \cdot & \BPiP! & \cdot & \cdot & \cdot & \cdot & \cdot & \cdot& \scriptstyle{3} \\
  &\scriptstyle{\pic{A1m}}  & \cdot & \cdot & \cdot & \cdot & \cdot & \cdot & \cdot & \cdot& \scriptstyle{3} \\
  &\scriptstyle{\pic{A2p}}  & \cdot & \cdot & \cdot & \cdot & \cdot & \cdot & \cdot & \cdot& \scriptstyle{5} \\
  &\scriptstyle{\pic{A2m}}  & \cdot & \cdot & \BPiP! & \cdot & \cdot & \cdot & \cdot & \cdot& \scriptstyle{5} \\
 &  & \scriptstyle{1}  & \scriptstyle{3}  & \scriptstyle{5}  & \scriptstyle{1}  & \scriptstyle{3}  & \scriptstyle{3}  & \scriptstyle{5}  & \scriptstyle{5} & 
\addpath{(2,1,3)uuuuuuuu}
\addpath{(10,1,3)uuuuuuuu}
\addpath{(5,6,3)l}
\addpath{(5,6,3)r}
\addpath{(5,6,3)u}
\addpath{(5,6,3)d}
\addpath{(2,6,.)rdlu}
\addpath{(5,9,.)rdlu}
\addpath{(3,4,.)rdlu}
\addpath{(7,8,.)rdlu}
\addpath{(4,3,.)rdlu}
\addpath{(8,7,.)rdlu}
\end{BMAT}

  \label{case-26}
\end{equation}
The elements of the matrix schematically represent the nonlinear Poisson brackets between the PiCs. The PiCs are labelled, along with their multiplicities, at the edges of the PPM. They are arranged so as to divide the matrix into translational and rotational blocks, separated by ($\begin{BMAT}{cc}{cc} & \\ & \addpath{(1,1,3)rludlrdu}\end{BMAT}$). All brackets are restricted to the PiC shell. Commuting PiCs are dentoted by ($\cdot$). Non-commuting PiCs denoted as ($\BPiP$) are strictly linear combinations of $\tensor{\hat{\pi}}{_{ij}^{\ovl{k}}}$ and $\tensor{\hat{\pi}}{_{i}^{\ovl{k}}}$ as detailed in \crefrange{comm1}{comm2}. Generally, these expressions can be quite lengthy, so henceforth we confine them to \cref{brackets}. Commutators depending on momenta which (as we shall shortly show) propagate in the final linear theory are denoted by ($\BPiP!$). These are significant as they are presumed to persist even when the full nonlinear Dirac--Bergmann algorithm is terminated, except possibly on any strongly coupled spacetimes which might be found away from Minkowski spacetime.
Constant terms only arise in brackets between conjugate PiCs ($\begin{BMAT}{cc}{cc} & \\ & \addpath{(0,0,.)rruulldd}\end{BMAT}$), and then only if both PiCs have non-vanishing mass parameters. Since all the PiC mass parameters vanish in \criticalcase{26}, no constant terms can arise. The linearised theory is sensitive only to these constant terms, but we see from \eqref{case-26} that the conjugate PiCs also commute in the nonlinear \criticalcase{26}.

Let us now consider the consistency of the PiCs, and implement the Dirac--Bergmann algorithm for the linearised theory~\cite{1958RSPSA.246..326D,1951PhRv...83.1018A,1955PhRv...98..531B}.
Within the PiC shell, we encounter the following SiCs
\begin{subequations}
\begin{align}
  \sicl[\ovl k]{B1m}& \approx -2\etaul{\ovl{ml}}\covderl{\ovl m} \PiPl[\ovl{kl}]{B1p},\\
  \sicl[\ovl{kl}]{A1p}& \approx 2\PiPl[\ovl{kl}]{B1p}-\frac{1}{6}\epsdl{\overline{klm}}\etaul{\ovl{mn}}\covderl{\ovl n}\PiPl{A0m},
  \label{<+label+>}
\end{align}
\end{subequations}
where quantities linearised on the Minkowski background\footnote{Note also that we use the linearised gauge covariant derivative $\covderl{\ovl{i}}$, even to replace the nonlinear coordinate derivative $\tensor{h}{^{\flat}_{\ovl{i}}^\mu}\tensor{\partial}{_\mu}$.} are denoted with ($\flat$).
Also within this shell, we find $\haml{mom0p}$ and $\haml{rot1m}$ already vanish weakly, while the linear super-momentum and vector part of the rotational super-momentum give further sSFCs 
\begin{subequations}
\begin{align}
  \haml[\alpha]{mom1m}& \approx -\hfl{\alpha}{\ovl j}\etaul{\ovl{kl}}\covderl{\ovl k}\PiPl[\ovl{jl}]{B1p},\\
  \haml{rot1p}& \approx 2\PiPl[\ovl{kl}]{B1p}-\frac{1}{6}\epsdl{\overline{klm}}\etaul{\ovl{mn}}\covderl{\ovl n}\PiPl{A0m}.
  \label{<+label+>}
\end{align}
\end{subequations}
The SiCs clearly vanish in the sSFC sub-shell, terminating the algorithm immediately. We find that $\haml{mom1m}$ is already implied by $\haml{rot1p}$, which constitutes a total of three sSFCs. The PiCs are all FC, and the total number of iPFCs can be read off from \eqref{case-26}. Recalling also the $10$ sPFCs, and counting all FCs twice, we find that there is only one propagating D.o.F, as expected from \cref{table-1}
\begin{align}
\begin{split}
  1&=\smash{\frac{1}{2}}\big(80-2\times 10[\text{sPFC}]-2\times 3[\text{sSFC}]\\
  &\phantom{=}\ \ -2\times(1+3+5+1+3+3+5+5)[\text{iPFC}]\big).
    \label{<+label+>}
\end{split}
\end{align}

So, what is this D.o.F? We know that there are $26$ undetermined multipliers, to match each of the iPFCs. Generically, this makes it very difficult to make sense of the equations of motion. However, we can make an educated guess by noticing that the functions $\pic{B1p}$ and $\pic{A0m}$ are not PiCs and in the end, it turns out to be the $0^-$ tordion which is propagating. An application of \eqref{consistency} allows us to find the velocity of the pseudoscalar part of the torsion
\begin{equation}
  \dot{\cTl{A0m}}\approx-\frac{1}{2\alp{3}}\PiPl{A0m}-\frac{3}{4\bet{3}{m_\text{p}}^2}\epsul{\ovl{jkl}}\covderl{\ovl{j}}\PiPl[\ovl{kl}]{B1p}.
  \label{<+label+>}
\end{equation}
Conveniently, we see that this quantity makes no reference to undetermined multipliers in the final shell. Moreover, the same can be said of the acceleration 
\begin{equation}
  \ddot{\cTl{A0m}}\approx -\etaul{\ovl{jk}}\covderl{\ovl{j}}\covderl{\ovl{k}}\cTl{A0m}-\frac{4\bet{3}}{\alp{3}}{m_\text{p}}^2\cTl{A0m},
  \label{<+label+>}
\end{equation}
which clearly describes a particle of mass
\begin{equation}
  m\equiv2\sqrt{\frac{|\bet{3}|}{|\alp{3}|}}{m_{\text{p}}},
\end{equation}
if $\bet{3}/\alp{3}>0$. The unitarity conditions in \cref{table-1} can now be decoded. The condition $\alp{3}<0$ clearly wards off a $0^-$ ghost by inspection of \eqref{ham26}, whereas $\bet{3}<0$ then prevents the $0^-$ from becoming tachyonic.

In the nonlinear theory, the PPM is no longer empty as shown in \eqref{case-26}. We anticipate that the emergent commutators will ultimately result in a fundamentally different particle spectrum.  
Particularly, we see from \cref{comm1b,comm2} that $\pic{B1m}$, $\pic{B2p}$, $\pic{A1p}$ and $\pic{A2m}$ are all demoted from iPFCs to iPSCs so long as $0^-$ is activated. Possibly, $0^-$ becomes strongly coupled on some other privileged surface within the final shell, but since the converse is unlikely to be true we conclude that an iPFC \emph{generally} becomes an iPSC in the nonlinear theory. According to Dirac's conjecture, the FCs are associated with gauge symmetries. More correctly, every PFC can be used to construct a nontrivial gauge generator using the Castellani algorithm~\cite{1982AnPhy.143..357C}. We therefore expect that a generator is generally broken.

To see one way in which this might affect the outcome, imagine that none of the sSFCs are degenerate in the full nonlinear theory, but that they still encode the iSFCs (which therefore need not appear in the final count). The nonlinear theory would then be expected to propagate two D.o.F
\begin{align}
\begin{split}
  2&\ifeq\smash{\frac{1}{2}}\big(80-2\times 10[\text{sPFC}]-2\times 10[\text{sSFC}]\\
  &\phantom{=}\ \ -2\times(1+1+3+5)[\text{iPFC}]\\
&\phantom{=}\ \ -(3+5+3+5)[\text{iPSC}]\big),
    \label{scenario}
\end{split}
\end{align}
suggesting that somehow one D.o.F from the $1^+$ sector (i.e. the only $J^P$ other than $0^-$ which is not primarily constrained), is generally activated, but becomes strongly coupled on Minkowski spacetime. It is not clear what this would look like, and we emphasise that the specific scenario in \eqref{scenario} is unlikely to be the one which is realised. The full picture can only be revealed by performing the nonlinear Dirac--Bergmann analysis, beginning from \eqref{case-26}. Following treatments of simpler cases of PGT\textsuperscript{q+} in~\cite{2002IJMPD..11..747Y}, we will not go this far. However, we think it likely that any activation of the $1^+$ sector will damage the unitarity of the theory, since we see from \eqref{ham26} that $\PiP{B1p}\PiPu{B1p}$ has a negative contribution to the energy, by the same condition $\bet{3}<0$ that upholds the unitarity of the $0^-$ mode. For further discussion of the `positive energy test', we direct the reader to \cref{sighost}.

Finally, \eqref{case-26} may also indicate that the nonlinear theory violates causality. We refer to the test based on the  tachyonic shock in the nonlinear Proca theory~\cite{1998AcPPB..29..961C}, and which was also implemented in~\cite{2002IJMPD..11..747Y}, whereby the PPM rank is required not to depend on the values of the fields and their momenta. 
The motivation for this requirement is as follows. It is easy to see from \eqref{consistency} that the multipliers $\tensor[^A]{u}{}$ and $\tensor[^B]{u}{}$ of a pair of PiCs $\tensor[^A]{\varphi}{}$ and $\tensor[^B]{\varphi}{}$ can be determined in the case that $\{\tensor[^A]{\varphi}{},\tensor[^B]{\varphi}{}\}\not\approx 0$ on the final shell. Moreover, $\tensor[^A]{u}{}$ will be nonvanishing if $\{\mathcal{  H}_\text{C},\tensor[^B]{\varphi}{}\}\not\approx 0$. Imagine that a dynamical trajectory intersected a surface $\Sigma$ on which $\{\tensor[^A]{\varphi}{},\tensor[^B]{\varphi}{}\}\to 0$. The multiplier $\tensor[^A]{u}{}$ had better not have any physical interpretation in that case, since it would diverge\footnote{The problem is somewhat analogous to one of strong coupling. If the prefactor to the kinetic term of a field vanishes (i.e. its mass becomes infinite) on some $\Sigma$, the Heisenberg principle suggests that quantum fluctuations will diverge on the approach to $\Sigma$.}. Unfortunately in the case of PGT\textsuperscript{q+}, the multipliers can be written in terms of the non-canonical velocities\footnote{We note a caveat here, that this interpretation is strictly true for theories with nonvanishing mass parameters; more careful investigation of the multiplier interpretation may be warranted for the cases at hand.} through the dual definitions of the PiC functions in \crefrange{PiCB0p}{PiCB2p} and \crefrange{PiCA0p}{PiCA2m}. The interpretation is then that a tachyonic excitation develops on the approach to $\Sigma$. In the case at hand, the nonlinear PPM in \eqref{case-26} is populated by momenta, and the linearised PPM is empty. Thus, Minkowski spacetime is just such a surface $\Sigma$.
More generally, when the linearised PPM is populated by constant mass parameters, the requirement becomes that the nonlinear PPM pseudodeterminant should be positive-definite within the final shell.

\subsection{\criticalcase{28}}
Since \criticalcase{28} has fewer PiCs than \criticalcase{26}, the kinetic part of the  Hamiltonian is more extensive
\begin{equation}
  \begin{aligned}
  \mathcal{  H}_\text{T}=\frac{b}{96}\Bigg(& \frac{6\big(\pic{A1m}\picu{A1m}+2\pic{A1p}\picu{A1p}\big)}{\alp{5}}\\
  &+\frac{18\pic{B1p}\picu{B1p}}{\bet{3}}-\frac{\pic{A0m}^2}{\alp{3}} \Bigg)+\text{fields},
  \label{ham28}
\end{aligned}
\end{equation}
while the PPM has fewer dimensions 
\begin{equation}
  \begin{BMAT}{rrccccccl}{ccccccccc}
 & & & & & & & & \\
 & & \scriptstyle{\pic{B0p}} & \scriptstyle{\pic{B1m}} & \scriptstyle{\pic{B2p}} & \scriptstyle{\pic{A0p}} & \scriptstyle{\pic{A2p}} & \scriptstyle{\pic{A2m}} & \\
  &\scriptstyle{\pic{B0p}}  & \cdot & \cdot & \cdot & \cdot & \cdot & \cdot& \scriptstyle{1} \\
  &\scriptstyle{\pic{B1m}}  & \cdot & \BPiP & \cdot & \BPiP & \BPiP & \BPiP& \scriptstyle{3} \\
  &\scriptstyle{\pic{B2p}}  & \cdot & \cdot & \BPiP & \cdot & \BPiP & \BPiP!& \scriptstyle{5} \\
  &\scriptstyle{\pic{A0p}}  & \cdot & \BPiP & \cdot & \cdot & \cdot & \cdot& \scriptstyle{1} \\
  &\scriptstyle{\pic{A2p}}  & \cdot & \BPiP & \BPiP & \cdot & \cdot & \cdot& \scriptstyle{5} \\
  &\scriptstyle{\pic{A2m}}  & \cdot & \BPiP & \BPiP! & \cdot & \cdot & \cdot& \scriptstyle{5} \\
 &  & \scriptstyle{1}  & \scriptstyle{3}  & \scriptstyle{5}  & \scriptstyle{1}  & \scriptstyle{5}  & \scriptstyle{5} & 
\addpath{(2,1,3)uuuuuu}
\addpath{(8,1,3)uuuuuu}
\addpath{(5,4,3)l}
\addpath{(5,4,3)r}
\addpath{(5,4,3)u}
\addpath{(5,4,3)d}
\addpath{(2,4,.)rdlu}
\addpath{(5,7,.)rdlu}
\addpath{(4,3,.)rdlu}
\addpath{(6,5,.)rdlu}
\end{BMAT}

  \label{case-28}
\end{equation}
Within the PiC shell, we find that $\picl{B0p}$ and $\picl{B2p}$ already weakly vanish, leaving the following SiCs
\begin{subequations}
\begin{align}
  \sicl[\ovl k]{B1m}& \approx -2\etaul{\ovl{ml}}\covderl{\ovl m} \PiPl[\ovl{kl}]{B1p},\\
  \sicl{A0p}& \approx -\etaul{\ovl{ml}}\covderl{\ovl m} \PiPl[\ovl l]{A1m},\\
  \sicl[\ovl{kl}]{A2p}& \approx \frac{1}{2}\covderl{\langle \ovl k} \PiPl[\ovl l \rangle]{A1m},\\
  \sicl[\ovl{klm}]{A2m}& \approx \frac{1}{2}\covderl{\ovl m}\PiPl[\ovl {kl}]{A1p}+\frac{1}{2}\covderl{{{[}\ovl{l}}}\PiPl[{\ovl{k}{]}\ovl{m}}]{A1p}\nonumber\\
  &\ \ \ +\frac{3}{4}\etadl{\ovl{m}[\ovl{k}|}\etaul{\ovl{ij}}\covderl{\ovl i}\PiPl[{|\ovl{l}{]}\ovl{j}}]{A1p},
  \label{<+label+>}
\end{align}
\end{subequations}
which do not give rise to any TiCs. Also within the PiC shell, the following sSFCs appear
\begin{subequations}
\begin{align}
  \haml[\alpha]{mom1m}& \approx -\hfl{\alpha}{\ovl j}\etaul{\ovl{kl}}\covderl{\ovl k}\PiPl[\ovl{jl}]{B1p},\\
  \haml{rot1p}& \approx 2\PiPl[\ovl{kl}]{B1p}-\frac{1}{6}\epsdl{\overline{klm}}\etaul{\ovl{mn}}\covderl{\ovl n}\PiPl{A0m}+\covderl{[\ovl{k}}\PiPl[{{\ovl{l}]}}]{A1m},\\
  \haml{rot1m}& \approx \etaul{\ovl{jl}}\covderl{\ovl j}\PiPl[\ovl{kl}]{A1p}.
  \label{<+label+>}
\end{align}
\end{subequations}
In this case it is easiest to restrict to sub-shells using the SiCs and sSFCs simultaneously. We first note that $\haml{rot1m}$ restricts $\PiPl{A1p}$ to be solenoidal, dual to the gradient of a scalar, and thus eliminates two D.o.F. The remaining D.o.F is eliminated by $\sicl{A2m}$. Similarly, $\sicl{A0p}$ restricts $\PiPl{A1m}$ to a solenoidal axial vector, removing one D.o.F. A further D.o.F is removed by substituting $\haml{rot1p}$ into $\haml{mom1m}$, and a final D.o.F is removed by $\sicl{A2p}$. Separately, $\haml{rot1p}$ removes three D.o.Fs. All the PiCs and SiCs are FC, and one D.o.F remains, as expected from \cref{table-1}
\begin{align}
\begin{split}
  1&=\smash{\frac{1}{2}}\big(80-2\times 10[\text{sPFC}]-2\times (1+3+2)[\text{sSFC}]\\
  &\phantom{=}\ \ \ -2\times(1+3+5+1+5+5)[\text{iPFC}]\\
&\phantom{=}\ \ \ -2\times (1+1+1)[\text{iSFC}]\big).
    \label{case-28_count}
\end{split}
\end{align}
As with \criticalcase{26}, the no-ghost condition $\alp{3}<0$ protects the $0^-$ mode in \eqref{ham28}. However, we note that the linearly-propagating $\PiP{A0m}$ again emerges at the nonlinear level in \eqref{case-28}, so that a linear gauge symmetry is broken and \eqref{case-28_count} is not valid sufficiently far from Minkowski spacetime. Whether or not an increase in the propagating D.o.F results in a ghost is not so clear in \criticalcase{28} as it was in \criticalcase{26}. From \eqref{ham28}, we see that an activation of $\PiP{B1p}$ would endanger nonlinear unitarity by the linear no-tachyon condition $\bet{3}<0$. However, if either of the vector tordions $\PiP{A1m}$ or $\PiP{A1p}$ were to propagate, positive-definite contributions to $\mathcal{  H}_\text{T}$ could be ensured by respectively fixing $\alp{5}<0$ or $\alp{5}>0$, since $\alp{5}$ does not serve to shore up the linearised unitarity. The key point here, as discussed in \cref{sighost}, is that with our `West Coast' signature every contraction on parallel indices introduces a factor of $-1$. Therefore, if \emph{both} vector tordions propate in the nonlinear theory, it would seem that negative kinetic energy contributions to $\mathcal{  H}_\text{T}$ are unavoidable. Whatever the status of ghosts, we observe that the nonlinear PPM has field-dependent rank.

\subsection{\criticalcase{25}}
The structure of \criticalcase{25} has many similarities with that of \criticalcase{28}.
The Hamiltonian takes the form
\begin{equation}
  \begin{aligned}
  \mathcal{  H}_\text{T}=\frac{b}{96}\Bigg(& \frac{4\big(\pic{B0p}^2+9\pic{B1m}\picu{B1m}\big)}{\bet{2}}\\
  &+\frac{18\pic{B1p}\picu{B1p}}{\bet{3}}-\frac{\pic{A0m}^2}{\alp{3}} \Bigg)+\text{fields},
  \label{ham25}
\end{aligned}
\end{equation}
while the nonlinear PPM is more sparesly populated:
\begin{equation}
  \begin{BMAT}{rrccccccl}{ccccccccc}
 & & & & & & & & \\
 & & \scriptstyle{\pic{B2p}} & \scriptstyle{\pic{A0p}} & \scriptstyle{\pic{A1p}} & \scriptstyle{\pic{A1m}} & \scriptstyle{\pic{A2p}} & \scriptstyle{\pic{A2m}} & \\
  &\scriptstyle{\pic{B2p}}  & \BPiP & \cdot & \cdot & \cdot & \cdot & \BPiP!& \scriptstyle{5} \\
  &\scriptstyle{\pic{A0p}}  & \cdot & \cdot & \cdot & \cdot & \cdot & \cdot& \scriptstyle{1} \\
  &\scriptstyle{\pic{A1p}}  & \cdot & \cdot & \cdot & \cdot & \cdot & \cdot& \scriptstyle{3} \\
  &\scriptstyle{\pic{A1m}}  & \cdot & \cdot & \cdot & \cdot & \cdot & \cdot& \scriptstyle{3} \\
  &\scriptstyle{\pic{A2p}}  & \cdot & \cdot & \cdot & \cdot & \cdot & \cdot& \scriptstyle{5} \\
  &\scriptstyle{\pic{A2m}}  & \BPiP! & \cdot & \cdot & \cdot & \cdot & \cdot& \scriptstyle{5} \\
 &  & \scriptstyle{5}  & \scriptstyle{1}  & \scriptstyle{3}  & \scriptstyle{3}  & \scriptstyle{5}  & \scriptstyle{5} & 
\addpath{(2,1,3)uuuuuu}
\addpath{(8,1,3)uuuuuu}
\addpath{(3,6,3)l}
\addpath{(3,6,3)r}
\addpath{(3,6,3)u}
\addpath{(3,6,3)d}
\addpath{(2,3,.)rdlu}
\addpath{(6,7,.)rdlu}
\end{BMAT}

  \label{case-25}
\end{equation}
Within the PiC shell, we have
\begin{subequations}
\begin{align}
  \sicl[\ovl{kl}]{B2p}& \approx -\covderl{\langle \ovl k}\PiPl[\ovl l \rangle]{B1m}\\
\sicl{A0p}& \approx \PiPl{B0p},\\
  \sicl[\ovl{kl}]{A1p}& \approx 2\PiPl[\ovl{kl}]{B1p}-\frac{1}{6}\epsdl{\overline{klm}}\etaul{\ovl{mn}}\covderl{\ovl n}\PiPl{A0m},\\
  \sicl[\ovl k]{A1m}& \approx 2\PiPl[\ovl k]{B1m},
  \label{<+label+>}
\end{align}
\end{subequations}
and this time, all $10$ sSFCs persist in the PiC shell
\begin{subequations}
\begin{align}
  \haml{mom0p}& \approx -\etaul{\ovl{kl}}\covderl{\ovl k}\PiPl[\ovl l]{B1m}\\
  \haml[\alpha]{mom1m}& \approx -\frac{1}{3}\hfl{\alpha}{\ovl k}\covderl{\ovl k}\PiPl{B0p}-\hfl{\alpha}{\ovl k}\etaul{\ovl{jl}}\covderl{\ovl j}\PiPl[\ovl{kl}]{B1p},\\
  \haml{rot1p}& \approx 2\PiPl[\ovl{kl}]{B1p}-\frac{1}{6}\epsdl{\overline{klm}}\etaul{\ovl{mn}}\covderl{\ovl n}\PiPl{A0m},\\
  \haml{rot1m}& \approx \PiPl[\ovl{k}]{B1m}.
\end{align}
\end{subequations}
We find that $\haml{rot1m}$ and $\haml{rot1p}$ each remove three D.o.F, while $\sicl{A0p}$ removes one D.o.F; the remaining sSFCs and SiCs are then implied, and the PiCs and SiCs are FC. Once again, one D.o.F remains as expected from \cref{table-1}
\begin{align}
\begin{split}
  1&=\smash{\frac{1}{2}}\big(80-2\times 10[\text{sPFC}]-2\times (3+3)[\text{sSFC}]\\
  &\phantom{=}\ \ \ -2\times(5+1+3+3+5+5)[\text{iPFC}]\\
&\phantom{=}\ \ \ -2\times 1[\text{iSFC}]\big).
    \label{<+label+>}
\end{split}
\end{align}
The discussion now proceeds in much the same way as with \criticalcase{28}, since PiC commutators linear in the propagating $\PiP{A0m}$ emerge away from Minkowski spacetime. This time, it is the tetrad momenta $\PiP{B0p}$ and $\PiP{B1m}$ which introduce extra uncertainty regarding ghosts. If only one of these momenta becomes activated, $\bet{2}$ may be used to ensure it has a positive contribution to $\mathcal{  H}_\text{T}$. Again, the nonlinear PPM rank is field-dependent.
\subsection{\criticalcase{24}}
\criticalcase{24} has only 16 PiCs, the fewest out of all the cases we consider. The kinetic part of the Hamiltonian is proportionally more complicated
\begin{equation}
  \begin{aligned}
  \mathcal{  H}_\text{T}=\frac{b}{96}\Bigg(& \frac{6\big(\pic{A1m}\picu{A1m}^2+2\pic{A1p}\picu{A1p}\big)}{\alp{5}}+\frac{18\pic{B1p}\picu{B1p}}{\bet{3}}\\
  &+\frac{4\big(\pic{B0p}^2+9\pic{B1m}\picu{B1m}\big)}{\bet{2}}-\frac{\pic{A0m}^2}{\alp{3}} \Bigg)+\text{fields},
  \label{ham24}
\end{aligned}
\end{equation}
while the PPM is extremely small:
\begin{equation}
  \begin{BMAT}{rrccccl}{ccccccc}
 & & & & & & \\
 & & \scriptstyle{\pic{B2p}} & \scriptstyle{\pic{A0p}} & \scriptstyle{\pic{A2p}} & \scriptstyle{\pic{A2m}} & \\
  &\scriptstyle{\pic{B2p}}  & \BPiP & \cdot & \BPiP & \BPiP!& \scriptstyle{5} \\
  &\scriptstyle{\pic{A0p}}  & \cdot & \cdot & \cdot & \cdot& \scriptstyle{1} \\
  &\scriptstyle{\pic{A2p}}  & \BPiP & \cdot & \cdot & \cdot& \scriptstyle{5} \\
  &\scriptstyle{\pic{A2m}}  & \BPiP! & \cdot & \cdot & \cdot& \scriptstyle{5} \\
 &  & \scriptstyle{5}  & \scriptstyle{1}  & \scriptstyle{5}  & \scriptstyle{5} & 
\addpath{(2,1,3)uuuu}
\addpath{(6,1,3)uuuu}
\addpath{(3,4,3)l}
\addpath{(3,4,3)r}
\addpath{(3,4,3)u}
\addpath{(3,4,3)d}
\addpath{(2,3,.)rdlu}
\addpath{(4,5,.)rdlu}
\end{BMAT}

  \label{case-24}
\end{equation}
Within the PiC shell, we have the following SiCs
\begin{subequations}
\begin{align}
  \sicl[\ovl{kl}]{B2p}& \approx -\covderl{\langle \ovl k}\PiPl[\ovl l \rangle]{B1m}\\
  \sicl{A0p}& \approx \PiPl{B0p}-\etaul{\ovl{kl}}\covderl{\ovl k}\PiPl[\ovl l]{A1m},\\
  \sicl[\ovl{kl}]{A2p}& \approx \frac{1}{2}\covderl{\langle\ovl k}\PiPl[\ovl{l}\rangle]{A1m},\\
  \sicl[\ovl{klm}]{A2m}& \approx \frac{1}{2}\covderl{\ovl m}\PiPl[\ovl {kl}]{A1p}+\frac{1}{2}\covderl{{{[}\ovl{l}}}\PiPl[{\ovl{k}{]}\ovl{m}}]{A1p}\nonumber\\
  &\ \ \ +\frac{3}{4}\etadl{\ovl{m}[\ovl{k}|}\etaul{\ovl{ij}}\covderl{\ovl i}\PiPl[{|\ovl{l}{]}\ovl{j}}]{A1p},
  \label{<+label+>}
\end{align}
\end{subequations}
and the following sPFCs
\begin{subequations}
\begin{align}
  \haml{mom0p}& \approx -\etaul{\ovl{kl}}\covderl{\ovl k}\PiPl[\ovl l]{B1m}\\
  \haml[\alpha]{mom1m}& \approx -\frac{1}{3}\hfl{\alpha}{\ovl k}\covderl{\ovl k}\PiPl{B0p}-\hfl{\alpha}{\ovl k}\etaul{\ovl{jl}}\covderl{\ovl j}\PiPl[\ovl{kl}]{B1p},\\
  \haml{rot1p}& \approx 2\PiPl[\ovl{kl}]{B1p}-\frac{1}{6}\epsdl{\overline{klm}}\etaul{\ovl{mn}}\covderl{\ovl n}\PiPl{A0m}+\covderl{[\ovl{k}}\PiPl[{{\ovl{l}]}}]{A1m},\\
  \haml{rot1m}& \approx \PiPl[\ovl{k}]{B1m}+\etaul{\ovl{jl}}\covderl{\ovl j}\PiPl[\ovl{kl}]{A1p}.
  \label{<+label+>}
\end{align}
\end{subequations}
It is clear from the PiC shell that \criticalcase{24} has much in common with \criticalcase{28}, and again we will implement the SiCs and sSFCs simultaneously. Firstly, we find that $\sicl{A2m}$ constitutes an overdetermined system in $\PiPl{A1p}$, which vanishes and takes with it three D.o.F. Consequently, from $\haml{rot1m}$, we see that $\PiPl{B1m}$ must vanish along with another three D.o.F, such that $\haml{mom0p}$ and $\sicl{B2p}$ vanish automatically. Similarly, $\sicl{A2p}$ is an overdetermined system in $\PiPl{A1p}$, which vanishes with another three D.o.F; $\sicl{A0p}$ then causes $\PiPl{B0p}$ to vanish with one D.o.F. As before, one D.o.F propagates in accordance with \cref{table-1}
\begin{align}
\begin{split}
  1&=\smash{\frac{1}{2}}\big(80-2\times 10[\text{sPFC}]-2\times(3+3)[\text{sSFC}]\\
    &\phantom{=}\ \ \ -2\times(5+1+5+5)[\text{iPFC}]\\
  &\phantom{=}\ \ \ -2\times(1+3+3)[\text{iSFC}]\big).
    \label{<+label+>}
\end{split}
\end{align}
It is clear from \eqref{ham24} that any inference of the nonlinear unitarity will just combine the discussions of \criticalcase{28} and \criticalcase{25}, while the PPM rank is again field dependent.
\subsection{\criticalcase{32}}
For the first time, we encounter non-vanishing mass parameters between the PiCs, specifically in $\picl{A1p}$ and $\picl{A2m}$. We anticipate the nonvanishing commutators even at the linear level $\Big\{\picl[\ovl{kl}]{A1p},\sicl[\ovl{ij}]{A1p}\Big\}\approx\mathcal{O}(1)$ -- noting that the natural conjugate $\picl{B1p}$ is not a PiC -- and $\Big\{\picl[\ovl{klm}]{A2m},\sicl[\ovl{ijn}]{A2m}\Big\}\approx\mathcal{O}(1)$. These PiCs and SiCs will be SC, allowing for the determination of their multipliers. The kinetic part of the Hamiltonian is
\begin{equation}
  \begin{aligned}
  \mathcal{  H}_\text{T}=\frac{b}{96}\Bigg(& \frac{6\big(3\pic{B1m}\picu{B1m}+2\pic{B2p}\picu{B2p}\big)}{\bet{1}}\\
  &+\frac{36\pic{B1p}\picu{B1p}}{\bet{1}+2\bet{3}}-\frac{\pic{A0m}^2}{\alp{3}} \Bigg)+\text{fields}.
  \label{ham32}
\end{aligned}
\end{equation}
In the PPM, we label the PiCs associated with nonvanishing mass parameters by ($\downarrow$), producing:
\begin{equation}
  \begin{BMAT}{rrccccccl}{ccccccccc}
 & & & & \scriptstyle{\downarrow} & & & \scriptstyle{\downarrow} & \\
 & & \scriptstyle{\pic{B0p}} & \scriptstyle{\pic{A0p}} & \scriptstyle{\pic{A1p}} & \scriptstyle{\pic{A1m}} & \scriptstyle{\pic{A2p}} & \scriptstyle{\pic{A2m}} & \\
  &\scriptstyle{\pic{B0p}}  & \cdot & \cdot & \cdot & \cdot & \cdot & \cdot& \scriptstyle{1} \\
  &\scriptstyle{\pic{A0p}}  & \cdot & \cdot & \cdot & \cdot & \cdot & \cdot& \scriptstyle{1} \\
\scriptstyle{\rightarrow}  &\scriptstyle{\pic{A1p}}  & \cdot & \cdot & \cdot & \cdot & \cdot & \cdot& \scriptstyle{3} \\
  &\scriptstyle{\pic{A1m}}  & \cdot & \cdot & \cdot & \cdot & \cdot & \cdot& \scriptstyle{3} \\
  &\scriptstyle{\pic{A2p}}  & \cdot & \cdot & \cdot & \cdot & \cdot & \cdot& \scriptstyle{5} \\
\scriptstyle{\rightarrow}  &\scriptstyle{\pic{A2m}}  & \cdot & \cdot & \cdot & \cdot & \cdot & \cdot& \scriptstyle{5} \\
 &  & \scriptstyle{1}  & \scriptstyle{1}  & \scriptstyle{3}  & \scriptstyle{3}  & \scriptstyle{5}  & \scriptstyle{5} & 
\addpath{(2,1,3)uuuuuu}
\addpath{(8,1,3)uuuuuu}
\addpath{(3,6,3)l}
\addpath{(3,6,3)r}
\addpath{(3,6,3)u}
\addpath{(3,6,3)d}
\addpath{(2,6,.)rdlu}
\addpath{(3,7,.)rdlu}
\end{BMAT}

  \label{case-32}
\end{equation}
Thus the PPM of this theory is remarkable, since it remains empty even in the nonlinear regime.
Within the PiC shell, we find the following SiCs
\begin{subequations}
\begin{align}
  \sicl{B0p}& \approx -\etaul{\ovl{kl}}\covderl{\ovl k}\PiPl[\ovl l]{B1m},\\
  \sicl[\ovl{kl}]{A1p}& \approx -\frac{\bet{1}+2\bet{3}}{\bet{1}-\bet{3}}\PiPl[\ovl{kl}]{B1p}\nonumber\\
  &\ \ \  -\frac{1}{6}\epsdl{\overline{klm}}\etaul{\ovl{mn}}\covderl{\ovl n}\PiPl{A0m}\nonumber\\
  &\ \ \ +\frac{9\bet{1}\bet{3}}{(\bet{1}-\bet{3})(\bet{1}+2\bet{3})}\picl[\ovl{kl}]{B1p},\\
  \sicl[\ovl k]{A1m}& \approx -\PiPl[\ovl k]{B1m},\\
  \sicl[\ovl{kl}]{A2p}& \approx \PiPl[\ovl{kl}]{B2p},\\
  \sicl[\ovl{klm}]{A2m}& \approx 4\bet{1}{m_\text{p}}^2\Tl[\ovl{klm}]{A2m}.
  \label{<+label+>}
\end{align}
\end{subequations}
Note the appearance of field strengths, specifically the torsion in $\picl{B1p}$ and $\Tl{A2m}$. Whilst these somewhat complicate the analysis, they naturally appear with the mass parameters. We also mark the first apparent instance of a TiC accompanying $\sicl[\ovl{kl}]{A2p}$. Using the notation $\zeta\equiv\dot\chi$, this may be written as
\begin{equation}
  \ticl[\ovl{kl}]{A2p} \approx \frac{4}{3}\etaul{\ovl{ij}}\covderl{\ovl i}\sicl[\langle\ovl{k}|\ovl{j}|\ovl{l}\rangle]{A2m},
  \label{TiC_32}
\end{equation}
which then vanishes in the SiC shell. The PiC shell contains the following sSFCs:
\begin{subequations}
\begin{align}
  \haml{mom0p}& \approx -\etaul{\ovl{kl}}\covderl{\ovl k}\PiPl[\ovl l]{B1m}\\
  \haml[\alpha]{mom1m}& \approx -\hfl{\alpha}{\ovl k}\etaul{\ovl{jl}}\covderl{\ovl j}\PiPl[\ovl{kl}]{B1p}-\hfl{\alpha}{\ovl k}\etaul{\ovl{jl}}\covderl{\ovl j}\PiPl[\ovl{kl}]{B2p},\\
  \haml{rot1p}& \approx 2\PiPl[\ovl{kl}]{B1p}-\frac{1}{6}\epsdl{\overline{klm}}\etaul{\ovl{mn}}\covderl{\ovl n}\PiPl{A0m},\\
  \haml{rot1m}& \approx \PiPl[\ovl{k}]{B1m}.
  \label{<+label+>}
\end{align}
\end{subequations}
Since two of the PiC chains are known to be self-terminating, the algorithm concludes quite quickly. As with \criticalcase{25}, $\haml{rot1m}$ and $\haml{rot1p}$ each eliminate three D.o.F. Another five D.o.F are then removed by $\sicl{A2p}$, with the remaining SiCs and sSFCs automatically satisfied. The one remaining D.o.F is again expected from \cref{table-1}
\begin{align}
\begin{split}
  1&=\smash{\frac{1}{2}}\big(80-2\times 10[\text{sPFC}]-2\times(3+3)[\text{sSFC}]\\
    &\phantom{=}\ \ \ -2\times(1+1+3+5)[\text{iPFC}]-(3+5)[\text{iPSC}]\\
  &\phantom{=}\ \ \ -2\times5[\text{iSFC}]-(3+5)[\text{iSSC}]\big).
    \label{<+label+>}
\end{split}
\end{align}
On the whole, the outlook for \criticalcase{32} appears more promising than for the previous cases, because the PPM retains its empty structure (and rank) when passing to the nonlinear regime. This is just the first hurdle, as the full nonlinear algorithm would still be required to determine whether further fields become activated. The implications of field activaton are slightly relaxed, compared to \criticalcase{25} or \criticalcase{28}. The linear tachyon condition $\bet{3}<0$ need not imply that a propagating $\PiP{B1p}$ contributes negative kinetic energy if $\bet{1}+2\bet{3}>0$. This can be realised even if $\PiP{B2p}$ is simultaneously activated. However for positive kinetic energy it seems $\PiP{B1m}$ must be activated on its own or not at all, since $\bet{1}<0$ would then be required.
\subsection{\criticalcase{20}}
The analysis of \criticalcase{20} is quite similar to \criticalcase{32}. Mass parameters again accompany the PiCs, and we expect $\picl{A1p}$, $\picl{A1m}$ and $\picl{A2m}$ to not commute with their respective SiCs on the final shell. The kinetic part of the Hamiltonian is
\begin{equation}
  \begin{aligned}
  &\mathcal{  H}_\text{T}=\frac{b}{96}\Bigg( \frac{4\pic{B0p}^2}{\bet{2}}+\frac{12\pic{B2p}\picu{B2p}}{\bet{1}}\\
  &+\frac{36\pic{B1p}\picu{B1p}}{\bet{1}+2\bet{3}}+\frac{36\pic{B1m}\picu{B1m}}{2\bet{1}+\bet{2}}-\frac{\pic{A0m}^2}{\alp{3}} \Bigg)+\text{fields},
  \label{ham20}
\end{aligned}
\end{equation}
and once again the PPM is empty both before and after linearisation:
\begin{equation}
  \begin{BMAT}{rrcccccl}{cccccccc}
 & & & \scriptstyle{\downarrow} & \scriptstyle{\downarrow} & & \scriptstyle{\downarrow} & \\
 & & \scriptstyle{\pic{A0p}} & \scriptstyle{\pic{A1p}} & \scriptstyle{\pic{A1m}} & \scriptstyle{\pic{A2p}} & \scriptstyle{\pic{A2m}} & \\
  &\scriptstyle{\pic{A0p}}  & \cdot & \cdot & \cdot & \cdot & \cdot& \scriptstyle{1} \\
\scriptstyle{\rightarrow}  &\scriptstyle{\pic{A1p}}  & \cdot & \cdot & \cdot & \cdot & \cdot& \scriptstyle{3} \\
\scriptstyle{\rightarrow}  &\scriptstyle{\pic{A1m}}  & \cdot & \cdot & \cdot & \cdot & \cdot& \scriptstyle{3} \\
  &\scriptstyle{\pic{A2p}}  & \cdot & \cdot & \cdot & \cdot & \cdot& \scriptstyle{5} \\
\scriptstyle{\rightarrow}  &\scriptstyle{\pic{A2m}}  & \cdot & \cdot & \cdot & \cdot & \cdot& \scriptstyle{5} \\
 &  & \scriptstyle{1}  & \scriptstyle{3}  & \scriptstyle{3}  & \scriptstyle{5}  & \scriptstyle{5} & 
\addpath{(2,1,3)uuuuu}
\addpath{(7,1,3)uuuuu}
\end{BMAT}

  \label{<+label+>}
\end{equation}
Within the PiC shell, we first the following SiCs
\begin{subequations}
\begin{align}
  \sicl{A0p}& \approx \PiPl[\ovl l]{B0p},\\
  \sicl[\ovl{kl}]{A1p}& \approx -\frac{\bet{1}+2\bet{3}}{\bet{1}-\bet{3}}\PiPl[\ovl{kl}]{B1p}\nonumber\\
  &\ \ \  -\frac{1}{6}\epsdl{\overline{klm}}\etaul{\ovl{mn}}\covderl{\ovl n}\PiPl{A0m}\nonumber\\
  &\ \ \ +\frac{9\bet{1}\bet{3}}{(\bet{1}-\bet{3})(\bet{1}+2\bet{3})}\picl[\ovl{kl}]{B1p},\\
  \sicl[\ovl k]{A1m}& \approx -\frac{\bet{1}+2\bet{2}}{\bet{1}-\bet{2}}\PiPl[\ovl k]{B1m}\nonumber\\
  &\ \ \ +\frac{9\bet{1}\bet{2}}{(\bet{1}-\bet{2})(2\bet{1}+\bet{2})}\picl[\ovl k ]{B1m},\\
  \sicl[\ovl{kl}]{A2p}& \approx \PiPl[\ovl{kl}]{B2p},\\
  \sicl[\ovl{klm}]{A2m}& \approx 4\bet{1}{m_\text{p}}^2\Tl[\ovl{klm}]{A2m}.
  \label{<+label+>}
\end{align}
\end{subequations}
This time, two TiCs appear, but uppon rearranging both may eventually be written in terms of the iSSCs, and are therefore satisfied automatically
\begin{subequations}
\begin{align}
  \ticl{A0p} & \approx \etaul{\ovl{ij}}\covderl{\ovl i}\sicl[\ovl{j}]{A1m},\\
  \ticl[\ovl{kl}]{A2p} & \approx \frac{4}{3}\etaul{\ovl{ij}}\covderl{\ovl i}\sicl[\langle\ovl{k}|\ovl{j}|\ovl{l}\rangle]{A2m}-\frac{1}{2}\covderl{\langle\ovl k}\sicl[\ovl l\rangle]{A1m}.
  \label{<+label+>}
\end{align}
\end{subequations}
The sSFC content in the PiC shell is largely the same as that of \criticalcase{32}, with the only difference marked in the linear super-momentum
\begin{subequations}
\begin{align}
  \haml[\alpha]{mom1m}& \approx -\frac{1}{3}\hfl{\alpha}{\ovl k}\covderl{\ovl k}\PiPl{B0p}-\hfl{\alpha}{\ovl k}\etaul{\ovl{jl}}\covderl{\ovl j}\PiPl[\ovl{kl}]{B1p}\nonumber\\
  &\ \ \ -\hfl{\alpha}{\ovl k}\etaul{\ovl{jl}}\covderl{\ovl j}\PiPl[\ovl{kl}]{B2p}.
  \label{<+label+>}
\end{align}
\end{subequations}
Aided by the additional conjugate pair of constraints, the algorithm terminates even faster than with \criticalcase{32}: we see that one and five D.o.F are removed by each of $\sicl{A0p}$ and $\picl{A2p}$. As before, the one propagating D.o.F is confirmed from \cref{table-1}
\begin{align}
\begin{split}
  1&=\smash{\frac{1}{2}}\big(80-2\times 10[\text{sPFC}]-2\times(3+3)[\text{sSFC}]\\
    &\phantom{=}\ \ \ -2\times(1+5)[\text{iPFC}]-(3+3+5)[\text{iPSC}]\\
  &\phantom{=}\ \ \ -2\times(1+5)[\text{iSFC}]-(3+3+5)[\text{iSSC}]\big).
    \label{<+label+>}
\end{split}
\end{align}
If positive kinetic energy is a requirement, it seems that the momenta $\PiP{B0p}$ and $\PiP{B1m}$ in combination with one or more of $\PiP{B2p}$ or $\PiP{B1p}$, should not all be activated at the same time.
\section{Massless results}\label{massless_theories}
\subsection{\criticalcase{17}}
Two theories in \cref{table-1} -- \criticalcase{17} and \criticalcase{3} -- admit a pair of massless modes according to the linearised analysis. Beginning with \criticalcase{17}, we find the Hamiltonian to have the structure
\begin{equation}
  \begin{aligned}
    \mathcal{  H}_\text{T}=\frac{b}{32}\Bigg(& \frac{2\big(\pic{A1m}\picu{A1m}+2\pic{A1p}\picu{A1p}\big)}{\alp{5}}\\
  &-\frac{3\pic{B1m}\picu{B1m}+2\pic{B2p}\picu{B2p}}{\bet{3}} \Bigg)+\text{fields},
  \label{ham17}
\end{aligned}
\end{equation}
As mentioned in \cref{poisson_section}, the evaluation of the PPM is complicated by the appearance of torsion in PiC $\pic{B1p}$ belonging to the translational sector.
In general, commutators between field strengths generate derivatives of the Dirac function. In many cases, these derivatives either happen to cancel, or they may be discarded up to a surface term within the PiC shell. In any case, we find that the full nonlinear PPM can be written purely in terms of the parallel momenta as before:
\begin{equation}
  \begin{BMAT}{rrccccccl}{ccccccccc}
 & & & \scriptstyle{\downarrow} & & \scriptstyle{\downarrow} & & \scriptstyle{\downarrow} & \\
 & & \scriptstyle{\pic{B0p}} & \scriptstyle{\pic{B1p}} & \scriptstyle{\pic{A0p}} & \scriptstyle{\pic{A0m}} & \scriptstyle{\pic{A2p}} & \scriptstyle{\pic{A2m}} & \\
  &\scriptstyle{\pic{B0p}}  & \cdot & \BPiP & \cdot & \cdot & \cdot & \cdot& \scriptstyle{1} \\
\scriptstyle{\rightarrow}  &\scriptstyle{\pic{B1p}}  & \BPiP & \cdot & \BPiP & \BPiP & \BPiP & \BPiP& \scriptstyle{3} \\
  &\scriptstyle{\pic{A0p}}  & \cdot & \BPiP & \cdot & \cdot & \cdot & \cdot& \scriptstyle{1} \\
\scriptstyle{\rightarrow}  &\scriptstyle{\pic{A0m}}  & \cdot & \BPiP & \cdot & \cdot & \cdot & \cdot& \scriptstyle{1} \\
  &\scriptstyle{\pic{A2p}}  & \cdot & \BPiP & \cdot & \cdot & \cdot & \cdot& \scriptstyle{5} \\
\scriptstyle{\rightarrow}  &\scriptstyle{\pic{A2m}}  & \cdot & \BPiP & \cdot & \cdot & \cdot & \cdot& \scriptstyle{5} \\
 &  & \scriptstyle{1}  & \scriptstyle{3}  & \scriptstyle{1}  & \scriptstyle{1}  & \scriptstyle{5}  & \scriptstyle{5} & 
\addpath{(2,1,3)uuuuuu}
\addpath{(8,1,3)uuuuuu}
\addpath{(4,5,3)l}
\addpath{(4,5,3)r}
\addpath{(4,5,3)u}
\addpath{(4,5,3)d}
\addpath{(2,5,.)rdlu}
\addpath{(4,7,.)rdlu}
\end{BMAT}

  \label{case-17}
\end{equation}
Due to the appearance of mass parameters, we will expect $\picl{B1p}$, $\picl{A0m}$ and $\picl{A2m}$ not to commute with their SiCs in the final shell. Within the PiC shell, we find the following SiCs 
\begin{subequations}
\begin{align}
  \sicl[\ovl{kl}]{B0p}& \approx -\etaul{\ovl{kl}}\covderl{\ovl k}\PiPl[\ovl l]{B1m},\\
  \sicl[\ovl{kl}]{B1p}& \approx -\frac{2\bet{3}}{\alp{5}}{m_\text{p}}^2\PiPl[\ovl{kl}]{A1p}-\covderl{{[\ovl{k}}}\PiPl[{\ovl{l}{]}}]{B1m}\nonumber\\
  &\ \ \ -8\bet{3}{m_\text{p}}^2\covderl{{[\ovl{k}}}\Tl[{\ovl{l}{]}}]{A1m},\\
  \sicl{A0p}& \approx -\etaul{\ovl{kl}}\covderl{\ovl k}\PiPl[\ovl l]{A1m},\\
  \sicl{A0m}& \approx 2\epsul{\ovl{jkl}}\covderl{\ovl j}\PiPl[\ovl{kl}]{A1p}-8\bet{3}{m_\text{p}}^2\Tl[\ovl{jkl}]{A0m},\\
  \sicl[\ovl{kl}]{A2p}& \approx \PiPl[\ovl{kl}]{B2p}+\frac{1}{2}\covderl{\langle\ovl k}\PiPl[\ovl{l}\rangle]{A1m},\\
  \sicl[\ovl{klm}]{A2m}& \approx \frac{1}{2}\covderl{\ovl m}\PiPl[\ovl {kl}]{A1p}+\frac{1}{2}\covderl{{{[}\ovl{l}}}\PiPl[{\ovl{k}{]}\ovl{m}}]{A1p}\nonumber\\
  &\ \ \ +\frac{3}{4}\etadl{\ovl{m}[\ovl{k}|}\etaul{\ovl{ij}}\covderl{\ovl i}\PiPl[{|\ovl{l}{]}\ovl{j}}]{A1p}\nonumber\\
  &\ \ \ -8\bet{3}{m_\text{p}}^2\Tl[\ovl{klm}]{A2m}.
  \label{<+label+>}
\end{align}
\end{subequations}
Among these, we note that a TiC accompanies $\sicl{A2p}$, but may be expressed in terms of $\sicl{A2m}$ by precisely \eqref{TiC_32}. Within the PiC shell, the sSFCs are
\begin{subequations}
\begin{align}
  \haml{mom0p}& \approx -\etaul{\ovl{kl}}\covderl{\ovl k}\PiPl[\ovl l]{B1m},\\
  \haml[\alpha]{mom1m}& \approx-\hfl{\alpha}{\ovl k}\etaul{\ovl{jl}}\covderl{\ovl j}\PiPl[\ovl{kl}]{B1p}-\hfl{\alpha}{\ovl k}\etaul{\ovl{jl}}\covderl{\ovl j}\PiPl[\ovl{kl}]{B2p},\\
  \haml{rot1p}& \approx 2\PiPl[\ovl{kl}]{B1p}+\covderl{[\ovl{k}}\PiPl[{{\ovl{l}]}}]{A1m},\\
  \haml{rot1m}& \approx \PiPl[\ovl{k}]{B1m}+\etaul{\ovl{jl}}\covderl{\ovl j}\PiPl[\ovl{kl}]{A1p}.
  \label{<+label+>}
\end{align}
\end{subequations}
The conjugate pairs together eliminate six, two and $10$ D.o.F before terminating. As with \criticalcase{28}, $\PiPl{A1m}$ becomes solenoidal due to $\sicl{A0p}$ and loses one D.o.F, while three D.o.F are lost by each of $\haml{rot1p}$ and $\haml{rot1m}$. In total, two propagating D.o.F remain as expected from \cref{table-1}
\begin{align}
\begin{split}
  2&=\smash{\frac{1}{2}}\big(80-2\times 10[\text{sPFC}]-2\times(3+3)[\text{sSFC}]\\
    &\phantom{=}\ \ \ -2\times(1+1+5)[\text{iPFC}]-(3+1+5)[\text{iPSC}]\\
  &\phantom{=}\ \ \ -2\times(1+5)[\text{iSFC}]-(3+1+5)[\text{iSSC}]\big).
    \label{<+label+>}
\end{split}
\end{align}

According to \cref{table-1}, these two D.o.F should be massless, and the power in the $2^+$ sector of the propagator invites speculation as to whether they constitute a graviton. In \cref{section_26} we were able to show that the one D.o.F of \criticalcase{26} belonged unambiguously to the $0^-$ sector, but our method cannot be so straightforwardly applied to \criticalcase{17}. This is ultimately related to the fact that the dependence of the PiC $\pic{B1p}$ on the parallel torsion $\tensor{\mathcal{  T}}{^i_{\ovl{kl}}}$ in \eqref{PiCB1p}, survives even when the defining constraints of \criticalcase{17} are imposed on the couplings. In the linear theory, this results in a conjugate SiC $\sicl[\ovl{kl}]{B1p}$ which depends on the gradient of the torsion $\covderl{{[\ovl{k}}}\Tl[{\ovl{l}{]}}]{A1m}$. This is problematic when it comes to determining the linear multiplier $\mull{B1p}$ through the consistency condition \eqref{consistency2}. The result is a PDE in the multiplier
\begin{equation}
  -8\bet{3}\covderl{[\ovl{k}|}\etaul{\ovl{ij}}\covderl{\ovl{i}}\mull[{|\ovl{l}{]}\ovl{j}}]{B1p}+16\frac{\bet{3}^2}{\alp{5}}\mull[\ovl{kl}]{B1p}\approx \tensor{X}{_{[\ovl{kl}]}},
\end{equation}
where the inhomogeneous part on the RHS results from a second-order Euler--Lagrange variation in the Poisson bracket, and likely involves gradients of the equal-time Dirac function. The remaining six determinable multipliers are not affected by this problem; we find with relative ease
\begin{subequations}
\begin{align}
  \iffalse
&\covderl{[\ovl{k}|}\etaul{\ovl{ij}}\covderl{\ovl{i}}\big(\PiPl[{|\ovl{l}{]}\ovl{j}}]{B2p}+\PiPl[{|\ovl{l}{]}\ovl{j}}]{B1p}\big)+\frac{\bet{3}}{\alp{5}}\covderl{[\ovl{k}}\PiPl[{\ovl{l}{]}}]{A1m}\nonumber\\
&-2\bet{3}\etaul{\ovl{ij}}\covderl{\ovl{i}}\covderl{\ovl{j}}\cTl[\ovl{kl}]{B1p}+\bet{3}\etaul{\ovl{ij}}\covderl{\ovl{i}}\big(7\tensor{\mathcal{  R}}{^{\flat}_{\perp\ovl{jkl}}}-8\cRl[\ovl{ikl}]{A2m}\big)\nonumber\\
&+16\frac{\bet{3}^2}{\alp{5}}\mull[\ovl{kl}]{B1p}-8\bet{3}\covderl{[\ovl{k}|}\etaul{\ovl{ij}}\covderl{\ovl{i}}\mull[{|\ovl{l}{]}\ovl{j}}]{B1p}=0,\\
\fi
  \mull{A0m}&\approx0,\\
  \mull{A2m}&\approx \frac{3}{4}\etadl{[\ovl{k}|\ovl{m}}\etaul{\ovl{ij}}\covderl{\ovl{i}}\PiPl[{|\ovl{l}{]}\ovl{j}}]{B2p}-\frac{3}{2}\covderl{[\ovl{k}}\PiPl[{\ovl{l}{]}\ovl{m}}]{B2p}\nonumber\\
&\ \ \ +\frac{3}{16}\etadl{[\ovl{k}|\ovl{m}}\etaul{\ovl{ij}}\big(\covderl{|\ovl{l}{]}}\covderl{\ovl{i}}\PiPl[\ovl{j}]{A1m}-\covderl{\ovl{i}}\covderl{\ovl{j}}\PiPl[{|\ovl{l}{]}}]{A1m}\big)\nonumber\\
&\ \ \ +\frac{3}{8}\covderl{\ovl{m}}\covderl{[\ovl{k}}\PiPl[\ovl{l}{]}]{A1m}+8\bet{3}\cRl{A2m}.
  \label{<+label+>}
\end{align}
\end{subequations}
The ambiguity of $\mull{B1p}$ is problematic, as this multiplier lingers in the equations of motion. Even worse, the eight indeterminate multipliers $\mull{B0p}$, $\mull{A0m}$ and $\mull{B2p}$ associated with the iPFCs also feature prominently. In order to discover the $J^P$ character of the propagating modes using Hamiltonian methods, one would have to fix the gauge.

We can draw some tentative conclusions just from the kinetic part of the Hamiltonian however. We see from \cref{table-1} that the linear theory is unitary if only $\alp{5}<0$. If unitarity is to be associated with the positive energy test, then the appearance of $\alp{5}$ in \eqref{ham17} would suggest that $\alp{5}<0$ serves to prevent the $1^-$ mode from becoming a ghost. By the same arguments, the $1^+$ mode should be strongly coupled within the final shell of the linearised theory, since it would otherwise enter with negative kinetic energy. It is reassuring to see from \cref{table-1} that the massless propagator does indeed have power in the $1^-$, but not the $1^+$ sectors. However, it also has power in the $2^+$ sector, possibly inviting speculation that the theory may contain a spin-two graviton akin to that of Einstein.
While $\PiP{B2p}$ does feature in \eqref{ham17}, it seems unlikely that this mode would independently propagate, since the unitarity of the theory does not depend on $\bet{3}$.
We reiterate that these conclusions may ultimately depend on the gauge choice.

Finally, without a definite understanding of the propagating $J^P$, we are unable to say concretely whether fields will be activated or the PPM rank be field-dependent in the nonlinear theory. It is quite likely that these phenomena will occur, since we find in \cref{brackets} that the commutator of $\pic{B1p}$ and $\pic{A2m}$ depends on $\PiP{A1m}$. This has precedent, since the $2^-$ commutator has spoiled all the theories in \cref{massive_only}, but due to the lingering gauge ambiguity we denote it with ($\BPiP$) rather than ($\BPiP!$) in \eqref{case-17}.
\subsection{\criticalcase{3}}
It should come as no surprise that \criticalcase{3} is a relaxation of \criticalcase{17}, which admits an extra D.o.F. The kinetic part of the Hamiltonian is given by \eqref{ham17}, in addition to the pseudoscalar term encountered in all the cases of \cref{massive_only}. This is the usual massive $0^-$ mode, and comes with the no-ghost condition $\alp{3}<0$. The extra condition $\bet{1}$ will prevent this mode from being tachyonic.
The nonlinear PPM is:
\begin{equation}
  \begin{BMAT}{rrcccccl}{cccccccc}
 & & & \scriptstyle{\downarrow} & & & \scriptstyle{\downarrow} & \\
 & & \scriptstyle{\pic{B0p}} & \scriptstyle{\pic{B1p}} & \scriptstyle{\pic{A0p}} & \scriptstyle{\pic{A2p}} & \scriptstyle{\pic{A2m}} & \\
  &\scriptstyle{\pic{B0p}}  & \cdot & \BPiP & \cdot & \cdot & \cdot& \scriptstyle{1} \\
\scriptstyle{\rightarrow}  &\scriptstyle{\pic{B1p}}  & \BPiP & \cdot & \BPiP & \BPiP & \BPiP!& \scriptstyle{3} \\
  &\scriptstyle{\pic{A0p}}  & \cdot & \BPiP & \cdot & \cdot & \cdot& \scriptstyle{1} \\
  &\scriptstyle{\pic{A2p}}  & \cdot & \BPiP & \cdot & \cdot & \cdot& \scriptstyle{5} \\
\scriptstyle{\rightarrow}  &\scriptstyle{\pic{A2m}}  & \cdot & \BPiP! & \cdot & \cdot & \cdot& \scriptstyle{5} \\
 &  & \scriptstyle{1}  & \scriptstyle{3}  & \scriptstyle{1}  & \scriptstyle{5}  & \scriptstyle{5} & 
\addpath{(2,1,3)uuuuu}
\addpath{(7,1,3)uuuuu}
\addpath{(4,4,3)l}
\addpath{(4,4,3)r}
\addpath{(4,4,3)u}
\addpath{(4,4,3)d}
\addpath{(2,4,.)rdlu}
\addpath{(4,6,.)rdlu}
\end{BMAT}

  \label{<+label+>}
\end{equation}
It is clear that $\picl{A0m}$ is no longer primarily constrained. The only change to the remaining SiCs of \criticalcase{17} is 
\begin{equation}
  \haml{rot1p} \approx 2\PiPl[\ovl{kl}]{B1p}-\frac{1}{6}\epsdl{\overline{klm}}\etaul{\ovl{mn}}\covderl{\ovl n}\PiPl{A0m}+\covderl{[\ovl{k}}\PiPl[{{\ovl{l}]}}]{A1m},
  \label{<+label+>}
\end{equation}
but $\haml{mom1m}$ is still satisfied. Overall, only the conjugate $\picl{A0m}$ and $\sicl{A0m}$ pair are removed, leaving three propagating D.o.F as expected from \cref{table-1}
\begin{align}
\begin{split}
  3&=\smash{\frac{1}{2}}\big(80-2\times 10[\text{sPFC}]-2\times(3+3)[\text{sSFC}]\\
    &\phantom{=}\ \ \ -2\times(1+1+5)[\text{iPFC}]-(3+5)[\text{iPSC}]\\
  &\phantom{=}\ \ \ -2\times(1+5)[\text{iSFC}]-(3+5)[\text{iSSC}]\big).
    \label{<+label+>}
\end{split}
\end{align}
We draw the same conclusions from \criticalcase{3} as from \criticalcase{17} regarding the \emph{vector} nature of the gravitational particle. This time however, we note the presence of $\PiP{A0m}$ in the nonlinear PPM, indicating that whatever the massless $J^P$, at least one gauge symmetry does not survive in the nonlinear regime.
\section{Phenomenology}\label{phenomenology}
The results of \cref{massless_theories} cast serious doubts on the health of even the massless theories considered here, on quite general grounds.
We can in fact rule these theories out more conclusively on the basis of their cosmology.
In general, this would be quite an arduous task, requiring a dedicated examination of all four equations of motion. However, we recently developed a mapping between the general quadratic torsion theory \eqref{lagrangian_soft} and a torsion-free biscalar-tensor theory, which immediately reveals the cosmological background~\cite{2020arXiv200603581B}. We begin with the spatially flat FRW line element 
\begin{equation}
  \mathrm{d}s^2=\mathrm{d}t^2-a^2\mathrm{d}\mathbf{x}^2,
  \label{<+label+>}
\end{equation}
where $a$ is the scale factor, normalised to the contemporary epoch, from which we define the Hubble number $H\equiv \dot{a}/a$. We now align the unit timelike normal $\tensor{n}{^k}$ to be perpendicular to the spatially flat slicing. Cosmological isotropy at the background level restricts only the $0^+$ and $0^-$ torsion modes to propagate. From these modes respectively we define a pair of scalar fields
\begin{equation}
  \phi\equiv \frac{2}{3}\tensor{\mathcal{  T}}{^{\ovl{k}}_{\perp\ovl{k}}}-2H, \quad \psi\equiv \frac{1}{6}\tensor{\epsilon}{_{\ovl{i}}^{\perp\ovl{jk}}}\tensor{\mathcal{  T}}{^{\ovl{i}}_{\ovl{jk}}}.
  \label{scaldef}
\end{equation}
These fields transform homogeneously and with the correct weight ${\phi\mapsto\Omega^{-1}\phi}$, ${\psi\mapsto\Omega^{-1}\psi}$ under changes of physical scale ${\tensor{b}{^i_\mu}\mapsto\Omega\tensor{b}{^i_\mu}}$. In the usual second-order formulation of gravity on the curved spacetime $\mathcal{  M}$, it can be shown that the theory
  \begin{subequations}
    \begin{align}
	L_{\text{G}}&\simeq\Big[ \bet{2}{m_{\text{p}}}^2+\frac{1}{4}(\alp{4}+\alp{6})\phi^2-\frac{1}{4}(\alp{2}+\alp{3})\psi^2 \Big]R\nonumber\\
	&\ \ +3(\alp{4}+\alp{6})X^{\phi\phi}-3(\alp{2}+\alp{3})X^{\psi\psi}+\sqrt{\vphantom{X^a_a}\smash[b]{|J_\mu J^\mu|}}\nonumber\\
	&\ \ +\frac{3}{4}(\alpha_{0}+2\bet{2}){m_{\text{p}}}^2\phi^2-\frac{3}{4}(\alpha_{0}+8\bet{3}){m_{\text{p}}}^2\psi^2\nonumber\\
	&\ \ +\frac{3}{8}(\alp{4}+\alp{6})\phi^4+\frac{3}{8}(\alp{4}+\alp{6})\psi^4\nonumber\\
	&\ \ -\frac{3}{4}\big((\alp{4}+\alp{6})+2(\alp{2}+\alp{3})\big)\phi^2\psi^2\\
	J_\mu&\equiv\big[(\alp{2}-\alp{3})-(\alp{4}-\alp{6})\big]\psi^3\tensor{\nabla}{_\mu}(\phi/\psi)\nonumber\\
	&\ \ -(\alpha_0+2\bet{2}){m_{\text{p}}}^2\tensor{\nabla}{_\mu}\phi,
    \end{align}
    \label{final}%
  \end{subequations}
  perfectly replicates the dynamics of the FRW background. We note that the Ricci scalar $R$ is derived from the Riemann curvature as defined in \eqref{riemann}, and while the biscalar-tensor theory is strictly torsion-free, the behaviour of the $0^+$ and $0^-$ modes is accurately replicated by the scalars. The theory \eqref{final} is known as the \emph{metrical analogue} (MA) of \eqref{lagrangian_soft}, and we have translated it here into the dimensionless couplings \eqref{diction} of \eqref{lagrangian}.

  We will restrict our attention to the massless theories. Following a reparameterisation to the weightless scalar ${\zeta=\sqrt{2}\phi/\psi}$, we find the MA of \criticalcase{3} becomes
\begin{equation}
  \begin{aligned}
  \tensor{L}{_{\text{G}}}\simeq-&3\alp{3}\tensor{X}{^{\psi\psi}}-\frac{1}{4}\alp{3}\psi^2 R+3\bet{1}{m_{\text{p}}}^2\psi^2\\
  +&\alp{3}\psi^3\sqrt{\big|\tensor{X}{^{\zeta\zeta}}\big|}-\frac{3}{4}\alp{3}\zeta^2\psi^4.
  \label{case3_ma}
\end{aligned}
\end{equation}
In this frame, we see that the MA can be partitioned into two. The first three terms in \eqref{case3_ma} describe a massive but \emph{conformally coupled} scalar $\psi$. The fourth and fifth terms describe a \emph{quadratic cuscuton} $\zeta$, which is conformally coupled by multiplication with the appropriate powers of $\psi$.

The quadratic cuscuton is itself remarkable for replicating the cosmological background of the Einstein--Hilbert term~\cite{2007PhRvD..75l3509A}
\begin{equation}
  c_1{m_\text{p}}^3\sqrt{\big|\tensor{X}{^{\zeta\zeta}}\big|}-c_2 {m_\text{p}}^4\zeta^2\simeq \frac{3{c_1}^2}{16c_2} {m_{\text{p}}}^2  R,
  \label{simple}
\end{equation}
This unlikely-looking relation may be verified by substituting the $\zeta$-equation into the $\tensor{g}{_{\mu\nu}}$-equation on the LHS of \eqref{simple}, and comparing with the Friedmann equations that follow from the $\tensor{g}{_{\mu\nu}}$-equation on the RHS. We find that the bizarre characteristics of the cuscuton can be taken further: when we replace the Planck mass with a dynamical scalar to obtain the conformally coupled quadratic cuscuton, we replicate the cosmological background of the same scalar, conformally coupled to gravity
\begin{equation}
  c_1\psi^3\sqrt{\big|\tensor{X}{^{\zeta\zeta}}\big|}-c_2\psi^4\zeta^2\simeq \frac{9{c_1}^2}{4c_2}\Big(\tensor{X}{^{\psi\psi}}+\frac{1}{12}\psi^2 R\Big).
  \label{simple2}
\end{equation}
This result is very satisfying, but has fatal implications for the massless theories under consideration. By applying \eqref{simple2}, we see that the fourth and fifth terms in \eqref{case3_ma} dynamically cancel with the first and second terms: the whole kinetic structure of the analogue theory vanishes! The same problem arises in \criticalcase{17}, since the extra condition $\alp{3}=0$ prevents the cancelling terms from appearing even at the level of \eqref{case3_ma}. In both cases, the gravitational Lagrangian responsible for the cosmological background is a pure $0^-$ mass, and so the theories are \emph{not viable}.

Notwithstanding the complete failure of the cases at hand, the result \eqref{simple2} suggests an interesting class of theories, of which \criticalcase{3} is a degenerate special case.
From the general quadratic torsion theory \eqref{lagrangian} we impose
\begin{equation}
  \alpha_0+2\bet{2}=\alp{4}+\alp{6}=0,
  \label{starob}
\end{equation}
noting from \eqref{PiCA0p} that the second constraint in \eqref{starob} results in the single $0^+$ PiC $\pic{A0p}\approx 0$. The cosmological analogue then becomes
\begin{equation}
  \begin{aligned}
    L_\text{G}&\simeq-\frac{1}{2}\alpha_0\planck^2R-3(\alp{2}+\alp{3})\Big(\tensor{X}{^{\psi\psi}}+\frac{1}{12}\psi^2R\Big)\\
    &\ \ \ +\big[(\alp{2}-\alp{3})-(\alp{4}-\alp{6})\big]\psi^3\sqrt{\big|\tensor{X}{^{\zeta\zeta}}\big|}\\
    &\ \ \ -\frac{3}{4}(\alp{2}+\alp{3})\zeta^2\psi^4-\frac{3}{4}(\alpha_0+8\bet{3})\planck^2\psi^2.
\end{aligned}
  \label{<+label+>}
\end{equation}
The interpretation of the first equality of \eqref{starob} is now clear: it forces the Einstein--Hilbert term to appear equally both in the torsion theory and the cosmological background analogue\footnote{Note that this is not \emph{generally} guaranteed for general choices of the coupling constants, as discussed in \cite{2020arXiv200603581B}.}. We can then set ${\alpha_0=1}$ to view these theories as additive modifications to the Einstein--Cartan or Einstein--Hilbert theories, respectively. In order to apply \eqref{simple2}, we will strictly require that ${\alp{2}+\alp{3}\neq0}$, i.e. that the $0^-$ mode is not primarily constrained according to \eqref{PiCA0m}. Under an appropriate rescaling of $\psi$ to $\xi$, the cosmological background becomes
\begin{equation}
  \tensor{L}{_{\text{G}}}\simeq-\frac{1}{2}{m_{\text{p}}}^2 R+\frac{1}{12}\xi^2 R+\tensor{X}{^{\xi\xi}}-\frac{1}{2}{m_\xi}^2\xi^2,
  \label{finalo}
\end{equation}
i.e. Einstein's gravity conformally coupled to a scalar $\xi$, whose mass is
\begin{equation}
  m_\xi^2\equiv-\frac{(1+8\bet{3})(\alp{2}+\alp{3})}{8(\alp{3}+\alp{4})(\alp{2}+\alp{6})}\planck^2.
  \label{supermass}
\end{equation}
The theory \eqref{finalo} is of course widely studied in the context of inflation~\cite{1993GReGr..25..855M,2013Galax...1...96F}. In Einstein's theory, a non-minimal scalar coupling will tend to run, with the conformal value of $1/12$ being a fixed point in the IR. This value is also used to preserve causality in a curved background, since it prevents a massive scalar from propagating along the light cone. We have shown that the cosmological background of the conformal scalar emerges as a consequence of the minimal constraints \eqref{starob} on the quadratic torsion theory, where the scalar is interpreted as the $0^-$ part of the torsion, and the $0^+$ part is primarily constrained.

We see also from \eqref{supermass} that the effect of the conformally coupled $0^-$ can be removed from the expansion history altogether. By setting ${\alp{3}+\alp{4}=0}$ or ${\alp{2}+\alp{6}=0}$, the mass $m_\xi$ becomes infinite and one is left with the cosmological background of the pure Einstein gravity in \eqref{GR}. By inspecting \crefrange{PiCA0p}{PiCA2m}, we see that these choices can be imposed without primarily constraining the torsion modes in the general theory, including the $0^-$ mode. This raises the interesting question of whether torsion theories allow the cosmological background to be altered independently of the the perturbations.
Note that \criticalcase{3} has just such a divergent mass, though the Einstein--Hilbert term never appears in the background analogue because of the universal constraint \eqref{bizarre} that appears to be required for power-counting renormalisability.

\section{Conclusions}\label{conclusions}
It was recently shown that among all parity-preserving theories of the form $L_\text{G}\sim\planck^2\mathcal{R}+\mathcal{R}^2+\planck^2\mathcal{T}^2$, there are 58 cases which are unitary and power-counting renormalisable when linearised in the absence of source currents~\cite{2019PhRvD..99f4001L,2020PhRvD.101f4038L}. The linearisation was done around a presumed Minkowski vacuum, since even a cosmological constant is excluded as a source. In this work we have considered \criticalcase{3}, \criticalcase{17}, \criticalcase{20}, \criticalcase{24}, \criticalcase{25}, \criticalcase{26}, \criticalcase{28} and \criticalcase{32} -- as detailed in \cref{table-1}. We have inspected their Hamiltonian structure under the same conditions, but in both the linear and \emph{nonlinear} regimes. Our principal findings may be summarised as follows;
\begin{enumerate}
  \item All eight cases (and indeed all the cases proposed in~\cite{2019PhRvD..99f4001L,2020PhRvD.101f4038L}) feature vanishing mass parameters. This greatly complicates the Hamiltonian analysis, compared to the `minimal' cases previously treated in the literature.
  \item The number of linear, propagating degrees of freedom are confirmed from~\cite{2019PhRvD..99f4001L,2020PhRvD.101f4038L} for all eight cases.
  \item With the exception of \criticalcase{17}, all eight cases linearly propagate a massive pseudoscalar mode, and the unitarity conditions from~\cite{2019PhRvD..99f4001L,2020PhRvD.101f4038L} correspond to the no-ghost and no-tachyon conditions on this mode.
  \item The two massless modes propagated by \criticalcase{3} and \criticalcase{17} are identified with \emph{vector}, rather than the hoped-for \emph{tensor} modes.
  \item With the possible exception of \criticalcase{20} and \criticalcase{32}, all eight cases feature primary constraints which transition from first to second class when moving to the nonlinear regime. This signals at least a broken gauge symmetry, and possibly acausal behaviour and/or activation of any of the primarily unconstrained spin-parity sectors.
  \item These primarily unconstrained spin-parity sectors include ghosts in all eight cases, according to the same conditions that ensure linearised unitarity.
  \item \criticalcase{3} and \criticalcase{17} are not viable theories of gravity despite their massless modes, because they do not support a dynamical FRW background.
\end{enumerate}

These findings come with various caveats. Principally, while we implement the linearised Dirac--Bergmann algorithm to completion in all cases, we do not prosecute the nonlinear algorithm beyond the second set of links in the constraint chains. This level of analysis at least matches the earlier treatment of less complicated theories, in which all couplings are set to zero except those absolutely necessary to propagate whichever mode is under investigation~\cite{2002IJMPD..11..747Y}. Consequently, we cannot say for certain if the strongly coupled sectors and the ghost sectors coincide. 

Separately, our definition of ghost sectors as set out in \cref{sighost} is based on the relevant quadratic momenta appearing as negative contributions to the Hamiltonian. We do not go so far as to quantise the theory and confirm that there are corresponding \emph{physical} states which violate the unitarity of the S-matrix. Additional steps would presumably be required to draw completely safe conclusions, such as adding terms to fix the Poincar{\'e} gauge (and any other case-specific symmetries), and good ghosts to cancel the anomalies~\cite{Henneaux:1992ig}. Meanwhile at the classical level, we mention that negative kinetic energy does not always imply instability. 

We have also interpreted acausal behaviour, which is linked to the phenomenon of constraint bifurcation or field-dependent constraint structure~\cite{1998AcPPB..29..961C}, as a pathology. This need not always follow, as has been demonstrated for some special theories in recent decades~\cite{2009PhRvD..79d3525M}. For example, the characteristic surface of a degree of freedom is allowed to lie outside the light cone if it can be shown that the field does not carry information~\cite{2007PhRvD..75h3513A}.

Even bearing these caveats in mind, the outlook for the remaining new torsion theories is not substantially improved by our results. Of the 58 novel theories in~\cite{2019PhRvD..99f4001L,2020PhRvD.101f4038L}, only 19 propagate the two massless degrees of freedom. Four of these additionally propagate a massive $0^-$ mode, while three instead propagate a massive $2^-$ mode. Of the remaining theories, 23 propagate only a massive $0^-$ mode. The selection in \cref{table-1} thus appears reasonably representative of the linearised particle spectra. Since fundamental changes to the constraint structure are observed throughout most of the sample, we do not find new cause for optimism in the current study.
Possibly, the admission of primary constraints dependent on the Riemann--Cartan curvature will miraculously remedy the various problems. Certainly, such constraints will complicate the analysis. We have already seen in \cref{massless_theories} that field-dependent primary constraints can invoke derivatives of the equal-time Dirac function. Ultimately, our findings are consistent with the predictions of Yo and Nester, who anticipate that generalising the quadratic torsion theory \eqref{lagrangian_soft} beyond very minimal test cases (most of which also fail) serves only to protract the calculations~\cite{1999IJMPD...8..459Y,2002IJMPD..11..747Y}. Even so, it might seem prudent to attempt to quantify the chances of future success: we provide a heuristic discussion along these lines in \cref{outlook}.

The tentative vector nature of the massless modes in \criticalcase{3} and \criticalcase{17} is potentially problematic. We recall that Poincar{\'e} invariance prohibits a matter amplitude involving soft gravitons of spin $J>2$, while $J=0$ gravitons are ruled out by matter coupling~\cite{VanDam:1974te}. Odd $J$ are supposed to give rise to \emph{repulsive} long-range forces, leading to the expectation of a tensor graviton~\cite{blagojevic2002gravitation}. 
Plausibly, the $J^P$ character will be gauge dependent, but it is difficult to see how this might change the sign of the Green's function.
We will not speculate as to whether this troubling feature is generic to the remaining massless cases.

Finally, we observed that the theories with massless modes could be written off instantly using the scalar-tensor analogue theory which replicates the background cosmology. As a by-product, our analysis suggested an interesting new class of quadratic torsion theories which mimic the background of the conformal inflaton, though not motivated by unitarity or renormalisability. It must be emphasised that the catastrophic failure of \criticalcase{3} and \criticalcase{17} is \emph{not} common to the remaining theories in~\cite{2019PhRvD..99f4001L,2020PhRvD.101f4038L}. We mention in particular \criticalcase{2}, which propagates two massless modes and the massive pseudoscalar, and \criticalcase{16}, a special case in which the pseudoscalar is non dynamical. These theories form a complementary pair to \criticalcase{3} and \criticalcase{17} in many respects, but they have an \emph{excellent} cosmological background. Not only does the cuscuton force the evolution towards a flat Friedmann solution, but the option exists to tune the early expansion history through an effective dark radiation component~\cite{2020arXiv200302690B}. Moreover in \criticalcase{2} the mass of the propagating pseudoscalar acts as a dark energy term (albeit hierarchical, i.e. not resolving the cosmological constant problem)~\cite{2020arXiv200603581B}. Other exact solutions to \criticalcase{2} and \criticalcase{16} include the Schwarzschild vacuum and plane gravitational waves. These cases call for a more dedicated Hamiltonian analysis, and will be among the remaining theories to be addressed in the companion paper.

\begin{acknowledgments}
  We are grateful to Emine \c{S}eyma Kutluk and Wei Chen Lin for very profitable conversations, and also to Yun-Cherng Lin for valuable computational advice. We would also like to thank Ignacy Sawicki for detailed and helpful correspondence. This manuscript was improved by the kind suggestions of Amel Durakovi\'{c}. WEVB is supported by the Science and Technology Facilities Council -- STFC under Grant ST/R504671/1, and WJH by a Royal Society University Research Fellowship.
\end{acknowledgments}

\appendix

\section{Irreducible decomposition of the fields}\label{irreducible_decomposition}
It is necessary to construct a complete set of idempotent and orthogonal projection operators for the irreducible parts of the field strengths. For general tensors, this can be done with the appropriate $\mathrm{O}(3)$ Young tableaux, following the methods of~\cite{1989gtap.book.....H}.
The three projections of the torsion are
\begin{subequations}
  \begin{align}
    \tensor[^1]{\mathcal{  P}}{_{ijk}^{mnq}}\tensor{\caligT}{_{mnq}}&=\frac{2}{3}\tensor{\caligT}{_{ijk}}+\frac{2}{3}\tensor{\caligT}{_{[j|i|k]}}+\frac{2}{3}\tensor{\eta}{_{i[j}}\tensor{\caligT}{_{k]}},\\
    \tensor[^2]{\mathcal{  P}}{_{ijk}^{mnq}}\tensor{\caligT}{_{mnq}}&=-\frac{2}{3}\tensor{\eta}{_{i[j}}\tensor{\caligT}{_{k]}},\\
    \tensor[^3]{\mathcal{  P}}{_{ijk}^{mnq}}\tensor{\caligT}{_{mnq}}&=\frac{1}{6}\tensor{\epsilon}{_{ijkl}}\tensor{\epsilon}{^{lmnq}}\tensor{\caligT}{_{mnq}}.
  \end{align}
\end{subequations}
The six projections of the Riemann--Cartan curvature are
\begin{subequations}
  \begin{align}
    \tensor[^1]{\mathcal{  P}}{_{ijkl}^{mnqp}}\tensor{\caligR}{_{mnqp}}&=\frac{1}{3}\tensor{\caligR}{_{ijkl}}+\frac{1}{3}\tensor{\caligR}{_{klij}}+\frac{2}{3}\tensor{\caligR}{_{[i\rVert[k|\lVert j]|l]}}\nonumber\\
    -\tensor{\eta}{_{[i|[k\rVert}}\tensor{\caligR}{_{|j]\lVert l]}}&-\tensor{\eta}{_{[i|[k\rVert}}\tensor{\caligR}{_{|l]\lVert j]}}+\frac{1}{3}\tensor{\eta}{_{i[k|}}\tensor{\eta}{_{j|l]}}\caligR,\\
    \tensor[^2]{\mathcal{  P}}{_{ijkl}^{mnqp}}\tensor{\caligR}{_{mnqp}}&=\frac{1}{2}\tensor{\caligR}{_{ijkl}}-\frac{1}{2}\tensor{\caligR}{_{klij}}-\tensor{\eta}{_{[i|[k\rVert}}\tensor{\caligR}{_{|j]\lVert l]}}\nonumber\\
    &\ \ \ +\tensor{\eta}{_{[i|[k\rVert}}\tensor{\caligR}{_{|l]\lVert j]}},\\
    \tensor[^3]{\mathcal{  P}}{_{ijkl}^{mnqp}}\tensor{\caligR}{_{mnqp}}&=-\frac{1}{24}\tensor{\epsilon}{_{ijkl}}\tensor{\epsilon}{^{mnop}}\tensor{\caligR}{_{mnop}},\\
    \tensor[^4]{\mathcal{  P}}{_{ijkl}^{mnqp}}\tensor{\caligR}{_{mnqp}}&=\tensor{\eta}{_{[i|[k\rVert}}\tensor{\caligR}{_{|j]\lVert l]}}+\tensor{\eta}{_{[i|[k\rVert}}\tensor{\caligR}{_{|l]\lVert j]}}\nonumber\\
    &\ \ \ -\frac{1}{2}\tensor{\eta}{_{i[k|}}\tensor{\eta}{_{j|l]}}\caligR,\\
    \tensor[^5]{\mathcal{  P}}{_{ijkl}^{mnqp}}\tensor{\caligR}{_{mnqp}}&=\tensor{\eta}{_{[i|[k\rVert}}\tensor{\caligR}{_{|j]\lVert l]}}-\tensor{\eta}{_{[i|[k\rVert}}\tensor{\caligR}{_{|l]\lVert j]}},\\
    \tensor[^6]{\mathcal{  P}}{_{ijkl}^{mnqp}}\tensor{\caligR}{_{mnqp}}&=\frac{1}{6}\tensor{\eta}{_{i[k|}}\tensor{\eta}{_{j|l]}}\caligR.
  \end{align}
\end{subequations}
Replicating the numbering used in~\cite{2002IJMPD..11..747Y,1999IJMPD...8..459Y}, our original na\"ive couplings in \eqref{lagrangian_soft} are expressible in terms of their more meaningful irreducible counterparts according to
\begin{equation}
  \begin{gathered}
    \alpha_1\equiv\alp{4}+\frac{1}{2}\alp{5}+\alp{6}, \quad \alpha_2\equiv\alp{4}-\alp{6},\\
    \alpha_3\equiv\alp{4}-\alp{5}+\alp{6},\\
    \alpha_4\equiv\frac{1}{2}\alp{2}+\frac{1}{2}\alp{3}+\alp{4}+\frac{1}{2}\alp{5}+\alp{6},\\
    \alpha_5\equiv\frac{1}{2}\alp{2}-\frac{1}{2}\alp{3}+\alp{4}-\alp{6},\\
    \alpha_6\equiv 6\alp{1}+\frac{3}{2}\alp{2}+\frac{3}{2}\alp{3}+\alp{4}+\frac{1}{2}\alp{5}+\alp{6},\\
    \beta_1\equiv\bet{1}+\frac{1}{2}\bet{2}, \quad \beta_2\equiv\bet{1}+\frac{1}{2}\bet{2}+\frac{3}{2}\bet{3},\\
    \beta_3\equiv\bet{1}-\bet{2},
  \end{gathered}
  \label{diction}
\end{equation}
where $\alp{I}$ and $\bet{I}$ multiply the $I$th irreducible quadratic invariants of curvature and torsion in \eqref{lagrangian}.

\section{Ghosts, ranks and signatures}\label{sighost}

In this appendix, we attempt to elaborate on the motivation of the `positive kinetic energy test', which was tacitly employed in the previous Hamiltonian treatment of Poincar{\'e} gauge theories~\cite{2002IJMPD..11..747Y}.

Consider the free, vector $\mathrm{U}(1)$ theory on $\check{\mathcal{  M}}$, without any coupling to gravity (and with Cartesian coordinates ${\tensor{\gamma}{_{\mu\nu}}\equiv\tensor{\eta}{_{\mu\nu}}}$), fixed to the Feynman gauge
\begin{equation}
  L=-\frac{1}{4}\tensor{F}{^{\mu\nu}}\tensor{F}{_{\mu\nu}}-\frac{1}{2}(\tensor{\partial}{_\mu}\tensor{A}{^\mu})^2,
  \label{feynman}
\end{equation}
where we have $\tensor{F}{_{\mu\nu}}\equiv 2\tensor{\partial}{_{[\mu}}\tensor{A}{_{\nu]}}$. Up to a surface term, \eqref{feynman} is of course equivalent to
\begin{equation}
  L=-\frac{1}{2}\tensor{\partial}{_\mu}\tensor{A}{_\nu}\tensor{\partial}{^\mu}\tensor{A}{^\nu},
  \label{feynman_equiv}
\end{equation}
which safely propagates four massless polarisations, without developing any classical instability
\begin{equation}
  \Box \tensor{A}{_\mu}=0.
  \label{shellu}
\end{equation}
Notwithstanding this reasonable behaviour, we see that the Hamiltonian of \eqref{feynman_equiv} is unbounded from below
\begin{equation}
  \mathcal{  H}=-\frac{1}{2}\tensor{\pi}{_\mu}\tensor{\pi}{^\mu}+\frac{1}{2}\tensor{\partial}{_\alpha}\tensor{A}{_\mu}\tensor{\partial}{^\alpha}\tensor{A}{^\mu},
  \label{mixed_ham}
\end{equation}
where the  momentum is ${\tensor{\pi}{_\mu}\equiv-\tensor{\partial}{_0}\tensor{A}{_\mu}}$, since the independent timelike polarisation will have a strictly \emph{negative} contribution.
This is naturally revealed in the $3+1$ picture, which we construct by defining a constant unit timelike normal $\tensor{n}{^\mu}\tensor{n}{_\mu}\equiv 1$, and (extending our previous overbar notation to holonomic indices) decomposing quantities into the $0^+$ and $1^-$ irreps
\begin{equation}
  \tensor{A}{_\mu}\equiv \tensor{A}{_\perp}\tensor{n}{_\mu}+\tensor{A}{_{\ovl{\mu}}}, \quad\tensor{\pi}{_\mu}\equiv \tensor{\pi}{_\perp}\tensor{n}{_\mu}+\tensor{\pi}{_{\ovl{\mu}}}.
  \label{<+label+>}
\end{equation}
The Hamiltonian then separates into 
\begin{equation}
  \begin{aligned}
  \mathcal{  H}&=-\frac{1}{2}\tensor{\pi}{_\perp}^2+\frac{1}{2}\tensor{\partial}{_\alpha}\tensor{A}{_\perp}\tensor{\partial}{^\alpha}\tensor{A}{_\perp}\\
  &\ \ \ -\frac{1}{2}\tensor{\pi}{_{\ovl{\mu}}}\tensor{\pi}{^{\ovl{\mu}}}+\frac{1}{2}\tensor{\partial}{_\alpha}\tensor{A}{_{\ovl{\mu}}}\tensor{\partial}{^\alpha}\tensor{A}{^{\ovl{\mu}}},
  \label{hamsep}
\end{aligned}
\end{equation}
where the first and last pairs of terms are respectively negative and positive-definite on the null shell defined by \eqref{shellu}.
The physical consequence is a loss of unitary: the timelike states have negative norm. In the $\mathrm{U}(1)$ theory, this is usually fixed by imposing a Gupta--Bleuler condition on the physical states, which is acceptable since the gauge-fixing term in \eqref{feynman} was added by hand anyway. However, in the theories of gravity under consideration, the validity of a Gupta--Bleuler condition is not certain. We note that in the kinetic Hamiltonia of \cref{ham26,ham28,ham25,ham24,ham32,ham20,ham17}, we encounter mixed quadratic forms in the momenta, just as we do with the first and third terms of \eqref{hamsep}. If such terms are negative-definite and propagating, we tentatively identify them with a loss of unitarity. We note that without full knowledge of both the nonlinear shell and the remaining field parts of the Hamiltonian (c.f. second and fourth terms in \eqref{hamsep}), this is quite dangerous. Moreover, as is evident from \eqref{shellu}, such negative-energy sectors do not necessarily correspond to classical ghosts. 

We also mention that the sign of quadratic momenta in the $3+1$ formulation is robust against the choice of signature (as indeed it should be). Recall that throughout this article we have used the `West Coast' signature $(+,-,-,-)$.
The sign of each such term may then be inferred by the tensor rank of the momentum irrep, since every contraction on parallel indices introduces a factor of $-1$. Had we chosen the `East Coast' signature $(-,+,+,+)$, these factors would not arise. Instead, we would have $\tensor{n}{^\mu}\tensor{n}{_\mu}\equiv -1$, whose powers would conspire in the $\mathrm{O}(3)$ decomposition of momenta to have the same effect up to an overall sign in the kinetic Hamiltonian. This final sign is changed by hand in the kinetic part of the Lagrangian, as is customary when changing signature.

\section{Nonlinear Poisson brackets}\label{brackets}
\paragraph*{\criticalcase{28}}
In \crefrange{comm1}{comm2} we provide the nonlinear commutators of \criticalcase{26}. In this appendix we list the emergent commutators of the other seven theories under consideration.
The commutators of \criticalcase{28} read
\begin{subequations}
  \begin{align}
    \Big\{\pic[\ovl{i}]{B1m},\pic[\ovl{l}]{B1m}\Big\}&\approx \text{RHS of \eqref{comm1}},\\
    \Big\{\pic[\ovl{i}]{B1m},\pic{A0p}\Big\}&\approx-\frac{1}{J^2}\PiP[\ovl{i}]{A1m}\delta^3,\\
    \Big\{\pic[\ovl{i}]{B1m},\pic[\ovl{lm}]{A2p}\Big\}&\approx\frac{1}{2J^2}\etad{\ovl{i}{\langle}\ovl{l}}\PiP[\ovl{m}{\rangle}]{A1m}\delta^3,\\
    \Big\{\pic[\ovl{i}]{B1m},\pic[\ovl{lmn}]{A2m}\Big\}&\approx\frac{1}{2J^2}\bigg[\etad{\ovl{in}}\PiP[\ovl{lm}]{A1p}-\frac{1}{2}\etad{\ovl{i}{[}\ovl{l}|}\PiP[|\ovl{m}{]}\ovl{n}]{A1p}\nonumber\\
    &\ \ \ \ \ \ \ \ \ \ \ \ -\frac{3}{4}\etad{{[}\ovl{l}|\ovl{n}}\PiP[\ovl{i}|\ovl{m}{]}]{A1p}\bigg]\delta^3,\\
    \Big\{\pic[\ovl{ij}]{B2p},\pic[\ovl{lm}]{B2p}\Big\}&\approx \text{RHS of \eqref{comm1b}},\\
    \Big\{\pic[\ovl{ij}]{B2p},\pic[\ovl{lm}]{A2p}\Big\}&\approx \text{RHS of \eqref{comm1b}},\\
    \Big\{\pic[\ovl{ij}]{B2p},\pic[\ovl{lmn}]{A2m}\Big\}&\approx\frac{1}{J^2}\bigg[\frac{1}{12}\epsd{\langle\ovl{i}|{[}\ovl{l}\rVert\ovl{n}}\etad{|\ovl{j}\rangle\lVert\ovl{m}{]}}\PiP{A0m}\nonumber\\
      +\frac{1}{12}&\epsd{\langle\ovl{i}|\ovl{lm}}\etad{|\ovl{j}\rangle\ovl{n}}\PiP{A0m}+\frac{3}{8}\etad{\langle\ovl{i}|{[}\ovl{l}}\etad{\ovl{m}{]}\ovl{n}}\PiP[|\ovl{j}\rangle]{A1m}\nonumber\\
    &\ \ \ \ \ \ \ \ \ \ -\frac{3}{4}\etad{\langle\ovl{i}|\ovl{n}}\etad{|\ovl{j}\rangle{[}\ovl{l}}\PiP[\ovl{m}{]}]{A1m}\bigg]\delta^3.\label{symmimpose}
  \end{align}
\end{subequations}
In the RHS of \eqref{symmimpose}, we see that the linearly propagating $\PiP{A0m}$ appears, signalling a definite change in the constraint structure when passing from linear to nonlinear regimes.
\paragraph*{\criticalcase{25}}
The nonlinear commutators of \criticalcase{25} have been encountered before
\begin{subequations}
  \begin{align}
    \Big\{\pic[\ovl{ij}]{B2p},\pic[\ovl{lm}]{B2p}\Big\}&\approx \text{RHS of \eqref{comm1b}},\\
    \Big\{\pic[\ovl{ij}]{B2p},\pic[\ovl{lmn}]{A2m}\Big\}&\approx \text{RHS of \eqref{comm2}}.\label{comm_25_bad}
  \end{align}
\end{subequations}
Again we see that at least \eqref{comm_25_bad} is expected to persist on the final shell.
\paragraph*{\criticalcase{24}}
Similarly for \criticalcase{24} we find 
\begin{subequations}
  \begin{align}
    \Big\{\pic[\ovl{ij}]{B2p},\pic[\ovl{lm}]{B2p}\Big\}&\approx \text{RHS of \eqref{comm1b}},\\
  \Big\{\pic[\ovl{ij}]{B2p},\pic[\ovl{lm}]{A2p}\Big\}&\approx \frac{1}{J^2}\etad{(\ovl{i}\rVert{(}\ovl{l}|}\PiP[\lVert\ovl{j})|\ovl{m}{)}]{B1p}\delta^3,\\
  \Big\{\pic[\ovl{ij}]{B2p},\pic[\ovl{lmn}]{A2m}\Big\}&\approx\text{RHS of \eqref{symmimpose}}.\label{comm_24_bad}
  \end{align}
\end{subequations}
Again we see that at least \eqref{comm_24_bad} is expected to persist on the final shell.
\paragraph*{\criticalcase{3}}
The nonlinear commutators of \criticalcase{3} are all new
\begin{subequations}
  \begin{align}
    \Big\{\pic{B0p},\pic[\ovl{lm}]{B1p}\Big\}&\approx \frac{1}{J^2}\PiP[\ovl{lm}]{B1p}\delta^3,\label{comm3a}\\
    \Big\{\pic[\ovl{ij}]{B1p},\pic{A0p}\Big\}&\approx -\frac{1}{J^2}\PiP[\ovl{ij}]{A1p}\delta^3,\label{comm3b}\\
    \Big\{\pic[\ovl{ij}]{B1p},\pic[\ovl{lm}]{A2p}\Big\}&\approx \frac{1}{J^2}\etad{{[}\ovl{i}\rVert{\langle}\ovl{l}|}\PiP[\lVert\ovl{j}{]}|\ovl{m}{\rangle}]{A1p}\delta^3,\label{comm3c}\\
    \Big\{\pic[\ovl{ij}]{B1p},\pic[\ovl{lmn}]{A2m}\Big\}&\approx\frac{1}{J^2}\bigg[\frac{1}{12}\epsd{ {[}\ovl{i}|{[}\ovl{l}\rVert\ovl{n}}\etad{|\ovl{j}{]}\lVert\ovl{m}{]}}\PiP{A0m}\nonumber\\
      +\frac{1}{12}&\epsd{ {[}\ovl{i}|\ovl{lm}}\etad{|\ovl{j}{]}\ovl{n}}\PiP{A0m}-\frac{1}{8}\epsd{\ovl{ij}{[}\ovl{l}}\etad{\ovl{m}{]}\ovl{n}}\PiP{A0m}\nonumber\\
      -\frac{3}{8}&\etad{ {[}\ovl{i}|{[}\ovl{l}}\etad{\ovl{m}{]}\ovl{n}}\PiP[|\ovl{j}{]}]{A1m}-\frac{1}{4}\etad{ {[}\ovl{i}|\ovl{n}}\etad{|\ovl{j}{]}{[}\ovl{l}}\PiP[\ovl{m}{]}]{A1m}\nonumber\\
  &\ \ \ \ \ \ \ \ \ \ +\frac{1}{4}\etad{\ovl{i}{[}\ovl{l} }\etad{\ovl{m}{]}\ovl{j} }\PiP[\ovl{n}]{A1m} \bigg]\delta^3.\label{comm3d}
  \end{align}
\end{subequations}
Since \eqref{comm3d} also depends on $\PiP{A0m}$, we believe that it will also persist on the final shell. Note that \eqref{comm3d} is also linear in $\PiP{A1m}$, which we suspect will contribute the massless modes in the linear theory.
\paragraph*{\criticalcase{17}}
The nonlinear commutators of \criticalcase{17} are of course mostly the same as \criticalcase{3} 
\begin{subequations}
  \begin{align}
    \Big\{\pic{B0p},\pic[\ovl{lm}]{B1p}\Big\}&\approx \text{RHS of \eqref{comm3a}},\\
    \Big\{\pic[\ovl{ij}]{B1p},\pic{A0p}\Big\}&\approx \text{RHS of \eqref{comm3b}},\\
    \Big\{\pic[\ovl{ij}]{B1p},\pic{A0m}\Big\}&\approx -\frac{1}{J^2}\etau{\ovl{kl}}\epsd{\ovl{ijk}}\PiP[\ovl{l}]{A1m}\delta^3,\\
    \Big\{\pic[\ovl{ij}]{B1p},\pic[\ovl{lm}]{A2p}\Big\}&\approx \text{RHS of \eqref{comm3c}},\\
    \Big\{\pic[\ovl{ij}]{B1p},\pic[\ovl{lmn}]{A2m}\Big\}&\approx-\frac{1}{J^2}\bigg[\frac{3}{8}\etad{ {[}\ovl{i}|{[}\ovl{l}}\etad{\ovl{m}{]}\ovl{n}}\PiP[|\ovl{j}{]}]{A1m}\nonumber\\
      +\frac{1}{4}\etad{ {[}\ovl{i}|\ovl{n}}&\etad{|\ovl{j}{]}{[}\ovl{l}}\PiP[\ovl{m}{]}]{A1m}-\frac{1}{4}\etad{\ovl{i}{[}\ovl{l} }\etad{\ovl{m}{]}\ovl{j} }\PiP[\ovl{n}]{A1m} \bigg]\delta^3.\label{comm_17_bad}
  \end{align}
\end{subequations}
Note that \eqref{comm_17_bad} is linear in $\PiP{A1m}$, the momentum of the vector graviton.

\section{Heuristic outlook}\label{outlook}
In this appendix we attempt to quantify the chances of future success, in light of our present results.
Let $k$ viable theories be found in a sample of $n=8$, drawn from a population of $N=58$ theories. We may model the probability of there being a grand total of $K$ viable theories in the parent population as
\begin{equation}
  P(K|k,n,N)\equiv \frac{n+1}{N+1}P_{\text{hyp}}(k|K,n,N),
  \label{handley}
\end{equation}
where the probability $P_{\text{hyp}}(k|K,n,N)$ of drawing $k$ given $K$ follows the standard hypergeometric distribution
\begin{equation}
  P_{\text{hyp}}(k|K,n,N)\equiv \frac{{{K}\choose{k}}{{N-K}\choose{n-k}}}{{{N}\choose{n}}}.
\end{equation}
Note that we have assumed a uniform prior on $K$, ${P(K|N)\equiv (N+1)^{-1}}$, which may or may not be justified. The pessimistic interpretation of our study would be $k=0$, but in that case the probability that $K=0$ is found to be only $0.15$ according to \eqref{handley}. Rather, we would expect $K=5\pm 4.9$. Moreover, the pessimistic interpretation is not necessarily the most conservative, since \criticalcase{20} and \criticalcase{32} are not ruled out at the level of the PPM: we would expect $K=11\pm 6.6$ and $K=17\pm 7.6$ for $k=1$ and $k=2$ respectively. This outlook is more promising, but still assumes a uniform prior which might be improved by considering the methods used to obtain the cases, from a theoretical perspective. In any case, it is clear that further study of the remaining theories will be necessary to draw firm conclusions.

\bibliographystyle{apsrev4-1}
\bibliography{bibliography}

\end{document}